\documentclass[preprint,aps]{revtex4}
\usepackage{amsmath}
\numberwithin{equation}{section}
\usepackage{amsmath}
\usepackage{amssymb}
\usepackage{stmaryrd}

\usepackage[dvipdfmx]{graphicx}
\usepackage{bm}
\usepackage{color}
\usepackage{braket}
\begin{document}

\title{Theory of proximity effect in $s+p$-wave superconductor junctions}
\author{Yukio Tanaka$^{1}$, Tim Kokkeler$^{2,3}$ 
and Alexander Golubov$^{2}$}
\affiliation{$^1$Department of Applied Physics, 
Nagoya University, Nagoya, 464-8603,
Japan \\ 
$^2$ MESA+ Institute for Nanotechnology, University of Twente, 7500 AE Enschede, The Netherlands \\
$^3$
Donostia International Physics Center (DIPC), 20018 Donostia-San Sebastian, 
Spain }
\date{\today}
\begin{abstract}
We derive a boundary condition for the Nambu Keldysh Green's function 
in diffusive normal metal / unconventional superconductor junctions 
applicable for mixed parity pairing. 
Applying this theory to a 1d model of $s+p$-wave superconductor, 
we calculate LDOS in DN and charge conductance of 
DN / $s+p$-wave superconductor junctions. 
When the $s$-wave component of the pair potential 
is dominant, LDOS has a gap like structure at zero energy 
and the dominant pairing in DN is even-frequency spin-singlet 
$s$-wave. On the other hand, when the $p$-wave component is dominant, 
the resulting LDOS has a zero energy peak and 
the dominant pairing in DN is odd-frequency spin-triplet $s$-wave. 
We show the robustness of the quantization of the 
conductance when the magnitude of $p$-wave 
component of the pair potential is larger than 
that of $s$-wave one. These results show the robustness of the 
anomalous proximity effect specific to spin-triplet superconductor junctions. 
\end{abstract}


\maketitle

\section{Introduction}
Superconducting proximity effect is one of the 
most fundamental problems in the physics of superconductivity, 
where a Cooper pair penetrates into normal metal attached to 
superconductor \cite{deGennes1,deGennes2}. 
In diffusive normal metal (DN) / superconductor junctions, 
the total resistance of the junction is seriously influenced by the 
penetrating Cooper pair in DN 
\cite{vanWees,Kastalsky,Larkin1977,Volkov1993,Nazarov1994,Yip1995}. 
This problem has been discussed by quasiclassical Green's function method 
with the Usadel equation \cite{Usadel,kopnin}. 
To calculate charge conductance, the 
boundary condition of the Green's function becomes a key ingredient. 
Kupriyanov and Lukichev (KL) derived a boundary condition (KL boundary condition) for spin-singlet $s$-wave 
superconductor junction with low transmissivity at the interface \cite{KLboundary}. 
The obtained bias voltage dependent charge conductance 
has a distinctive behaviour 
as compared to that by Blonder Tinkham Klapwijk  (BTK) theory \cite{BTK82}
in ballistic junctions. 
Later, KL boundary condition 
was extended by Nazarov by taking account of the mesoscopic ballistic 
region near the interface \cite{Nazarov1999}. 
This theory can reproduce  KL theory and BTK theory as limiting cases. 
The correction to KL boundary condition 
due to finite transparency has been studied 
\cite{Lambert,Laikhtman}. 
\par
In order to extend charge transport theory available for 
unconventional superconductor junctions, 
one of the authors Y.T. $et$ $al.$ have developed a theory for a boundary condition 
(TN boundary condition) of Nambu Keldysh Green's function \cite{Proximityd,Proximityd2}. 
They have calculated the charge conductance between DN / unconventional superconductor junctions both for spin-singlet and spin-triplet superconductors.  
The most remarkable feature of unconventional superconductor is the 
generation of zero energy surface Andreev bound states (ZESABS) due to the sign change of the pair potential on the Fermi surface 
\cite{ABS,Hara,ABSR2,Bruder90,Hu94,TK95,Kashiwaya00}. 
The merit of TN boundary condition is that it can naturally take into account the effect of ZESABS \cite{Proximityd}. 
It has been clarified that ZESABS in spin-singlet $d$-wave superconductor can 
not penetrate into DN. This means that 
proximity effect and ZESABS are competing 
each other in spin-singlet $d$-wave superconductor junctions 
\cite{Proximityd,Proximityd2}. 
On the other hand, in spin-triplet $p$-wave case, 
ZESABS can penetrate into DN and the resulting density of states in DN has a 
zero energy peak (ZEP) \cite{Proximityp,Proximityp2,Proximityp3}. 
This property is by contrast to the conventional proximity effect with spin-singlet $s$-wave superconductor junction, where the quasiparticle density of states has a zero energy gap \cite{Golubov88,Belzig96}. 
In the extreme case, 
if spin-triplet $p$-wave pairing has a $p_{x}$-wave symmetry, where the lobe direction of $p$-wave pair potential is perpendicular to the interface, the total resistance of the junction at zero voltage does not depend on the resistances in DN and that at the interface\cite{Proximityp}. 
In other words, the zero voltage conductance is quantized 
\cite{Asano2013,Ikegaya2016}. 
These exotic feature specific to spin-triplet superconductor junction is 
called the anomalous proximity effect 
\cite{Proximityp,tanaka12,Asano2013,Suzuki2019}. \par

In order to understand the physical origin of the anomalous proximity effect, 
the symmetry of Cooper pair has been elucidated \cite{odd1}. 
It has been understood that the symmetry of the Cooper pair 
in DN is not a spin-singlet $s$-wave but spin-triplet $s$-wave \cite{odd1}. 
The latter symmetry belongs to the so called 
odd-frequency pairing where pair amplitude in DN 
has a sign change with the exchange of time of two electrons forming a Cooper 
pair \cite{Berezinskii,Efetov2,tanaka12,LinderBalatsky,Cayao2020,Triola2020}. 
Near the interface, 
it has been shown that spin-triplet $s$-wave pairing is generated  due to the breakdown of the translational invariance 
\cite{odd3,odd3b,Eschrig2007}. 
The induced  pairing symmetry belongs to the so called 
odd-frequency pairing where pair amplitude  
has a sign change with the exchange of time of two electrons forming a Cooper 
pair \cite{Berezinskii,Balatsky,Balatsky2,Coleman,Vojta,Fuseya,Efetov1,
Efetov2,tanaka12,LinderBalatsky,Cayao2020,Triola2020}. 
This odd-frequency spin-triplet $s$-wave pairing can penetrate into diffusive normal metal by the anomalous proximity effect \cite{odd1}.
Thus, the anomalous proximity effect has a significant importance for the 
condensed matter physics \cite{tanaka12}. 
To detect ZEP of LDOS in DN, T-shaped junction has been proposed 
\cite{Asano2007PRL}. It is noted that there is 
a relevant experimental report detecting a zero bias conductance peak in  
${\mathrm{CoSi}}_{2}/{\mathrm{TiSi}}_{2}$ heterostructures \cite{Lin2021}.  
\par
Although the anomalous proximity effect has originated from the 
TN boundary condition [eq. (2) in \cite{Proximityd}], 
this boundary condition shows a general relation of 
Nambu-Keldysh Green's function at the interface rather symbolically. 
In the actual process to obtain LDOS and charge conductance, we must go through rather long and complicated calculations of retarded and Keldysh part of 
the Green's function as shown in  Ref. \cite{Proximityd2,Proximityp2}.
Only the cases with spin-triplet even-parity or 
spin-singlet odd-parity pair potential have been studied where the 
parity of the superconductor is a good quantum number. 
Recently, one of the authors Y.T. has found that 
eq. (2) of Ref. \cite{Proximityd}
can be expressed more compactly \cite{TextTanaka2021}. 
Then, it becomes more transparent to show the derivation of the LDOS and charge conductance. Also, we can challenge more complicated situation 
where spin-triplet and spin-singlet pair potentials are mixed. \par
On the other hand, to clarify the pairing symmetry and superconducting property of non-centrosymmetric (NCS) superconductor 
has become a hot topic in this two decades \cite{Bauer}. 
In NCS superconductors, since 
the spatial inversion symmetry is broken, 
spin-singlet pairing and spin-triplet one can mix each other
and the resulting pair potential can have both 
spin-singlet even-parity and spin-triplet odd-parity components 
\cite{Gorkov,Frigeri,Fujimoto1,VVE08}. 
It has been revealed that in ballistic normal metal /$s+p$-wave 
superconductor junction with helical $p$-wave pairing, the 
tunneling conductance has a qualitatively different behaviour depending on
whether $p$-wave component is dominant or not. 
If spin-triplet $p$-wave component is dominant, 
the resulting conductance has a zero bias conductance peak. 
On the other hand, when spin-singlet component is dominant 
gap like structure appears at zero voltage 
\cite{TYBN09}. 
The present difference has been understood by the topological phase transition. If the $p$-wave component of the pair potential $\Delta_{p}$ is larger than 
the $s$-wave one $\Delta_{s}$, $s+p$-wave superconductor is in the 
topological phase with SABS. On the other hand, if $\Delta_{s}>\Delta_{p}$ is satisfied, it is in the non-topological phase without SABS \cite{TYBN09}. 
Topological phase transition occurs at  $\Delta_{s}=\Delta_{p}$, where 
the bulk energy gap of $s+p$-wave superconductor closes \cite{TYBN09}. 
Similar feature has been predicted for LDOS in DN of DN/NCS 
superconductor junction \cite{Gaetano} and 
charge conductance in T-shaped junction \cite{Mishra}.
Based on these backgrounds, it is useful to derive the 
more compact boundary condition of the retarded and Keldysh part of the 
Nambu-Keldysh Green's function applicable for mixed parity pairing case. 

In this paper, we revisit the boundary condition of 
the Nambu-Keldysh Green's function in 
DN/unconventional superconductor junctions derived in ref.\cite{Proximityd}. 
We derive the more compact formula of the boundary condition. 
We show that it is consistent with 
the formal boundary condition of Green's function by Zaitsev \cite{Zaitsev1984}. We derive both the retarded part and the Keldysh part of the boundary conditions in a more compact way 
applicable for general situation including
mixed parity case. We further show a clearer and shorter way to 
derive the charge conductance of the junction as compared to the 
previous derivation in Ref. \cite{Proximityd2,Proximityp2}. 
We apply this new formula to mixed parity $s+p$-wave 
one-dimensional superconductor model and calculate LDOS, pair amplitude and 
charge conductance. 
When $s$-wave component of the pair potential 
is dominant, the dominant pairing in DN is even-frequency spin-singlet 
$s$-wave and local density of states (LDOS) of quasiparticle have a gap like structure. On the other hand, when spin-triplet $p$-wave component is dominant, the dominant pairing in DN is odd-frequency spin-triplet $s$-wave \cite{odd1,tanaka12}. 
We show the robustness of the quantization of the 
conductance when the magnitude of $p$-wave 
component of the pair potential is larger than 
that of the $s$-wave one. These results show the robustness of the 
anomalous proximity effect against the inclusion of spin-singlet $s$-wave 
component of the pair potential when the $p$-wave component is dominant.

\section{Boundary condition of Green's function}
\label{sec:Nazarovboundarycondition}
In this section, we revisit the outline of the derivation 
of the boundary condition of the Nambu-Keldysh Green's function 
in diffusive normal metal / unconventional superconductor junctions. 
We show a more compact expression of the boundary condition of 
retarded and Keldysh part of the Green's function. 
\par
\subsection{Conventional spin-singlet $s$-wave superconductor junctions} 
Nazarov has derived a boundary condition of Nambu-Keldysh 
Green's function at the interface in order to 
study charge transport in mesoscopic superconductor junctions 
\cite{Nazarov1999}. 
This theory reproduces the BTK theory when the resistance in the 
diffusive normal metal can be ignored. 
First, we explain the outline of the theory by Nazarov \cite{Nazarov1999}. We consider a diffusive normal metal (DN) ($x<0$) / superconductor (S) ($x>0$)
junction model in 2D 
where the length of DN $L$ is much larger than mean free path 
$\ell =v_{F}\tau _{imp}$ with impurity scattering time $\tau_{imp}$. 
In DN, thermal diffusion length $\xi _{1}=\sqrt{D/2\pi T}$
is chosen much larger than $\ell$. 
Here, $v_{F}$ is the Fermi velocity and $D$ is the diffusion constant 
in the normal metal.
As a model of the interface,  
$\delta$-function potential is used at $x=0$, where the transparency at the 
interface is given by $\sigma_{N}=\cos^{2}\theta /(\cos^{2}\theta +Z^{2})$
with barrier parameter $Z$ and an injection angle $\theta$. 
To make a boundary condition of the Green's function, 
we assume that the interface zone $-L_{1}<x<L_{1}$ 
is composed of diffusive region $-L_{1}<x<-L_{2}$ 
and ballistic one with $-L_{2}<x<L_{1}$.  
Here, $L_{1}$ and $L_{2}$ satisfy
$L_{1}, L_{2} \ll \xi_{1}$, 
We denote the envelope function of Green's function 
of N side (S side) at the interface by 
$\bar{g}_{1}$ ($\bar{g}_{2}$). 
Here $\bar{g}_{1}$ and $\bar{g}_{2}$ are Nambu-Keldysh Green's function
with directional space distinguishing quasiparticle with positive and 
negative group velocity. 
By using the interface matrix ${\bar{M}}$, 
$\bar{g}_{2}$ and $\bar{g}_{1}$ are related each other 
\begin{equation}
\bar{g}_{2}=\bar{M}^{\dagger }\bar{g}_{1}\bar{M}; \ \ 
\bar{M}= \left(
\begin{array}{cc}
\frac{1}{t^{*}}\check{\mathbb{I}} & \frac{r}{t}\check{\mathbb{I}} \\
-\frac{r}{t}\check{\mathbb{I}} & \frac{1}{t}\check{\mathbb{I}}
\end{array}
\right ).
\label{matrixboundarycondition}
\end{equation}
In general, $\check{\mathbb{I}}$ denotes a $8 \times 8$ unit matrix in
particle-hole, Keldysh and spin space. 
Here, $\bar{M}$ is defined for each injection channel
and $t$ and $r$ express the coefficients of transmission and reflection with 
\[
\mid t \mid^{2} + \mid r \mid^{2}=1, \  tr^{*} + t^{*} r =0. 
\]
Nazarov has shown that $\bar{g}_{2}$ is given by \cite{Nazarov1999}
\begin{equation}
\bar{g}_{2} = \left(\bar{Q}\bar{G}_{2} + \bar{G}_{1} \right)^{-1} 
\{ 2\bar{Q} + \left(\bar{G}%
_{1}-\bar{Q}\bar{G}_{2} \right)\bar{\Sigma}^{z} \},
\end{equation}
for a spin-singlet $s$-wave superconductor junction 
with 
\begin{equation}
\bar{G}_{1}= \left(
\begin{array}{cc}
\check{G}_{1} & \  \check{0} \\
\check{0} & \  \check{G}_{1}
\end{array}
\right ), \ \ 
\bar{\Sigma}^{z} = \left (
\begin{array}{cc}
\check{\mathbb{I}} & \ \check{0} \\
\check{0} & \ -\check{\mathbb{I}}
\end{array}
\right ), \ \ 
\bar{\mathbb{I}} = \left (
\begin{array}{cc}
\check{\mathbb{I}} & \ \check{0} \\
\check{0} & \  \check{\mathbb{I}}
\end{array}
\right ), 
\end{equation}
\begin{equation}
\bar{G}_{2} = \left (
\begin{array}{cc}
\check{G}_{2} & \  \check{0} \\
\check{0} & \  \check{G}_{2}
\end{array}
\right ), 
\ \ 
\check{G}_{2}=\check{G}(x=0_{+})
\end{equation}
\begin{equation}
\bar{Q}=\bar{M}^{\dagger}\bar{M}, \ 
\bar{\Sigma}^{z}=\bar{M}\bar{\Sigma}^{z}\bar{M}^{\dagger}, \ 
\bar{Q}^{-1}=\bar{\Sigma}^{z}\bar{Q}\bar{\Sigma}^{z}
\end{equation}
Here, $\check{G}(x)$ is a Nambu-Keldysh Green's function and 
$\check{G}_{1}$ denotes
\[
\check{G}_{1} = \check{G}\left(x=-L_{1} \right)
\sim \check{G}\left( x=0_{-} \right)
\]
When the decomposition of the Green's function into 
an each spin sector is possible, 
$\check{G}_{1}$, $\check{G}_{2}$ and 
unit matrix $\check{I}$ become $4 \times 4$ matricies. 
The boundary condition of the Green's function is given by 
\begin{equation}
\frac{L}{R_{d}}
\left( \check{G}\left(x \right)\frac{\partial}{\partial x} \check{G}\left(x \right)\right) 
\Big|_{x=0_{-}}
=-\frac{1}{R_{b}} \left<\check{I}\left(\theta \right)\right>
\label{KeldyshKLNazarov}
\end{equation}
where $<\check{I}\left(\theta \right)>$ denotes the angular averaged current 
given by 
\begin{equation}
\left<\check{I}\left(\theta \right) \right>=
\displaystyle\frac{\displaystyle
\int^{\pi/2}_{-\pi/2} d\theta
\cos\theta \check{I}\left(\theta \right)}
{\displaystyle\int^{\pi/2}_{-\pi/2} d\theta
\cos\theta \sigma_{N}(\theta)}
\label{Nazarovboundaryangular}
\end{equation}
Here, $R_{d}$ and $R_{b}$ are resistances in the DN and that at the interface.
Matrix current $\check{I}(\theta)$ is given by 
\begin{equation}
\check{I}(\theta)=\mathrm{Tr}[\bar{\Sigma}^{z}\bar{g}
_{1}]=\mathrm{Tr}[\bar{\Sigma}^{z}\bar{g}_{2}]. 
\label{matrixcurrent}
\end{equation}%
Here, $\rm{Tr}$ denotes the summation of channels with various
injection angles. 
In the actual calculation of $\check{I}(\theta)$, 
it is convenient to choose the basis where 
$\bar{Q}$ becomes a diagonalized matrix. In this basis, 
$\bar{\Sigma}^{z}$ is also transformed. 
$\bar{Q}$, $\bar{\Sigma}^{z}$ are given by 
\begin{equation}
\bar{Q}= \left(
\begin{array}{cc}
q_{n} \check{\mathbb{I}}& \check{0} \\
\check{0} & q_{n}^{-1}\check{\mathbb{I}}
\end{array}
\right ), \ \ \bar{\Sigma}^{z}= \left(
\begin{array}{cc}
\check{0} & \ \check{\mathbb{I}} \\
\check{\mathbb{I}} & \ \check{0}%
\end{array}
\right).
\end{equation}
Here, we denote the channel index $n$ corresponding to 
the injection angle $\theta$. 
Using the eigenvalue $q_{n}$, 
the transmissivity at the interface $\sigma_{N}$ is also expressed by 
\[
\sigma_{N}=\frac{4q_{n}}{(1 + q_{n})^{2}}
\]
As a result, $\bar{g}_{2}$ becomes \cite{Nazarov1999}
\begin{equation}
\bar{g}_{2} = \left(
\begin{array}{cc}
q_{n}\check{G}_{2} + \check{G}_{1} & \check{0} \\
\check{0} & q_{n}^{-1}\check{G}_{2} + \check{G}_{1}%
\end{array}
\right)^{-1} \left(
\begin{array}{cc}
2q_{n} \check{\mathbb{I}}& \check{G}_{1}-q_{n}\check{G}_{2} \\
\check{G}_{1} - q_{n}^{-1}\check{G}_{2} & 2q_{n}^{-1}\check{\mathbb{I}} %
\end{array}
\right).
\label{barg2}
\end{equation}
Finally, $\check{I}(\theta)=\check{I}_{n}$ is given by 
\begin{equation}
\check{I}_{n}
=\check{I}\left(\theta \right)=
\displaystyle\frac{2 \sigma_{N}\left[\check{G}_{2},\check{G}_{1} \right ]}
{\left(4 - 2 \sigma_{N}\right)\check{\mathbb{I}} + \sigma_{N}\left[\check{G}_{2},\check{G}_{1}\right]_{+}}
\label{matrixcurrentnazarov}
\end{equation}
with
\[
\check{G}^{2}_{1}=\check{G}^{2}_{2}=\check{\mathbb{I}}, \ \ 
\left[\check{G}_{2},\check{G}_{1}\right]_{+}
=\check{G}_{2}\check{G}_{1} + \check{G}_{1}\check{G}_{2}
\]
\subsection{Boundary condition of Unconventional superconductor junctions} 
Next, let us consider unconventional superconductor junctions.
One of the authors Y.T. has extended Nazarov's theory which is available for 
unconventional superconductor junctions 
\cite{Proximityd,Proximityd2,Proximityp2}. 
In this case, we must take into account the directional dependence 
of the Nambu Keldysh Green's function in $\bar{G}_{2}$, \par
\[
\bar{G}_{2} = \left (
\begin{array}{cc}
\check{G}_{2 +} & 0 \\
0 & \check{G}_{2 -}
\end{array}
\right ).
\]
\setlength{\abovedisplayshortskip}{0.5 \baselineskip}%
Here, $\check{G}_{2+}$ and $\check{G}_{2-}$ are 
Green's function for bulk state with different trajectory and 
they satisfy normalization condition 
\[
\check{G}^{2}_{2+}=\check{\mathbb{I}}, \ 
\check{G}^{2}_{2-}=\check{\mathbb{I}}. 
\]
Owing to the presence of two kinds of Nambu-Keldysh Green's function 
$\check{G}_{2+}$ and $\check{G}_{2-}$, 
surface Andreev bound states (SABS) can be naturally taken into account. 
It has been shown that $\bar{g}_{2}$ is given by \cite{Proximityd,Proximityd2}
\begin{equation}
\bar{g}_{2} = \left(
\begin{array}{cc}
q_{n}\check{H}_{+} + \check{G}_{1} & q_{n}\check{H}_{-} \\
q_{n}^{-1}\check{H}_{-} & q_{n}^{-1}\check{H}_{+} + \check{G}_{1}%
\end{array}
\right)^{-1} \left(
\begin{array}{cc}
q_{n}\left(2\check{\mathbb{I}} 
- \check{H}_{-} \right) & \ \check{G}_{1}-q_{n}\check{H}_{+} \\
\check{G}_{1} - q_{n}^{-1}\check{H}_{+} & \ q_{n}^{-1}
\left(2\check{\mathbb{I}} - \check{H}_{-} \right)%
\end{array}
\right)
\label{barg2unconventional}
\end{equation}
with 
\begin{equation}
\check{H}_{+} \equiv (\check{G}_{2+} + \check{G}_{2-})/2, \ 
\check{H}_{-} \equiv (\check{G}_{2+} - \check{G}_{2-})/2.
\label{definictionH+-}
\end{equation}
The relations
\begin{equation}
\check{H}_{+}^{2} + \check{H}_{-}^{2}=\check{\mathbb{I}}, \ \ 
\check{H}_{+}\check{H}_{-} + \check{H}_{-}\check{H}_{+}=\check{0}, \ \
\check{\mathbb{I}}-\check{H}_{-}^{-2}= 
\left(\check{H}_{-}^{-1}\check{H}_{+} \right)^{2}
\end{equation}
are satisfied. 
By using this relation, the matrix current (\ref{matrixcurrent})
has been calculated. 
The details of the derivation are shown in 
Appendix A.
$\check{I}_{n}$ is given by \cite{Proximityd}
\begin{equation}
\check{I}_{n}=
{\rm Trace}[\bar{\Sigma}^{z}\bar{g}_{2}]
=2\left[\check{G}_{1},\check{B} \right],
\label{matrixcurrentPRL2003basic}
\end{equation}
\begin{equation}
\check{B}=
\left(\check{H}_{-}^{-1}\check{H}_{+}
-\sigma_{1N}\left[\check{G}_{1},\check{H}_{-}^{-1} \right]
-\sigma_{1N}^{2}\check{G}_{1}\check{H}_{-}^{-1}\check{H}_{+}
\check{G}_{1}\right)^{-1}
\left(\sigma_{1N}\left(\check{\mathbb{I}}-\check{H}_{-}^{-1} \right)
+\sigma_{1N}^{2}\check{G}_{1}\check{H}_{-}^{-1}\check{H}_{+}\right),
\label{matrixcurrentPRL2003}
\end{equation}%
with
\begin{equation}
\sigma_{1N}\equiv\frac{\sigma_{N}}{2-\sigma_{N} + 2\sqrt{1 - \sigma_{N}}}.
\label{sigma1n}
\end{equation}
\par
It is noted that Eq. (\ref{matrixcurrentPRL2003basic}) 
can be expressed more compactly. 
First, we simplify $\check{B}$ in 
eq. (\ref{matrixcurrentPRL2003}) 
\cite{TextTanaka2021}. 
Using 
\begin{eqnarray}
&& 
\left( \check{\mathbb{I}} - \check{H}_{-}^{-1} + \sigma_{1N}\check{G}_{1}
\check{H}_{-}^{-1}\check{H}_{+} \right)
\left( \check{\mathbb{I}} + \check{H}_{-}^{-1} 
- \sigma_{1N}\check{G}_{1}\check{H}_{-}^{-1}\check{H}_{+} \right)
\nonumber
\\
&=&
\check{\mathbb{I}}- \check{H}_{-}^{-2} + \sigma_{1N}
\left( \check{H}_{-}^{-1}\check{G}_{1}\check{H}_{-}^{-1}\check{H}_{+}
+\check{G}_{1}\check{H}_{-}^{-1}\check{H}_{+}\check{H}_{-}^{-1} \right)
-\sigma_{1N}^{2}\check{G}_{1}\check{H}_{-}^{-1}\check{H}_{+}\check{G}_{1}
\check{H}_{-}^{-1}\check{H}_{+}
\nonumber
\end{eqnarray}
and 
\begin{equation}
 \check{\mathbb{I}} - \check{H}_{-}^{-2}=
\left(\check{H}_{-}^{-1}\check{H}_{+}\right)^{2},  \  \ 
\check{H}_{-}\check{H}_{+} =- \check{H}_{+}\check{H}_{-}
\label{Nazarovappendixrelation2}
\end{equation}
we derive 
\begin{eqnarray}
&& \left( \check{\mathbb{I}} - \check{H}_{-}^{-1} + \sigma_{1N}\check{G}_{1}
\check{H}_{-}^{-1}\check{H}_{+} \right)
\left(
\check{\mathbb{I}} + \check{H}_{-}^{-1} - \sigma_{1N}\check{G}_{1}\check{H}_{-}^{-1}\check{H}_{+} \right)
\nonumber
\\
&=&
\left(
\check{H}_{-}^{-1}\check{H}_{+} - 
\sigma_{1N}\left[ \check{G}_{1}, \check{H}_{-}^{-1} \right]
- \sigma_{1N}^{2} \check{G}_{1}\check{H}_{-}^{-1}\check{H}_{+}\check{G}_{1}
\right)
\check{H}_{-}^{-1}\check{H}_{+}.
\end{eqnarray}
Then, the denominator of $\check{B}$ is transformed into 
\begin{equation}
\left( \check{\mathbb{I}} - \check{A} \right)
\left( \check{\mathbb{I}} + \check{A} \right)\check{H}_{+}^{-1}\check{H}_{-}
\end{equation}
with 
\[
\check{A} \equiv \check{H}_{-}^{-1}-\sigma_{1N}\check{G}_{1}\check{H}_{-}^{-1}
\check{H}_{+}.
\]
On the other hand, the numerator of $\check{B}$ becomes 
\begin{equation}
\sigma_{1N}\left( \check{\mathbb{I}} - \check{A} \right). 
\end{equation}
As a result, $\check{B}$ is expressed as 
\begin{eqnarray}
\check{B} &=& 
\sigma_{1N}
\left[\left( \check{\mathbb{I}} - \check{A} \right)
\left( \check{\mathbb{I}} + \check{A} \right)\check{H}_{+}^{-1}\check{H}_{-}
\right]^{-1}
\left( \check{\mathbb{I}} - \check{A} \right)
\nonumber
\\
&=&
\sigma_{1N}\check{H}_{-}^{-1}\check{H}_{+} 
\left( \check{\mathbb{I}} + \check{A} \right)^{-1}
\nonumber
\\
&=&
-\sigma_{1N}
\left[
\check{H}_{+}^{-1}
\left( \check{\mathbb{I}} - \check{H}_{-}\right) + \sigma_{1N} \check{G}_{1} 
\right]^{-1}.
\label{Newboundarycondition}
\end{eqnarray}
If we define 
\begin{equation}
\check{C} \equiv \check{H}_{+}^{-1}
\left(  \check{\mathbb{I}} - \check{H}_{-}\right)
\end{equation}
from eq. (\ref{definictionH+-}), 
$\check{C}$ satisfies the following 
normalization condition
\[
\check{C}^{2} =  \check{\mathbb{I}}. 
\]
The generation of SABS is naturally taken into account in 
$\check{C}$. 
In the case of spin-singlet $s$-wave superconductor, 
\[
\check{G}_{2+}=\check{G}_{2-}=\check{G}_{2}
\]
is satisfied and $\check{C}$ is reduced to be $\check{G}_{2}$. 
Owing to the normalization condition of $\check{C}$, 
$\check{B}$ can be transformed as 
\begin{eqnarray}
\check{B} &=& -\sigma_{1N}
\left( \check{C} + \sigma_{1N} \check{G}_{1}\right)^{-1}
\nonumber
\\
&=&
-\sigma_{1N}
\left[ \left( 1 + \sigma_{1N}^{2} \right) \check{\mathbb{I}}
+ \sigma_{1N} [\check{G}_{1}, \check{C}]_{+}
\right]^{-1}
\left( \check{C} + \sigma_{1N} \check{G}_{1} 
\right).
\end{eqnarray}
By plugging this equation into 
eq. (\ref{matrixcurrentPRL2003basic}), we obtain 
\begin{eqnarray}
\check{I}_{n} &=& \check{I}\left( \theta \right) = 
2 
\left[\check{G}_{1},\check{B} \right]
\nonumber
\\
&=& 
-2 \sigma_{1N} 
\left[ 
\left(1 + \sigma_{1N}^{2} \right) \check{\mathbb{I}}
+ \sigma_{1N} 
\left[\check{G}_{1},\check{C} \right]_{+} 
\right]^{-1}
\left[
\check{G}_{1},\check{C}
\right]
\nonumber
\\
&=& 
2 \sigma_{N} 
\left[ \left(4 - 2\sigma_{N} \right) 
\check{\mathbb{I}}
+ \sigma_{N} 
\left[\check{C},\check{G}_{1} \right]_{+} 
\right]^{-1}
\left[\check{C},
\check{G}_{1}\right]. 
\label{matrixcurrentTanakanew}
\end{eqnarray}
Here, the following relation is used 
\[
\frac{\sigma_{1N}}{1 + \sigma^{2}_{1N}}
=\frac{\sigma_{N}}{2 \left( 2 - \sigma_{N}\right)}
.\]
\label{newcompactboundary}
This means that $\check{G}_{2}$ in eq.(\ref{matrixcurrentnazarov}) 
is replaced with $\check{C}$. \par
As shown in Appendix B, 
eq. (\ref{matrixcurrentTanakanew}) satisfies Zaitsev's boundary condition 
originally derived for spin-singlet $s$-wave 
superconductor junctions. 
We can calculate ${\rm Trace }[\bar{g}_{2}]$, 
${\rm Trace }[\bar{g}_{1}]$, ${\rm Trace }[\bar{\Sigma}^{z}\bar{g}_{1}]$ 
as shown in  eqs. (\ref{traceg2bar}), 
(\ref{matrixcurrentTanakaNazarov5}), and (\ref{traceg1bar}), respectively. 
From these functions, we can define the interface Green's function 
appearing in Zaitsev's boundary condition 
$\check{g}_{1}^{s}$, $\check{g}_{2}^{s}$, 
$\check{g}_{1}^{a}$, and $\check{g}_{2}^{a}$, 
as shown in eqs. (\ref{g1s}), (\ref{g2s}), (\ref{g1a}) and (\ref{g2a}), 
respectively. 
Then, it is shown in Appendix B
that the expressions
satisfy Zaitsev's boundary condition: 
\begin{eqnarray}
&&\check{g}^{a} \left[\left(1 -\sigma_{N} \right)
\left( \check{g}_{s+}\right)^{2} + \left(\check{g}_{s-}\right)^{2} \right] 
\nonumber
\\
&=& 4 \sigma_{N} \left(1 -\sigma_{N} \right) \left[\check{C}, \check{G}_{1} \right]
\left\{ 2\left( 2-\sigma_{N} \right) 
\check{\mathbb{I}}
+ \sigma_{N} 
\left[\check{C},\check{G}_{1} \right]_{+} 
\right\}^{-2} 
\nonumber
\\
&=&\sigma_{N} \check{g}_{s-}\check{g}_{s+}
\end{eqnarray}
It is noted that eq. (\ref{matrixcurrentTanakanew}) is 
greatly simplified as compared to the original equation in 
Ref. \cite{Proximityd} and is useful for the application to more complicated 
pair potentials. 
\par
In the above, $\check{H}_{+}$, $\check{H}_{-}$, $\check{C}$
are expressed as
\begin{align}
\check{H}_{+} =&\left(
\begin{array}{cc}
\hat{R}_{+} & \hat{K}_{+} \\
\hat{0} & \hat{A}_{+}%
\end{array}%
\right) , \;\; 
\check{H}_{-} =\left(
\begin{array}{cc}
\hat{R}_{-} & \hat{K}_{-} \\
\hat{0} & \hat{A}_{-}%
\end{array}%
\right), \; 
\check{C}=
\begin{pmatrix}
\hat{C}_{R} & \hat{C}_{K} \\
\hat{0} & \hat{C}_{A}
\end{pmatrix}
, 
\ \ 
\check{I}_n=\left(
\begin{array}{cc}
\hat{I}_{R} & \hat{I}_{K} \\
\hat{0} & \hat{I}_{A}%
\end{array}%
\right).
\label{NazarovTanakablock}
\end{align}
$\hat{R}_{+}$, $\hat{R}_{-}$, $\hat{C}_{R}$ and $\hat{I}_{R}$ 
are retarded parts, 
$\hat{K}_{+}$, $\hat{K}_{-}$ $\hat{C}_{K}$ and $\hat{I}_{K}$ 
are Keldysh ones, 
and $\hat{A}_{+}$, $\hat{A}_{-}$, $\hat{C}_{A}$ and $\hat{I}_{A}$ are 
advanced ones. 
On the other hand, the Green's function in DN $\check{G}_{1}(x)$
is expressed by 
\begin{equation}
\check{G}_{1}\left( x \right)=
\begin{pmatrix}
\hat{R}_{1}\left( x \right) & \hat{K}_{1}\left( x \right) \\
0 & \hat{A}_{1}\left( x \right)
\end{pmatrix}
\end{equation}
with retarded part $\hat{R}_{1}(x)$, Keldysh part $\hat{K}_{1}(x)$ and 
advanced one $\hat{A}_{1}(x)$. 
We denote $\hat{R}_{1}(x=0_{-})=\hat{R}_{1}$, 
$\hat{K}_{1}(x=0_{-})=\hat{K}_{1}$,
and $\hat{A}_{1}(x=0_{-})=\hat{A}_{1}$. 
Then, the retarded part $\hat{I}_{R}$ is expressed by 
\begin{equation}
\hat{I}_{R}\left(\theta \right)
=2 \sigma_{N}
\left[
2(2 -\sigma_{N})\hat{\mathbb{I}} + \sigma_{N}
\left(\hat{C}_{R}\hat{R}_{1} + \hat{R}_{1}\hat{C}_{R} \right) \right]^{-1}
\left[\hat{C}_{R},\hat{R}_{1} \right].
\label{matrixcurrentIR}
\end{equation}
with unit matrix $\hat{\mathbb{I}}$. 
Here, $\hat{C}_{R}$ is the retarded part of $\check{C}$
and is expressed by 
\begin{equation}
\hat{C}_{R}=\hat{R}_{+}^{-1}\left(\hat{\mathbb{I}}-\hat{R}_{-} \right),
\label{CRfunction}
\end{equation}
which satisfies $\hat{C}_{R}^{2}=\hat{\mathbb{I}}$ owing to the 
relation of $\hat{R}_{+}$ and $\hat{R}_{-}$. \par
In the following, we consider the situation where 
the decomposition of the Green's function into each spin sector is 
possible. 
Both $\hat{C}_{R}$ and $\hat{R}_{1}$ are linear combinations of 
$\hat{\tau}_{1}$, $\hat{\tau}_{2}$ and $\hat{\tau}_{3}$ and 
$\hat{C}_{R}\hat{R}_{1} + \hat{R}_{1}\hat{C}_{R}$ is proportional to 
$\hat{\mathbb{I}}$ since it is an anticommutator of $2 \times 2$ matricies. 
As a result,  the denominator of $\hat{I}_{R}$ is 
proportional to $\hat{\mathbb{I}}$. 
Then, $\hat{I}_{R}$ is given by 
\begin{equation}
\hat{I}_{R}=\frac{2 \sigma_{1N}}{d_{R}}
\left[\hat{C}_{R},\hat{R}_{1} \right]
\label{Tanakamatrixretarded}
\end{equation}
with 
\begin{equation}
d_{R} \hat{\mathbb{I}}
\equiv \left(1 + \sigma_{1N}^{2}\right) \hat{\mathbb{I}}
+\sigma_{1N}\left(\hat{C}_{R}\hat{R}_{1} + \hat{R}_{1}\hat{C}_{R} \right).
\end{equation}
\par
\par
First, let us discuss the boundary condition of the retarded part at $x=0$. 
In general, $\hat{R}_{1}(x)$ can be decomposed into 
\begin{eqnarray}
\hat{R}_{1}\left(x \right)
&=&
s_{1}\left( x \right)\hat{\tau}_{1}+s_{2}\left( x \right)\hat{\tau}_{2}
+s_{3}\left( x \right)\hat{\tau}_{3}
\nonumber
\\
&=&\cos \psi \sin \zeta \hat{\tau}_{1}
+ \sin \psi \sin \zeta \hat{\tau}_{2}
+ \cos \zeta \hat{\tau}_{3}, 
\label{parametrization}
\end{eqnarray}
by using Pauli matrices in electron-hole space. 
$\zeta$ and $\psi$ follow from
\begin{align}
&D\left[\frac{\partial^{2}}{\partial x^{2}}\zeta 
- \left(\frac{\partial \psi}{\partial x}\right)^{2}\cos \zeta 
\sin \zeta\right] 
+ 2i\varepsilon \sin\zeta =0,\\
&\frac{\partial}{\partial x}
\left[\sin^{2}\zeta \left(\frac{\partial \psi}{\partial x}\right)\right]=0
\end{align}
In the case  
$\sin^{2}\zeta \left(\frac{\partial \psi}{\partial x}\right) \neq 0$,  
supercurrent without dissipation can flow with zero voltage. 
Since we are considering the charge transport in normal electrode / DN/superconductor junction, 
it is natural to assume $\frac{\partial \psi}{\partial x}=0$. \par
The retarded part of the boundary condition in eq.(\ref{KeldyshKLNazarov})
is given by  
\begin{equation}
\frac{L}{R_{d}}
\left( \hat{R}_{1}\left(x \right)\frac{\partial}{\partial x} \hat{R}_{1}
\left(x \right)\right) 
\Big|_{x=0_{-}}
=-\frac{1}{R_{b}} \left<\hat{I}_{R}\left(\theta \right)\right>. 
\label{RetardedTanakaNazarov}
\end{equation}
The angular average by an injection angle $\theta$ is given in eq. 
(\ref{Nazarovboundaryangular}). 
By plugging $\frac{\partial \psi}{\partial x}=0$ into 
(\ref{RetardedTanakaNazarov}), 
the left side of eq. (\ref{RetardedTanakaNazarov}) is transformed into 
\begin{equation}
\left.\frac{L}{R_{d}}
\hat{R}_{1}\left(x \right)\frac{\partial}{\partial x} \hat{R}_{1}\left(x \right) \right|_{x=0_{-}}
=\left.\frac{Li}{R_{d}}[-\sin \psi \hat{\tau}_{1} + \cos \psi \hat{\tau}_{2}]
\left(\frac{\partial \theta}{\partial x}\right) \right|_{x=0_{-}}.
\label{retarded}
\end{equation}
It is noted that $\hat{\tau}_{3}$ component of 
$<\hat{I}_{R}\left(\theta \right)>$ 
is absent in eq. (\ref{RetardedTanakaNazarov}) due to the 
absence of supercurrent. 
This means 
\begin{equation}
\left< {\rm Trace}\left[\hat{I}_{R}\hat{\tau}_{3}\right] \right>=0. 
\label{ballancerelation}
\end{equation}
$\psi$ is determined from this equation. 
On the other hand, $\zeta(x)=\zeta$
satisfies Usadel equation
\begin{equation}
D \frac{\partial^{2}}{\partial x^{2}}\zeta\left( x \right) 
+ 2i\varepsilon \sin\zeta\left( x \right) =0. 
\label{Usadeleq}
\end{equation}
\par
Next, we calculate the Keldysh part of the matrix current 
$\check{I}(\theta)$ given by 
\begin{eqnarray}
\check{I}\left( \theta \right)
&=&2\sigma_{1N}
\begin{pmatrix}
\left( 1 + \sigma_{1N}^{2}\right) \hat{\mathbb{I}}
 + \sigma_{1N}\left[\hat{C}_{R},\hat{R}_{1} \right]_{+}
& \sigma_{1N}\left(\hat{D}_{1} + \hat{D}_{2} \right) \\
\hat{0} & 
\left( 1 + \sigma_{1N}^{2} \right)\hat{\mathbb{I}} 
+ \sigma_{1N}\left[\hat{C}_{A},\hat{A}_{1} \right]_{+}
\end{pmatrix}
\nonumber
\\
&\times&
\begin{pmatrix}
\left[\hat{C}_{R},\hat{R}_{1} \right]
& 
\hat{D}_{1} - \hat{D}_{2} \\
\hat{0} & 
\left[ \hat{C}_{A},\hat{A}_{1} \right]
\end{pmatrix}
\label{matrixcurrentnewTanaka}
\end{eqnarray}
with
\begin{equation}
\hat{D}_{1} \equiv \hat{C}_{R}\hat{K}_{1} + \hat{C}_{R}\hat{A}_{1}, \ 
\hat{D}_{2} \equiv \hat{R}_{1}\hat{C}_{K} + \hat{K}_{1}\hat{C}_{A}.
\label{D1D2andK1}
\end{equation}
The relation between the retarded and the advanced part of 
the Green's function is given by 
\[
\hat{C}_{A}=-\hat{\tau}_{3}\hat{C}_{R}^{\dagger}\hat{\tau}_{3}
\ \ 
\hat{A}_{1}=-\hat{\tau}_{3}\hat{R}_{1}^{\dagger}\hat{\tau}_{3}
\]
and 
\[
\left(1 + \sigma_{1N}^{2}\right)\hat{\mathbb{I}}
+\sigma_{1N}\left(\hat{C}_{A}\hat{A}_{1} + \hat{A}_{1}\hat{C}_{A} 
\right)
=d^{*}_{R}\hat{\mathbb{I}}.
\]
The Keldysh component of $\hat{I}_{K}$ is given by 
\begin{equation}
\hat{I}_{K}=
\frac{ 2\sigma_{1N}}
{\mid d_{R} \mid^{2}}
\left[
d_{R}^{*}\left(\hat{D}_{1} - \hat{D}_{2} \right) 
- \sigma_{1N}\left(\hat{D}_{1} + \hat{D}_{2} \right)
\left[ \hat{C}_{A}, \hat{A}_{1} \right]
\right]
\end{equation}
\[
\hat{K}_{1}=\left(
\hat{R}_{1}-\hat{A}_{1} \right) 
f_{0N}\left(x \right)
+
\left( \hat{R}_{1} \hat{\tau}_{3} - 
\hat{\tau}_{3}\hat{A}_{1}
\right)
f_{3N}\left(x \right)
\]
with $\hat{K}_{1}$ in eq. (\ref{D1D2andK1}) 
and $\hat{C}_{K}$ given by 
\[
\hat{C}_{K}=
\left( \hat{C}_{R}-\hat{C}_{A} \right)
f_{S}\left(x \right)
\]
with distribution function $f_{3N}(x)$,
$f_{0N}(x)$ and $f_{S}(x)$.

By using above relations, $\hat{I}_{K}$ becomes
\begin{eqnarray}
\hat{I}_{K}&=&
\frac{ 2\sigma_{1N}}
{\mid d_{R} \mid^{2}}
\left[
\left(1 + \sigma_{1N}^{2} \right)
\hat{\Lambda}_{1} 
+ 2\sigma_{1N} \hat{\Lambda}_{2}
\right],
\label{Keldysh1}
\\
\hat{\Lambda}_{1}&=&
\left[
\hat{C}_{R}\hat{K}_{1} + \hat{C}_{K}\hat{A}_{1}
- \hat{R}_{1}\hat{C}_{K} - \hat{K}_{1}\hat{C}_{A}
\right],
\label{Keldysh2}
\\
\hat{\Lambda}_{2}&=&
\left[
\left(\hat{C}_{R}\hat{K}_{1} + \hat{C}_{K}\hat{A}_{1} \right)
\hat{A}_{1}\hat{C}_{A} 
- \left(\hat{R}_{1}\hat{C}_{K} + \hat{K}_{1} \hat{C}_{A}\right)
\hat{C}_{A}\hat{A}_{1}
\right].
\label{Keldysh3}
\end{eqnarray}
The explicit form of 
$\hat{\Lambda}_{1}$ and $\hat{\Lambda}_{2}$ are given by 
eqs. (\ref{lambda1a}) and (\ref{lambda2a}) in Appendix C. 
To obtain the charge current, 
we focus on the boundary condition given by eq. (\ref{KeldyshKLNazarov}). 
From eq. (\ref{KeldyshKLNazarov}), the 
$\hat{\tau}_{3}$ component of the boundary condition of the Keldysh
component is given by 
\begin{equation}
\frac{L}{R_{d}}
\left.
{\rm Trace}
\left[
\left(\hat{R}\frac{\partial}{\partial x}\hat{K} + 
\hat{K}\frac{\partial }{\partial x} \hat{A} \right)
\hat{\tau}_{3}
\right] \right|_{x=0_{-}}
=
-\frac{1}{R_{b}} \left< {\rm Trace}
\left[ \hat{I}_{K} \hat{\tau}_{3} \right]
\right>.
\end{equation}
By using eq. (\ref{derivationKeldysha}) in Appendix C, 
following relation is satisfied  
\begin{equation}
\frac{L}{R_{d}}
\left.
{\rm Trace}
\left[
\left(\hat{R}\frac{\partial}{\partial x}\hat{K} + 
\hat{K}\frac{\partial }{\partial x} \hat{A} \right)
\hat{\tau}_{3}
\right] \right|_{x=0_{-}}
=\left.
4 \left(
\frac{\partial f_{3N}\left( x \right) }
{\partial x}
\right)
{\rm cosh}^{2} \zeta_{im} \right|_{x=0_{-}}.
\end{equation}
with the imaginary part of $\zeta$ denoted by $\zeta_{im}$. 
Then, the boundary condition is expressed by 
\begin{equation}
\left.
4 \left(
\frac{L}{R_{d}}
\right)
\left(
\frac{ \partial f_{3} \left(x \right)}
{\partial x}
\right)
{\rm cosh}^{2} \zeta_{im} \right|_{x=0_{-}}
=
-\frac{1}{R_{b}}
\left< 
{\rm Trace}
\left[ \hat{I}_{K}\hat{\tau}_{3}
\right] 
\right>.
\end{equation}

Below, we calculate 
\begin{equation}
\left.
\left(
\frac{L}{R_{d}}
\right)
\left(
\frac{ \partial f_{3} \left(x \right)}
{\partial x}
\right)
{\rm cosh}^{2} \zeta_{im} \right|_{x=0_{-}}
=
-\frac{1}{4R_{b}}
\left< 
{\rm Trace}
\left[ \hat{I}_{K}\hat{\tau}_{3}
\right] 
\right>
=-\frac{1}{R_{b}}
\left< I_{K} \right>.
\end{equation}
From eqs.(\ref{Keldysh1}), (\ref{Keldysh2}),
(\ref{Keldysh3}), (\ref{Tanakamatrixcurrentcharge1a}), 
and (\ref{Tanakamatrixcurrentcharge2a}), 
${\rm Trace}[\hat{I}_{K}\hat{\tau}_{3}]$ becomes 
\begin{eqnarray}
{\rm Trace}\left[\hat{I}_{K}\hat{\tau}_{3}\right]
&=&
\frac{2 \sigma_{1N}}{\mid d_{R} \mid^{2}}
\biggl[
{\rm Trace}
\left[ \left\{
\left( \hat{C}_{R} + \hat{C}^{\dagger}_{R} \right)\hat{R}_{1}
+ 
\hat{R}^{\dagger}_{1}
\left( \hat{C}_{R} + \hat{C}^{\dagger}_{R} \right) \right\}
\hat{\tau}_{3} \right]
\left( 1 + \sigma_{1N}^{2}\right) f_{0N}\left(x \right)
\nonumber
\\
&+& 
2{\rm Trace}
\left[ \left\{
\hat{\mathbb{I}} + \hat{C}^{\dagger}_{R}\hat{C}_{R} 
+ \hat{R}^{\dagger}_{1}
\left( \hat{\mathbb{I}} + \hat{C}^{\dagger}_{R} \hat{C}_{R} \right) 
\hat{R}_{1}
\right\}
\hat{\tau}_{3} \right]
\sigma_{1N}f_{0N}\left(x \right)
\nonumber
\\
&-&
{\rm Trace}
\left[ \left\{
\left( \hat{R}_{1} + \hat{R}^{\dagger}_{1} \right)\hat{C}_{R}
+ \hat{C}^{\dagger}_{R}
\left( \hat{R}_{1} + \hat{R}^{\dagger}_{1} \right) \right\}
\hat{\tau}_{3} \right]
\left( 1 + \sigma_{1N}^{2}\right) f_{S}\left(x \right)
\nonumber
\\
&-& 
2{\rm Trace}
\left[ \left\{
\hat{\mathbb{I}} + \hat{R}^{\dagger}_{1}\hat{R}_{1} 
+ \hat{C}^{\dagger}_{R}
\left( \hat{\mathbb{I}} + \hat{R}^{\dagger}_{1} \hat{R}_{1} \right) 
\hat{C}_{R}
\right\}
\hat{\tau}_{3} \right]
\sigma_{1N}f_{S}\left(x \right)
\nonumber
\\
&+&
{\rm Trace}
\left[ 
\left( \hat{R}_{1} + \hat{R}^{\dagger}_{1} \right)
\left( \hat{C}_{R} + \hat{C}^{\dagger}_{R} \right) 
\right]
\left( 1 + \sigma_{1N}^{2}\right) f_{3N}\left(x \right)
\nonumber
\\
&+&
2{\rm Trace}
\left[
\left( 
\hat{\mathbb{I}} + \hat{R}^{\dagger}_{1}\hat{R}_{1} 
\right)
\left( \hat{\mathbb{I}}
+ \hat{C}^{\dagger}_{R}\hat{C}_{R} \right)
\right]
\sigma_{1N}f_{3N}\left(x \right)
\biggr].
\label{Tanakamatrixcurrentcharge3}
\end{eqnarray}
Eqs.(\ref{Tanakamatrixretarded}) and 
(\ref{Tanakamatrixcurrentcharge3}) 
are available for general cases of pair potentials. 
In the following section, we will use these equations 
for the 
specific model of unconventional superconductor junctions.

\section{Superconductor with Various Parity}
In this section, we revisit the case of 
superconductors with spatial inversion symmetry 
where the pairing symmetry of superconductor is 
spin-singlet even-parity or spin-triplet odd-parity with 
time reversal symmetry. We consider spin-triplet paring where 
$z$ component of the spin momentum of Cooper pair is $S_{z}=0$. 
In that case, 
we calculate Nambu Keldysh Green's function and derive the 
charge conductance of the 
junctions in a more compact way as compared to previous papers 
\cite{Proximityp2,Proximityd2}. 
In the present case, we can choose the gauge of the pair potential 
so that $\hat{R}_{\pm}$ is expressed by 
\begin{equation}
\hat{R}_{+}=g_{+}\hat{\tau}_{3} + f_{+}\hat{\tau}_{2}, \ 
\hat{R}_{-}=g_{-}\hat{\tau}_{3} + f_{-}\hat{\tau}_{2}, \ 
\end{equation}
with $g_{\pm}=\varepsilon/\Omega_{\pm}$,  
$f_{\pm}=i \Delta_{\pm}/\Omega_{\pm}$, 
and 
\begin{equation}
\Omega_{\pm}
\equiv \lim_{\delta \rightarrow 0} 
\sqrt{\left(\varepsilon + i\delta \right)^{2} - \mid \Delta_{\pm} \mid^{2}}
= \left \{
\begin{array}{ll}
\sqrt{\varepsilon^{2} - \mid \Delta_{\pm} \mid^{2}} & \varepsilon  \geq \mid \Delta_{\pm} \mid \\
i \sqrt{\mid \Delta_{\pm} \mid^{2}  - \varepsilon^{2}} &  -\mid \Delta_{\pm} \mid \leq \varepsilon \leq  \mid \Delta_{\pm} \mid \\ 
-\sqrt{\varepsilon^{2} - \mid \Delta_{\pm} \mid^{2}} 
& \varepsilon  \leq -\mid \Delta_{\pm} \mid
\end{array}
\right.
\label{Omegapm}.
\end{equation}

We denote $\hat{C}_{R}$ as 
\[
\hat{C}_{R}=c_{1}\hat{\tau}_{1}+ c_{2}\hat{\tau}_{2} + c_{3}\hat{\tau}_{3}, 
\]
where $c_{1}$, $c_{2}$ and $c_{3}$ are expressed by 
\begin{equation}
c_{1}=\frac{i\left(f_{+}g_{-}-g_{+}f_{-}\right)}{1+f_{+}f_{-}+g_{+}g_{-}}, \ 
c_{2}=\frac{f_{+}+f_{-}}{1+f_{+}f_{-}+g_{+}g_{-}}, \ 
c_{3}=\frac{g_{+}+g_{-}}{1+f_{+}f_{-}+g_{+}g_{-}}. 
\label{coefficient1}
\end{equation}
$g_{\pm}=g_{\pm}(\theta)$, and 
$f_{\pm}=f_{\pm}(\theta)$ are given by 
\begin{equation}
g_{+}\left(\theta \right)=g_{-}\left(-\theta \right),  \ 
f_{\pm}\left(\theta \right)=f_{\mp}\left(-\theta \right)
\label{evenparity}
\end{equation}
for spin-singlet even-parity superconductors and 
\begin{equation}
g_{+}\left(\theta \right)=g_{-}\left(-\theta \right), \ 
f_{\pm}\left(\theta \right)=-f_{\mp}\left(-\theta \right)
\label{oddparity}
\end{equation}
for spin-triplet odd-parity superconductors \cite{Proximityp2}. 
Then, $c_{1}=c_{1}(\theta)$, $c_{2}=c_{2}(\theta)$, 
and $c_{3}=c_{3}(\theta)$ 
satisfy
\begin{equation}
c_{1}(\theta)=-c_{1}(-\theta), \ 
c_{2}(\theta)=c_{2}(-\theta), \ 
c_{3}(\theta)=c_{3}(-\theta)
\label{coefficients2}
\end{equation}
for spin-singlet even-parity superconductor and 
\begin{equation}
c_{1}(\theta)=c_{1}(-\theta), \ 
c_{2}(\theta)=-c_{2}(-\theta), \ 
c_{3}(\theta)=c_{3}(-\theta)
\label{coefficients3}
\end{equation}
for spin-triplet odd-parity superconductor. 
From Eq. (\ref{ballancerelation}), we can determine 
the relation between $s_{1}$ and $s_{2}$ 
written by 
\begin{equation}
\left<
\frac{
\sigma_{1N}
\left( 
c_{1}\left(\theta \right)s_{2}
-c_{2}\left(\theta \right)s_{1}
\right)
}
{2 -\sigma_{1N} + \sigma_{1N}
\left(
c_{2}\left(\theta \right)s_{2} + c_{3}\left(\theta \right)s_{3} 
+ c_{1}\left(\theta \right)s_{1} \right)
}
\right>=0
\label{angularaverage1}
\end{equation}
with $\sigma_{1N}=\sigma_{1N}(\theta)=\sigma_{1N}(-\theta)$.\par
We decompose the denominator 
of eq. (\ref{angularaverage1})
by the summation of $d_{e}(\theta)$ and 
$d_{o}(\theta)$ 
with
\begin{equation}
d_{e}\left(\theta \right)=d_{e}\left(-\theta \right), \ \ 
d_{o}\left(\theta \right)=-d_{o}\left(-\theta \right). 
\end{equation}
For spin-singlet even-parity case, 
using eq. (\ref{coefficients2})
$d_{e}(\theta)$ and $d_{o}(\theta)$ are given by 
\begin{equation}
d_{e}(\theta)=\left(2-\sigma_{1N}\right)+ 
\sigma_{1N}\left[c_{2}\left(\theta\right)s_{2}
+ c_{3}\left(\theta\right) \right], \ \ 
d_{o}\left(\theta\right)
=\sigma_{1N}c_{1}\left(\theta\right).
\end{equation}
Then, 
\begin{eqnarray}
&&
\left<\frac{\sigma_{1N}
\left(c_{1}\left(\theta \right)s_{2}-c_{2}\left(\theta \right)
s_{1}\right)}
{d_{e}\left(\theta\right) + d_{o}\left(\theta \right)s_{1}}
\right>
\nonumber
\\
&=&
\frac{1}{2}
\left[
\left<
\frac{\sigma_{1N}
\left[c_{1}\left(\theta \right)s_{2}-c_{2}\left(\theta \right)
s_{1}\right]}
{d_{e}\left(\theta\right) + d_{o}\left(\theta \right)s_{1}}
\right>
+
\left<\frac{\sigma_{1N}
\left[-c_{1}\left(\theta \right)s_{2}-c_{2}\left(\theta \right)
s_{1}\right]}
{d_{e}\left(\theta\right) - d_{o}\left(\theta \right)s_{1}}
\right>
\right]
\nonumber
\\
&=&
-\left<
\frac{
\sigma_{1N}
\left[ d_{e}\left(\theta\right)c_{2}\left(\theta\right)
+d_{o}\left(\theta\right)c_{1}\left(\theta\right)s_{2}
\right]}
{d_{e}^{2}\left(\theta \right) -s_{1}^{2}d_{o}^{2}(\theta)}
\right>s_{1}=0.
\end{eqnarray}
From this relation, we obtain $s_{1}=0$ \cite{Proximityp2}. \par
On the other hand, for spin-triplet odd-parity pairing case, 
using eq. (\ref{coefficients3}), 
$d_{e}(\theta)$ and $d_{odd}(\theta)$ are given by 
\begin{equation}
d_{e}(\theta)=\left(2-\sigma_{1N}\right)+ 
\sigma_{1N}\left[c_{1}\left(\theta\right)s_{1}
+ c_{3}\left(\theta\right) \right], \ \ 
d_{o}\left(\theta\right)
= \sigma_{1N}c_{2}\left(\theta\right)
\end{equation}
\begin{eqnarray}
&&
\left<\frac{\sigma_{1N}
\left(c_{1}\left(\theta \right)s_{2}-c_{2}\left(\theta \right)
s_{1}\right)}
{d_{e}\left(\theta\right) + d_{o}\left(\theta \right)s_{2}}
\right>
\nonumber
\\
&=&
\frac{1}{2}
\left[
\left<\frac{\sigma_{1N}
\left[c_{1}\left(\theta \right)s_{2}-c_{2}\left(\theta \right)
s_{1}\right]}
{d_{e}\left(\theta\right) + d_{o}\left(\theta \right)s_{2}}
\right>
+
\left<\frac{\sigma_{1N}
\left[c_{1}\left(\theta \right)s_{2}+c_{2}\left(\theta \right)
s_{1}\right]}
{d_{e}\left(\theta\right) - d_{o}\left(\theta \right)s_{2}}
\right>
\right]
\nonumber
\\
&=&
\left<
\frac{
\sigma_{1N}
\left[ d_{e}\left(\theta\right)c_{1}\left(\theta\right)
+d_{o}\left(\theta\right)c_{2}\left(\theta\right)s_{1}
\right]}
{d_{e}^{2}\left(\theta \right) -s_{2}^{2}d_{o}^{2}(\theta)}
\right>s_{2}=0.
\end{eqnarray}
From this relation, we obtain $s_{2}=0$ \cite{Proximityp2}. \par
This means $\cos \psi=0$ for spin-singlet even-parity pairing 
and $\sin \psi=0$ for spin-triplet odd-parity pairing, respectively. 
To summarise 
$\hat{R}_{1}(x)$ becomes 
\begin{equation}
\hat{R}_{1}\left( x \right)
=s_{2}\left( x \right)
\hat{\tau}_{2} + s_{3}\left( x \right)\hat{\tau}_{3}
\label{R1structure1}
\end{equation}
for a spin-singlet  superconductor and 
\begin{equation}
\hat{R}_{1}\left(x\right)
=s_{1}\left( x \right)\hat{\tau}_{1} 
+ s_{3}\left( x \right)\hat{\tau}_{3}
\label{R1structure2}
\end{equation}
for a spin-triplet one consistent with previous results 
\cite{Proximityp,Proximityp2,Proximityd2}. \par
Here, let us discuss about this physical meaning of the symmetry of a Cooper pair. In the DN, only $s$-wave pairing is possible due to the 
impurity scattering. 
Since there is no spin flip scattering at the interface, the symmetry of spin structure in DN is equivalent to that in the superconductor. 
It is noted that for the spin-singlet superconductor case, 
$\hat{R}_{1}(x)$ is expressed by 
$\hat{\tau}_{2}$ and $\hat{\tau}_{3}$ similar to 
bulk superconductor. This means that the 
symmetry of the Cooper pair in the DN is equivalent to that of the bulk,  
where the pairing symmetry is spin-singlet even-parity.  
On the other hand, for the spin-triplet superconductor case, 
$\hat{R}_{1}$ is expressed by $\hat{\tau}_{1}$ and $\hat{\tau}_{3}$,   
different from $\hat{R}_{2\pm}$. 
This implies that the different symmetry of Cooper pair, $i.e.$, an odd-frequency spin-triplet $s$-wave pair, is generated in the DN \cite{odd1,tanaka12}. \par
The boundary condition of $\zeta(x)$ is given by 
\cite{Proximityd,Proximityd2,Proximityp,Proximityp2,Proximityp3}
\begin{equation}
\frac{L}{R_{d}}\left. \left( \frac{\partial \zeta\left(x\right) }{\partial x}\right) 
\right|_{x=0_{-}}=
\frac{\langle F_{1}\rangle }{R_{b}},  
\label{usadeleq.2}
\end{equation}%
\begin{equation}
F_{1}=\frac{2\sigma_{N}\left(f_{S}\cos \zeta _{N}-g_{S}\sin \zeta _{N} \right)}{%
2-\sigma_{N}+\sigma_{N}\left(\cos \zeta_{N}g_{S}+\sin \zeta_{N}f_{S} \right)}
\label{boundary},
\end{equation}
with $\zeta(x=0_{-})=\zeta_{N}$. 
If we denote $F_{1}=F_{1}(\theta)$, 
the angular average is expressed by 
\begin{equation}
\langle F_{1}\left(\theta \right)\rangle 
=\frac{ \int_{-\pi /2}^{\pi /2}d\theta 
	\cos \theta
	F_{1}\left(\theta \right)}{
	\int_{-\pi /2}^{\pi /2}d\theta \sigma_{N}\cos \theta}. 
\label{usadelboundaryaverage}
\end{equation}%

Here $g_{S}$, and $f_{S}$ are 
\begin{equation}
g_{S}=
\left\{
\begin{array}{cc}
\left(g_{+}+g_{-} \right)
/\left(1+g_{+}g_{-}+f_{+}f_{-}\right) & {\rm spin-triplet} \\
\left(g_{+}+g_{-} \right)/\left(1+g_{+}g_{-}+f_{+}f_{-}\right) & 
{\rm spin-singlet} 
\end{array}
\right.
\label{gs}.
\end{equation}
\begin{equation}
f_{S}=
\left\{
\begin{array}{cc}
i\left(f_{+}g_{-}-f_{-}g_{+} \right)/\left(1+g_{+}g_{-}+f_{+}f_{-}\right) & 
{\rm spin-triplet} \\
\left(f_{+}+f_{-}\right)/\left(1+g_{+}g_{-}+f_{+}f_{-}\right) & 
{\rm spin-singlet} 
\end{array}
\right. 
\label{fs}.
\end{equation}
Next, let us calculate ${\rm Trace}(\hat{I}_{K}\hat{\tau}_{3})$ 
by using $\zeta(x=0_{-})$, $s_{1}=s_{1}(x=0_{-})$, 
$s_{2}=s_{2}(x=0_{-})$, and 
$s_{3}=s_{3}(x=0_{-})$. 

The details of the calculation are shown in 
Appendix D.
${\rm Trace}(\hat{I}_{K}\hat{\tau}_{3})$ is given by 
\begin{eqnarray}
\left<{\rm Trace}\left[\hat{I}_{K}\hat{\tau}_{3}\right]\right>
&=&
\left<
\frac{2 \sigma_{1N}}{\mid d_{R} \mid^{2}}
{\rm Trace}
\left[ 
\left( \hat{R}_{1} + \hat{R}^{\dagger}_{1} \right)
\left( \hat{C}_{R} + \hat{C}^{\dagger}_{R} \right) 
\right]
\left( 1 + \sigma_{1N}^{2}\right) 
\right>
f_{3N}\left(x \right)
\nonumber
\\
&+&
\left<
\frac{4 \sigma_{1N}}{\mid d_{R} \mid^{2}}
{\rm Trace}
\left[
\left( 
\hat{\mathbb{I}} + \hat{R}^{\dagger}_{1}\hat{R}_{1} 
\right)
\left( \hat{\mathbb{I}}
+ \hat{C}^{\dagger}_{R}\hat{C}_{R} \right)
\right]
\sigma_{1N}
\right>f_{3N}\left(x \right).
\label{Tanakamatrixcurrentcharge3parity}
\end{eqnarray}
If we define $I_{K}$ as  
\[
I_{K}=\frac{1}{4}
{\rm Trace}\left[\hat{I}_{K}\hat{\tau}_{3}\right]
\]
the total resistance of the junction 
$R$ is given by \cite{Proximityd2,Proximityp2}
\begin{equation}
R=\frac{R_{b}}{\left<I_{K} \right>}
+\frac{R_{d}}{L}\int_{-L}^{0}\frac{dx}{\cosh
^{2}\zeta_{\rm{im}}\left(x \right)},
\label{resistance}
\end{equation}
\begin{align}
\langle I_{K} \rangle =&\left\langle
\frac{\sigma_{N}}{2}\frac{C_{0}}
{\mid (2-\sigma_{N}) + \sigma_{N} (\cos \zeta_{N}g_{S}
+\sin\zeta_{N}f_{S}) \mid ^{2}} \right\rangle,
\label{ib0}\\
C_{0}=&\sigma_{N} \left(1+\mid \cos \zeta_{N}\mid ^{2}+\mid \sin \zeta_{N}\mid ^{2}\right)
\left[ \mid g_{S}\mid ^{2}+\mid f_{S}\mid ^{2}+1
+\mid \bar{f}_{S}\mid ^{2}\right] \nonumber\\
&+4\left(2-\sigma_{N} \right)
\left[\mathrm{Real}\left(g_{S} \right)\mathrm{Real}\left(\cos\zeta_{N} \right) 
+ \mathrm{Real}\left(f_{S} \right)\mathrm{Real}\left(\sin\zeta_{N} \right) \right]\nonumber \\
&+ 4\sigma_{N} \mathrm{Imag}\left(f_{S}g_{S}^{*} \right)
\mathrm{Imag}\left(\cos\zeta_{N}\sin\zeta_{N}^{*} \right),
\label{C0} 
\end{align}
\cite{Proximityp2}. 
Here, $\bar{f}_{S}$ is given by 
\begin{equation}
\bar{f}_{S}=
\left\{
\begin{array}{cc}
\left(f_{+}+f_{-}\right)/\left(1+g_{+}g_{-}+f_{+}f_{-}\right) & 
{\rm spin-triplet} \\
i\left(f_{+}g_{-}- g_{+}f_{-}\right)/\left(1+g_{+}g_{-}+f_{+}f_{-}\right) & 
{\rm spin-singlet} 
\end{array}
\right. 
\label{bfs}.
\end{equation}
\par
Using $\Gamma_{\pm}$ defined by 
\begin{equation}
\Gamma_{\pm}=\frac{\Delta_{\pm}}{\varepsilon + \Omega_{\pm}}
\label{definitionGamma},
\end{equation}
$g_{S}$, $f_{S}$, $\bar{f}_{S}$ are given by
\begin{equation}
g_{S}=
\left\{
\begin{array}{cc}
\left(1 + \Gamma_{+}\Gamma_{-} \right)
/\left(1 - \Gamma_{+}\Gamma_{-} \right) & {\rm spin-triplet} \\
\left(1 + \Gamma_{+}\Gamma_{-} \right)
/\left(1 - \Gamma_{+}\Gamma_{-} \right)
& {\rm spin-singlet} 
\end{array}
\right. 
\label{gsgamma},
\end{equation}
\begin{equation}
f_{S}=
\left\{
\begin{array}{cc}
\left(\Gamma_{-} -  \Gamma_{+}\right)/\left( 1 -  \Gamma_{+}\Gamma_{-} \right)
& {\rm spin-triplet} \\
i\left(\Gamma_{+} +  \Gamma_{-}\right)/\left( 1 - \Gamma_{+}\Gamma_{-} \right)
& {\rm spin-singlet} 
\end{array}
\right. 
\label{fsgamma},
\end{equation}
\begin{equation}
\bar{f}_{S}=
\left\{
\begin{array}{cc}
i\left(\Gamma_{+} +  \Gamma_{-}\right)/\left( 1 - \Gamma_{+}\Gamma_{-} \right)
& {\rm spin-triplet} \\
\left(\Gamma_{-} -  \Gamma_{+}\right)/\left( 1 - \Gamma_{+}\Gamma_{-} \right)
& 
{\rm spin-singlet} 
\end{array}
\right. 
\label{bfsgamma}.
\end{equation}
Then, $I_{K}$ is given by 
\begin{align}
I_{K}&=\frac{\sigma_{N}}{2}\times
\nonumber \\
&\frac{C_{1}}
{\left| 1 - \left(1 - \sigma_{N} \right) \Gamma_{+}\Gamma_{-}
+ \sigma_{N} \sin \left(\frac{\zeta_{N}}{2}\right)
\left[-\left(1 +\Gamma_{+}\Gamma_{-} \right)
\sin\left( \frac{\zeta_{N}}{2} \right)
+ i \left( \Gamma_{+} + \Gamma_{-} \right)
\cos\left( \frac{\zeta_{N}}{2} \right)
\right]
\right|^{2}
}
\end{align}
\begin{eqnarray}
C_{1} &=& 2\left[ 1 + \sigma_{N} \mid \Gamma_{+}\mid^{2} + 
\left(\sigma_{N}-1 \right)\mid \Gamma_{+} \Gamma_{-}\mid^{2} \right] 
\nonumber \\
& + & 
\sigma_{N} \sinh^{2}\left(\zeta_{Ni}\right)
\left(1 + \mid \Gamma_{+} \mid^{2} \right)
\left(1 + \mid \Gamma_{-} \mid^{2} \right) 
\nonumber \\
& + &
\left(2 - \sigma_{N} \right)\left(1 - \mid \Gamma_{+} \Gamma_{-}\mid^{2} \right)
\left[ \cos \zeta_{Nr} \cosh \zeta_{Ni} - 1 \right] 
\nonumber \\
& - &
\left(2 - \sigma_{N} \right) \cosh \zeta_{Ni} \sin \zeta_{Nr} 
{\rm Imag}\left[ \left(\Gamma_{+} + \Gamma_{-} \right)\left(1 - \Gamma^{*}_{+}\Gamma^{*}_{-}\right) 
\right] 
\nonumber \\
& - & 
\sigma_{N}\cosh \zeta_{Ni} \sinh \zeta_{Ni} {\rm Real}
\left[ \left(\Gamma_{+} + \Gamma_{-} \right)
\left(1 + \Gamma^{*}_{+}\Gamma^{*}_{-}\right)  
\right], 
\label{Ib0spinsinglet}
\end{eqnarray}
for a spin-singlet superconductor and 
\begin{align}
I_{K}&=\frac{\sigma_{N}}{2}\times
\nonumber \\
&\frac{C_{1}}
{\left| 1 - \left(1 - \sigma_{N} \right) \Gamma_{+}\Gamma_{-}
	+ \sigma_{N} \sin \left(\frac{\zeta_{N}}{2}\right)
	\left[-\left(1 +\Gamma_{+}\Gamma_{-} \right)
	\sin\left( \frac{\zeta_{N}}{2} \right)
	-\left( \Gamma_{+} - \Gamma_{-} \right)
	\cos\left( \frac{\zeta_{N}}{2} \right)
	\right]
	\right|^{2}
}
\end{align}
\begin{eqnarray}
C_{1} &=& 2\left[ 1 + \sigma_{N} \mid \Gamma_{+}\mid^{2} + 
\left(\sigma_{N}-1 \right)\mid \Gamma_{+} \Gamma_{-}\mid^{2} \right]
\nonumber \\
& + & 
\sigma_{N} \sinh^{2}\left(\zeta_{Ni}\right)
\left(1 + \mid \Gamma_{+} \mid^{2} \right)
\left(1 + \mid \Gamma_{-} \mid^{2} \right)
\nonumber \\
& + &
\left(2 - \sigma_{N} \right)\left(1 - \mid \Gamma_{+} \Gamma_{-}\mid^{2} \right)
\left[ \cos \zeta_{Nr} \cosh \zeta_{Ni} - 1 \right]
\nonumber \\
& - &
\left(2 - \sigma_{N} \right) \cosh \zeta_{Ni} \sin \zeta_{Nr} 
{\rm Real}\left[ \left(\Gamma_{+} - \Gamma_{-} \right)\left(1 - \Gamma^{*}_{+}\Gamma^{*}_{-}\right) 
\right]
\nonumber \\
& - & 
\sigma_{N}\cosh \zeta_{Ni} \sinh \zeta_{Ni} {\rm Imag}
\left[ \left(\Gamma^{*}_{+} - \Gamma^{*}_{-} \right)
\left(1 + \Gamma_{+}\Gamma_{-}\right) \
\right]
\label{Ib0spintriplet}
\end{eqnarray}
for a spin-triplet superconductor. 
Here, $\zeta_{Nr}$, $\zeta_{Ni}$ denote the 
real and imaginary part of 
$\zeta_{N}$, respectively. 
We have used 
$\Gamma_{+}(\theta)=\Gamma_{-}(-\theta)$ for spin-singlet superconductor
and $\Gamma_{+}(\theta)=-\Gamma_{-}(-\theta)$ for spin-triplet supercopnductor 
with $\Gamma_{+}=\Gamma(\theta)$, $\Gamma_{-}=\Gamma(\pi-\theta)$ and the 
injection angle $\theta$. 
In the case for $R_{d}=0$, 
$\zeta_{N}=0$, $\zeta_{Nr}=0$, and $\zeta_{Ni}=0$ are satisfied. 
Then, $I_{K}$ reproduces the formula obtained in  
ballistic normal metal / 
unconventional superconductor junctions \cite{TK95,KT96}. 
\par
\section{Charge conductance in non-centrosymmetric superconductor junctions}
Since we get a more compact expression of the matrix current 
$\check{I}(\theta)$ as shown in eqs. (\ref{Tanakamatrixretarded}) 
and (\ref{Tanakamatrixcurrentcharge3}) 
as compared to the previous one 
\cite{Proximityd2,Proximityp2}, 
it is possible to challenge a more complicated system. 
In this section, we apply our boundary condition to a 
mixed parity superconductor junction. 
Mixed parity state like $s+p$-wave pairing is possible in 
non-centrosymmetric superconductors. 
Here, we assume that $S_{z}=0$ for the spin-triplet pair potential where the $d$-vector of spin-triplet pair potential is along the $z$-direction. 
In this case, we can discuss the retarded part of the Green's function by a
$2 \times 2$ matrix denoting the 
spin index. 
To elucidate the charge conductance and LDOS based on analytical calculation 
in the limiting case, we focus on $s+p$-wave superconductor model in 1D.  
It is an interesting issue to clarify whether the anomalous proximity effect 
predicted in spin-triplet superconductor junction 
\cite{Proximityp2,Proximityd2} 
is robust with the 
inclusion of the additional $s$-wave component. \par
Here, $\Delta_{+}$ and $\Delta_{-}$ are given by 
\[
\Delta_{+}=\Delta_{s} + \Delta_{p}, \ 
\Delta_{-}=\Delta_{s} - \Delta_{p}
\]
for up-spin sector 
and 
these are given by 
\[
\Delta_{+}=-\Delta_{s} + \Delta_{p}, \ 
\Delta_{-}=-\Delta_{s} - \Delta_{p}
\]
for down-spin one. 
If we denote quasiclassical Green's function for up and down spin sector as 
\[
g_{\pm\uparrow\left(\downarrow \right)}
\hat{\tau}_{3} + f_{\pm\uparrow\left(\downarrow \right)}\hat{\tau}_{2},
\]
$g_{\pm\uparrow(\downarrow)}$
and $f_{\pm\uparrow(\downarrow)}$
are given by 
\begin{eqnarray}
g_{+ \uparrow}
&=&g_{- \downarrow} = g_{+}, \ 
g_{- \uparrow}
=g_{+ \downarrow} = g_{-} ,
\nonumber
\\ 
f_{+ \uparrow}
&=&-f_{- \downarrow} = f_{+}, \ 
f_{- \uparrow}
=-f_{+ \downarrow} = f_{-} ,
\end{eqnarray}
with 
\begin{equation}
f_{\pm}=\frac{\Delta_{\pm}}{\sqrt{\Delta^{2}_{\pm}-\varepsilon^{2}}}, 
\ 
g_{\pm}=\frac{\varepsilon}{\sqrt{\varepsilon^{2} - \Delta^{2}_{\pm}}} .
\end{equation}
We define $\hat{C}_{R\uparrow(\downarrow)}$ for up(down) spin sector
as 
\[
C_{R\uparrow\left(\downarrow\right)}=c_{1\uparrow(\downarrow)}\hat{\tau}_{1} + 
c_{2\uparrow(\downarrow)}\hat{\tau}_{2} 
+ c_{3\uparrow(\downarrow)}\hat{\tau}_{3}.
\]
By using coefficients defined in eq. (\ref{coefficient1}), 
we derive
\begin{eqnarray}
c_{1\uparrow}&=&c_{1\downarrow}=c_{1}, \ 
c_{2\uparrow}=-c_{2\downarrow}=c_{2}, 
\nonumber
\\ 
c_{3\uparrow}&=&c_{3\downarrow}=c_{3}.
\label{coefficientsCR}
\end{eqnarray}
In DN side, we write 
Green's function of Usadel equation 
as 
\[
\hat{R}_{1\uparrow(\downarrow)}\left( x \right)
=s_{1\uparrow(\downarrow)}\hat{\tau}_{1} + 
s_{2\uparrow(\downarrow)}\hat{\tau}_{2} 
+ s_{3\uparrow(\downarrow)}\hat{\tau}_{3}
\]
with
\begin{equation}
s_{1\uparrow}=\cos\psi_{\uparrow}\sin\zeta_{\uparrow}, \ 
s_{2\uparrow}=\sin\psi_{\uparrow}\sin\zeta_{\uparrow}, \ 
s_{3\uparrow}=\cos\zeta_{\uparrow}
\end{equation}
and
\begin{equation}
s_{1\downarrow}=\cos\psi_{\downarrow}\sin\zeta_{\downarrow}, \ 
s_{2\downarrow}=\sin\psi_{\downarrow}\sin\zeta_{\downarrow}, \ 
s_{3\downarrow}=\cos\zeta_{\downarrow}.
\end{equation}
Due to the absence of a supercurrent, 
$\psi_{\uparrow\left(\downarrow \right)}$ 
is independent of $x$. 
The relation of the 
$\hat{\tau}_{3}$ component of the 
boundary condition of eq. (\ref{angularaverage1}), 
is greatly simplified in 1d model case with
\[
c_{1\uparrow}\sin\psi_{\uparrow}=
c_{2\uparrow}\cos\psi_{\uparrow} \ \ 
c_{1\downarrow}\sin\psi_{\downarrow}=
c_{2\downarrow}\cos\psi_{\downarrow}.
\]
Then, we obtain 
\begin{eqnarray}
\cos \psi_{\uparrow}
&=& 
\frac{c_{1\uparrow}}
{\sqrt{c^{2}_{1\uparrow}+c^{2}_{2\uparrow}}}
=\frac{c_{1}}{\sqrt{c^{2}_{1}+c^{2}_{2}}}=\cos \psi ,
\nonumber
\\ 
\cos \psi_{\downarrow}
&=&
\frac{c_{1\downarrow}}
{\sqrt{c^{2}_{1\downarrow}+c^{2}_{2\downarrow}}}
=\frac{c_{1}}{\sqrt{c^{2}_{1}+c^{2}_{2}}}=\cos \psi .
\end{eqnarray}

\begin{eqnarray}
\sin \psi_{\uparrow}
&=& 
\frac{c_{2\uparrow}}{\sqrt{c^{2}_{1\uparrow}+c^{2}_{2\uparrow}}}
=\frac{c_{2}}{\sqrt{c^{2}_{1}+c^{2}_{2}}}=\sin \psi 
\nonumber
\\
\sin \psi_{\downarrow}
&=&
\frac{c_{2\uparrow}}{\sqrt{c^{2}_{1\downarrow}+c^{2}_{2\downarrow}}}
=-\frac{c_{2}}{\sqrt{c^{2}_{1}+c^{2}_{2}}}=-\sin \psi. 
\end{eqnarray}
If we define $\alpha_{\uparrow(\downarrow)}$ as follows 
\begin{eqnarray}
\hat{C}_{R\uparrow\left(\downarrow \right)}
\hat{R}_{1\uparrow\left(\downarrow \right)}
+ \hat{R}_{1\uparrow\left(\downarrow \right)}
\hat{C}_{R\uparrow\left(\downarrow \right)}
=2\alpha_{\uparrow\left(\downarrow \right)}\hat{\mathbb{I}},
\end{eqnarray}
$\alpha_{\uparrow}$ and $\alpha_{\downarrow}$ are given by 
\begin{eqnarray}
\alpha_{\uparrow}
&\equiv&
\left(c_{1\uparrow}\cos\psi_{\uparrow} 
+ c_{2\uparrow} \sin\psi_{\uparrow} \right)
\sin \zeta_{\uparrow} + c_{3\uparrow} \cos \zeta_{\uparrow}
\nonumber
\\
&=&
\left(c_{1}\cos\psi + c_{2} \sin\psi \right)
\sin \zeta_{\uparrow} + c_{3} \cos \zeta_{\uparrow}
\nonumber
\\
&=&
\sqrt{c_{1}^{2}+c_{2}^{2}}\sin\zeta_{\uparrow} + 
c_{3}\cos\zeta_{\uparrow},
\end{eqnarray}
\begin{eqnarray}
\alpha_{\downarrow}
&\equiv&
\left(c_{1\downarrow}\cos\psi_{\downarrow} 
+ c_{2\downarrow} \sin\psi_{\downarrow} \right)
\sin \zeta_{\downarrow} + c_{3\downarrow} \cos \zeta_{\downarrow} 
\nonumber
\\
&=&
\left(c_{1}\cos\Psi + c_{2} \sin\psi \right)
\sin \zeta_{\downarrow} + c_{3} \cos \zeta_{\downarrow} 
\nonumber
\\
&=&
\sqrt{c_{1}^{2}+c_{2}^{2}}\sin\zeta_{\downarrow} + 
c_{3}\cos\zeta_{\downarrow}.
\end{eqnarray}
In the following, 
we denote $\hat{R}_{1\uparrow(\downarrow)}(x=0_{-})
=R_{1\uparrow(\downarrow)}$, 
$\hat{K}_{1\uparrow(\downarrow)}(x=0_{-})=K_{1\uparrow(\downarrow)}$,
$\hat{A}_{1\uparrow(\downarrow)}(x=0_{-})=A_{1\uparrow(\downarrow)}$, and 
$\zeta_{\uparrow(\downarrow)}(x=0_{-})=\zeta_{N\uparrow(\downarrow)}$. 
Then, the boundary condition of $\zeta_{\uparrow(\downarrow)}(x)$
becomes 
\begin{equation}
\left.
L \left( \frac{\partial \zeta_{\uparrow}\left(x \right)}
{\partial x}
\right) \right|_{x=0_{-}}
= \frac{2 R_{d}}{R_{b}}
\frac{\sqrt{c_{1}^{2}+c_{2}^{2}}\cos\zeta_{N\uparrow} 
-c_{3}\sin\zeta_{N\uparrow}}
{2 - \sigma_{N} + \sigma_{N}
\left(
\sqrt{c_{1}^{2}+c_{2}^{2}}\sin\zeta_{N\uparrow} 
+ c_{3}\cos\zeta_{N\uparrow}
\right)
}
\label{boundarys+p1}
\end{equation}

\begin{equation}
\left.
L \left( \frac{\partial \zeta_{\downarrow}\left(x \right)}
{\partial x}
\right) \right|_{x=0_{-}}
= \frac{2 R_{d}}{R_{b}}
\frac{\sqrt{c_{1}^{2}+c_{2}^{2}}\cos\zeta_{N\downarrow} 
-c_{3}\sin\zeta_{N\downarrow}}
{2 - \sigma_{N} + \sigma_{N}
\left( \sqrt{c_{1}^{2}+c_{2}^{2}}\sin\zeta_{N\downarrow} 
+ c_{3}\cos\zeta_{N\downarrow}
\right)
}
\label{boundarys+p2}
\end{equation}
and
\begin{equation}
\zeta_{\uparrow}\left(x=-L \right)
=
\zeta_{\downarrow}\left(x=-L \right)
=0.
\end{equation}
Both $\zeta_{\uparrow}$ and $\zeta_{\downarrow}$ 
satisfy 
\begin{equation}
D \frac{\partial^{2}}{\partial x^{2}}\zeta_{\uparrow\left(\downarrow\right)}
\left( x \right) 
+ 2i\varepsilon \sin\zeta_{\uparrow\left(\downarrow\right)}
\left( x \right) =0. 
\label{Usadeleqspin}
\end{equation}
Then, we obtain 
\begin{equation}
\zeta_{\uparrow}\left(x \right) 
=\zeta_{\downarrow}\left(x \right)=\zeta
\end{equation}
and
\begin{equation}
s_{1\uparrow}=s_{1\downarrow}=s_{1}, \ \ 
s_{2\uparrow}=-s_{2\downarrow}=s_{2}, \ \ 
s_{3\uparrow}=s_{3\downarrow}=s_{3},
\label{relations1no1}
\end{equation}
with
\begin{equation}
s_{1}=\cos\psi\sin\zeta, \ s_{2}=\sin\psi\sin\zeta, 
s_{3}=\cos \zeta.
\label{relations1no2}
\end{equation}
Here, $s_{1}$ and $s_{2}$ express the spin-triplet pair amplitude and the spin-singlet one, respectively. Since only $s$-wave pairing 
is possible in DN, 
$s_{1}$ and $s_{2}$ correspond to the odd-frequency and even-frequency pair amplitude, respectively. Here, we show calculated results of normalized local density of states by its value in the normal state 
$\rho(\varepsilon)$ and pair amplitudes $s_{1}$ and $s_{2}$, 
where $\rho(\varepsilon)$ is given by  
\begin{equation}
\rho(\varepsilon) = {\rm Real}[\cos \zeta ]. 
\end{equation}
Here, we focus on $s_{1}$, $s_{2}$ and $\rho(\varepsilon)$ at DN/S interface 
$x=0$. We choose $E_{Th}=0.02\Delta_{0}$, $R_{d}/R_{b}=0.5$, and $Z=0.75$
in the following calculations. 
$\Delta_{s}$ and $\Delta_{p}$ are set to be 
$\Delta_{s}+\Delta_{p}=\Delta_{0}$ 
and changing their ratio. \par
\begin{figure}[t]
\begin{center}
\includegraphics[width=8cm,clip]{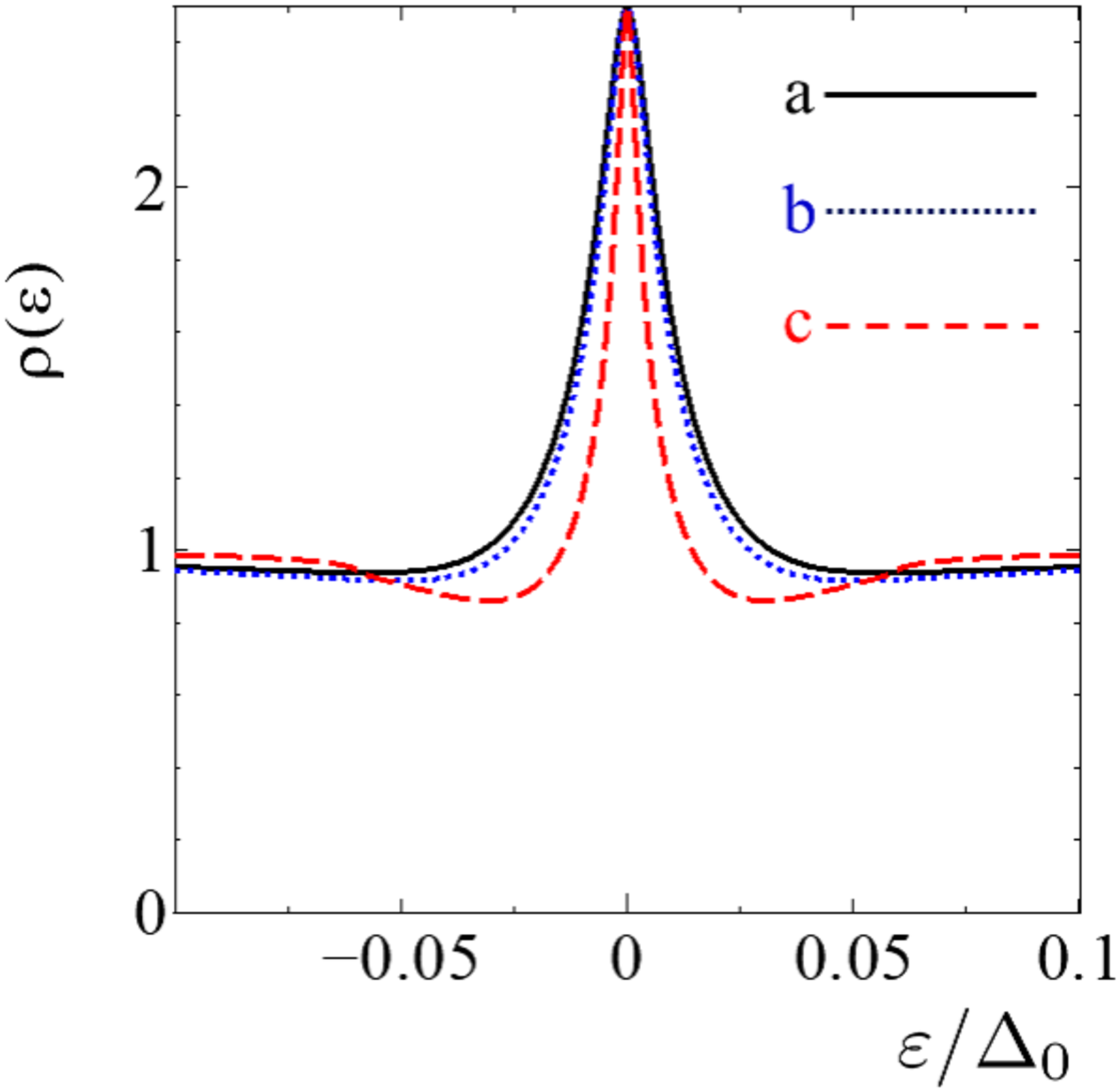}
\end{center}
\caption{Normalized local density of states $\rho(\varepsilon)$ at $x=0$ by its value in the normal state is plotted as a function of $\varepsilon$. 
$E_{Th}=0.02\Delta_{0}$, $R_{d}/R_{b}=0.5$, and $Z=0.75$.  
(a)$\Delta_{p}=\Delta_{0}$ and $\Delta_{s}=0$, 
(b)$\Delta_{p}=0.7\Delta_{0}$ and $\Delta_{s}=0.3\Delta_{0}$,
and (c)$\Delta_{p}=0.53\Delta_{0}$ and $\Delta_{s}=0.47\Delta_{0}$.
}
\label{Fig1}
\end{figure}
In Fig. \ref{Fig1}, $\rho(\varepsilon)$ is plotted for 
$\Delta_{p}>\Delta_{s}$.  
The resulting $\rho(\varepsilon)$ always has a ZEP irrespective of the 
value of $\Delta_{s}$. It is noted that $\rho(\varepsilon=0)$ is 
independent of the value of $\Delta_{s}$ for $\Delta_{p}>\Delta_{s}$. 
We can show that $c_{1}$ and $c_{3}$ in 
eqs. (\ref{boundarys+p1}) and (\ref{boundarys+p2}) are proportional to 
$1/\varepsilon$ around $\varepsilon=0$ and satisfy $c_{1}=ic_{3}$. 
Then, $\zeta$ is analytically obtained as  
\begin{equation}
\zeta=\frac{2R_{d}i}{\sigma_{N}R_{b}L}\left(x + L \right),
\label{analyticalzeta}
\end{equation}
independent of the magnitude of $\Delta_{s}$. 
The peak width of $\rho(\varepsilon)$ becomes narrower 
only in the regime where $\Delta_{s}$ becomes the same order to that of 
$\Delta_{p}$ (curve (c) in Fig. \ref{Fig1}). \par
\begin{figure}[t]
\begin{center}
\includegraphics[width=12cm,clip]{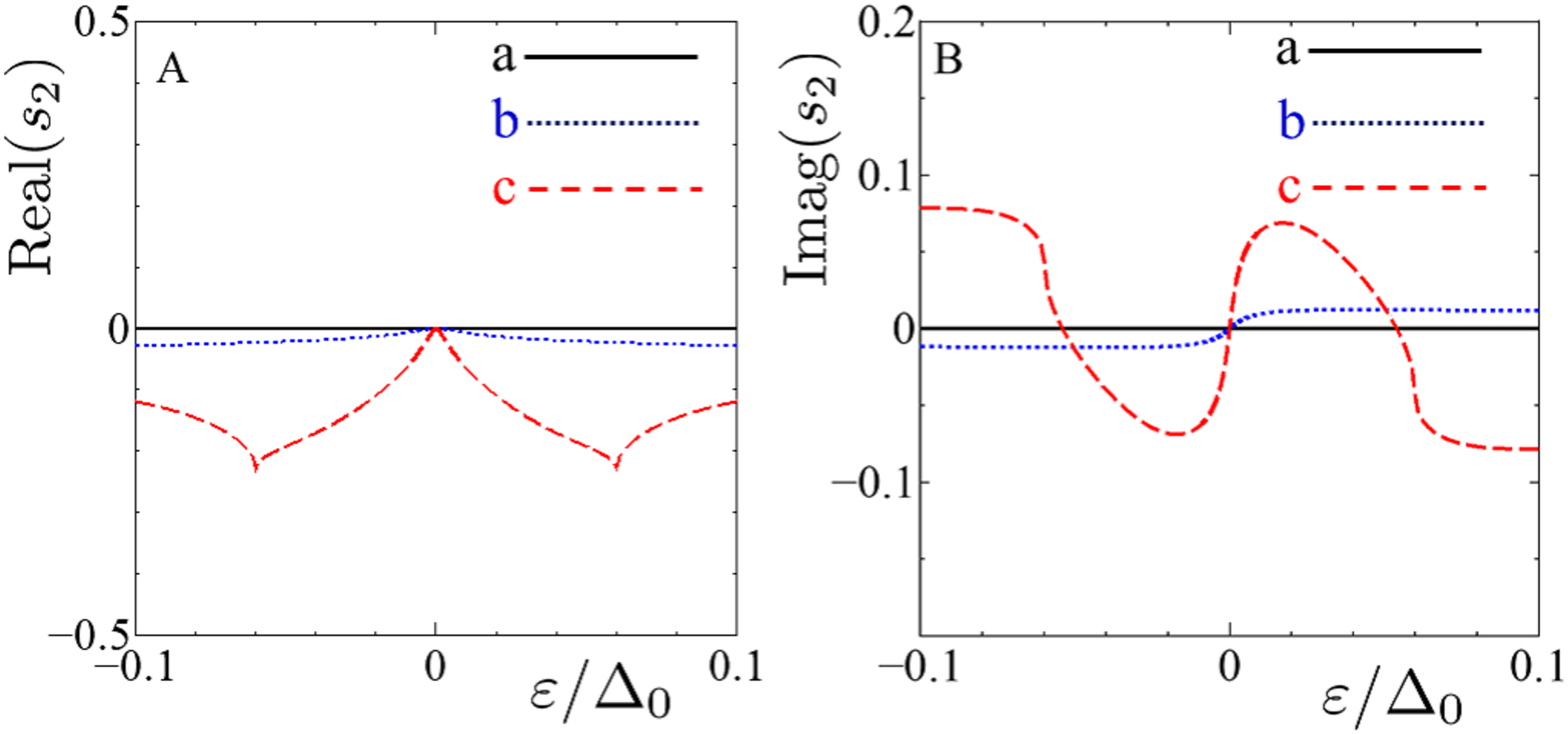}
\end{center}
\caption{Real and Imaginary parts of 
even-frequency pair amplitude  $s_{2}$ at $x=0$ is plotted as a function 
of $\varepsilon$ in Figs.2A and 2B, respectively. 
$E_{Th}=0.02\Delta_{0}$, $R_{d}/R_{b}=0.5$, and $Z=0.75$.  
(a)$\Delta_{p}=\Delta_{0}$ and $\Delta_{s}=0$, 
(b)$\Delta_{p}=0.7\Delta_{0}$ and $\Delta_{s}=0.3\Delta_{0}$,
and (c)$\Delta_{p}=0.53\Delta_{0}$ and $\Delta_{s}=0.47\Delta_{0}$.
}
\label{Fig2}
\end{figure}
In Fig.\ref{Fig2}, 
the even-frequency pair amplitude $s_{2}$ is plotted for the same parameters 
used in Fig. \ref{Fig1}. 
The real (imaginary) part of $s_{2}$ is an even (odd) function of 
$\varepsilon$. This $\varepsilon$ dependence 
is consistent with DN/$s$-wave or DN/$d$-wave superconductor junctions 
\cite{Proximityd2,odd1}. 
For $\Delta_{s}=0$, $s_{2}$ becomes zero [curves (a) in Figs.\ref{Fig2}A and 
Fig.\ref{Fig2}B] and 
the magnitude of $s_{2}$ is also suppressed for $\Delta_{s}=0.3\Delta_{0}$
[curves (b) in Figs.\ref{Fig2}A and Fig.\ref{Fig2}B]. 
Only when the magnitude of $\Delta_{s}$ becomes the same order with that of 
$\Delta_{p}$, $s_{2}$ is a little bit enhanced at nonzero $\varepsilon$ 
[curves (c) in Figs.\ref{Fig2}A and 
Fig.\ref{Fig2}B]. \par
\begin{figure}[t]
\begin{center}
\includegraphics[width=12cm,clip]{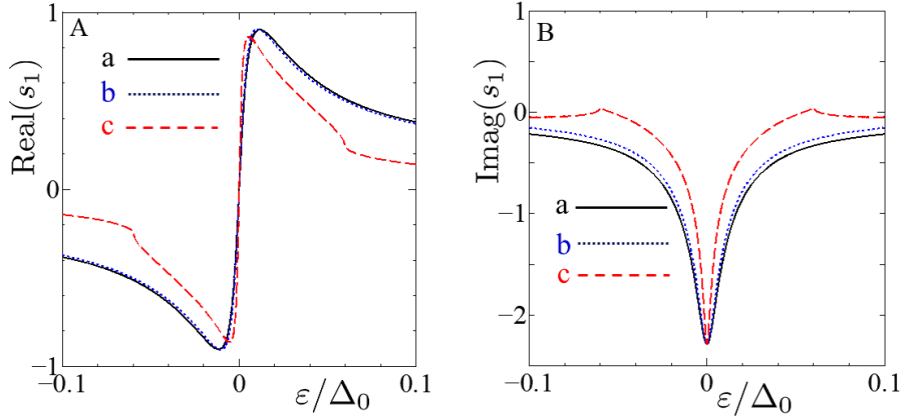}
\end{center}
\caption{Real and Imaginary parts of 
odd-frequency pair amplitude  $s_{1}$ at $x=0$ is plotted as a 
function of $\varepsilon$ in Figs.3A and 3B, respectively. 
$E_{Th}=0.02\Delta_{0}$, $R_{d}/R_{b}=0.5$, and $Z=0.75$. 
(a)$\Delta_{p}=\Delta_{0}$ and $\Delta_{s}=0$, 
(b)$\Delta_{p}=0.7\Delta_{0}$ and $\Delta_{s}=0.3\Delta_{0}$,
and (c)$\Delta_{p}=0.53\Delta_{0}$ and $\Delta_{s}=0.47\Delta_{0}$.
}
\label{Fig3}
\end{figure}
The corresponding odd-frequency pair amplitude $s_{1}$ is shown in 
Fig.\ref{Fig3}. The obtained real (imaginary) part of $s_{1}$ 
is an odd (even) function of $\varepsilon$.
These features are consistent with that in DN/p-wave 
superconductor junctions \cite{Proximityp,odd1}. 
The real part of $s_{1}$ is enhanced around $\varepsilon=0$ for all three cases  (Fig. \ref{Fig3}A). On the other hand, the magnitude of the imaginary part of $s_{1}$ has a sharp zero energy peak (ZEP) at $\varepsilon=0$. 
The value of $s_{1}$ at $\varepsilon=0$ 
is independent of $\Delta_{s}$. The peak width becomes narrower with the 
increase of $\Delta_{s}$ (Fig. \ref{Fig3}B). 
These features are quite similar to $\rho(\varepsilon)$. 
Since the magnitude of $s_{1}$ exceeds that of $s_{2}$ 
for all cases, proximity effect in this parameter region is 
governed by the odd-frequency 
pairing even in the presence of $s$-wave pair potential.   
\par
\begin{figure}[t]
\begin{center}
\includegraphics[width=8cm,clip]{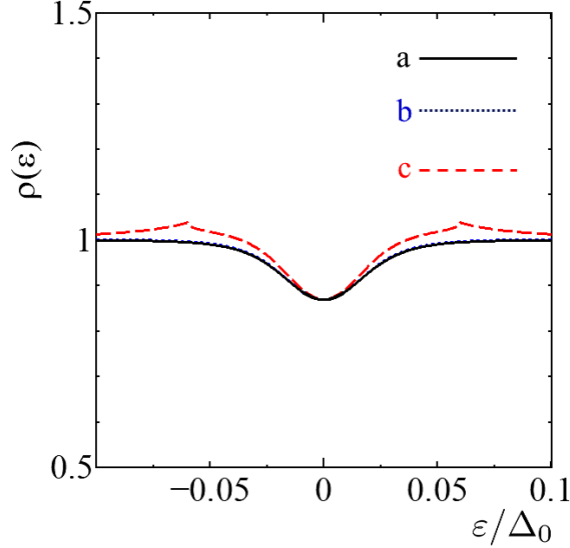}
\end{center}
\caption{Normalized local density of states $\rho(\varepsilon)$ by 
its value in the normal state is plotted as a function of $\varepsilon$. 
$E_{Th}=0.02\Delta_{0}$, $R_{d}/R_{b}=0.5$, and $Z=0.75$.  
(a)$\Delta_{s}=\Delta_{0}$ and $\Delta_{p}=0$, 
(b)$\Delta_{s}=0.7\Delta_{0}$ and $\Delta_{p}=0.3\Delta_{0}$,
and (c)$\Delta_{s}=0.53\Delta_{0}$ and $\Delta_{p}=0.47\Delta_{0}$.
}
\label{Fig4}
\end{figure}
Next, we look at the case for $\Delta_{s}>\Delta_{p}$. 
In Fig. \ref{Fig4}, $\rho(\varepsilon)$ is plotted for 
(a)$\Delta_{s}=\Delta_{0}$ and $\Delta_{p}=0$, 
(b)$\Delta_{s}=0.7\Delta_{0}$ and $\Delta_{p}=0.3\Delta_{0}$,
and (c)$\Delta_{s}=0.53\Delta_{0}$ and $\Delta_{p}=0.47\Delta_{0}$. 
$\rho(\varepsilon)$ always has a dip structure around $\varepsilon=0$. 
The curves (a) and (b) almost overlap each other within this energy window. 
The line shapes of $\rho(\varepsilon)$ is consistent  with standard proximity effect in 
DN/$s$-wave superconductor junctions. \par
\begin{figure}[t]
\begin{center}
\includegraphics[width=12cm,clip]{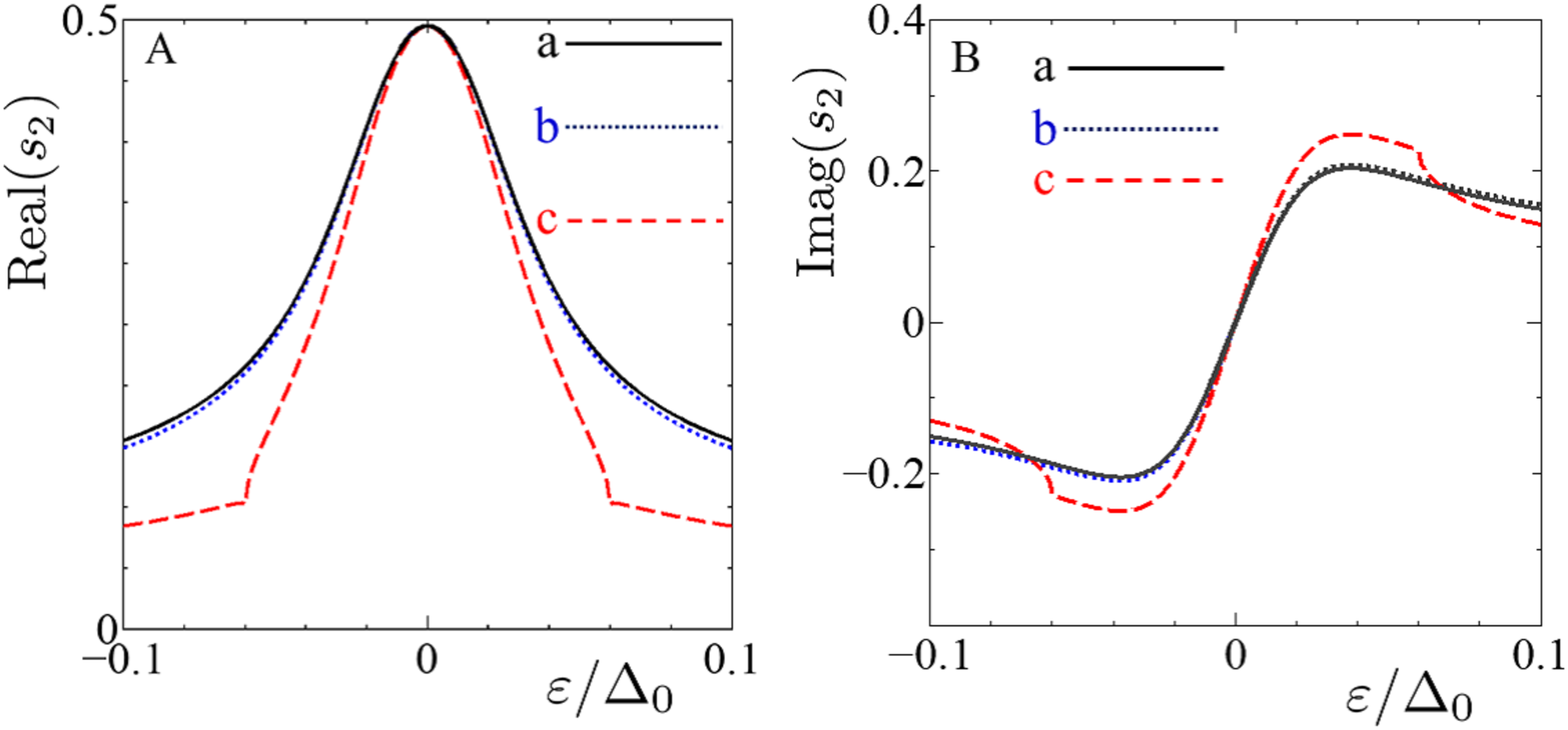}
\end{center}
\caption{Real and Imaginary parts of 
even-frequency pair amplitude  $s_{2}$ is plotted as a function 
of $\varepsilon$ in Figs.5A and 5B, respectively. 
$E_{Th}=0.02\Delta_{0}$, $R_{d}/R_{b}=0.5$, and $Z=0.75$.
(a)$\Delta_{s}=\Delta_{0}$ and $\Delta_{p}=0$, 
(b)$\Delta_{s}=0.7\Delta_{0}$ and $\Delta_{p}=0.3\Delta_{0}$,
and (c)$\Delta_{s}=0.53\Delta_{0}$ and $\Delta_{p}=0.47\Delta_{0}$.
}
\label{Fig5}
\end{figure}
In Fig.\ref{Fig5}, 
even-frequency pair amplitude $s_{2}$ is plotted for the same parameters 
used in the calculation of $\rho(\varepsilon)$ in Fig. \ref{Fig4}. 
The real (imaginary) part of $s_{2}$ is an even(odd) 
function of $\varepsilon$  similar to the case of Fig. \ref{Fig2}. 
As compared to the $p$-wave dominant case (Fig. \ref{Fig2}), 
the magnitude of $s_{2}$ is enhanced. 
The real part of $s_{2}$ has a peak at $\varepsilon=0$. 
\begin{figure}[t]
\begin{center}
\includegraphics[width=12cm,clip]{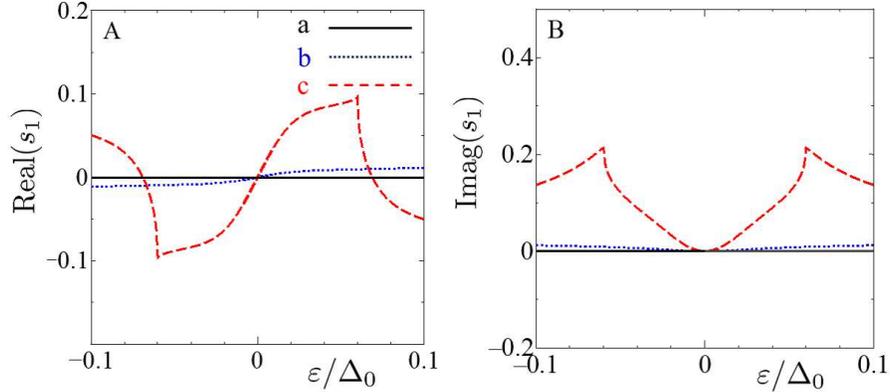}
\end{center}
\caption{Real and Imaginary parts of 
odd-frequency pair amplitude $s_{1}$ is plotted as a function of 
$\varepsilon$ in Figs.6A and 6B, respectively. 
$E_{Th}=0.02\Delta_{0}$, $R_{d}/R_{b}=0.5$, and $Z=0.75$.  
(a)$\Delta_{s}=\Delta_{0}$ and $\Delta_{p}=0$, 
(b)$\Delta_{s}=0.7\Delta_{0}$ and $\Delta_{p}=0.3\Delta_{0}$,
and (c)$\Delta_{s}=0.53\Delta_{0}$ and $\Delta_{p}=0.47\Delta_{0}$.
}
\label{Fig6}
\end{figure}
The corresponding odd-frequency pair amplitude $s_{1}$ is shown in 
Fig.\ref{Fig6}. 
The obtained real (imaginary) part of $s_{1}$ is an odd (even) function of 
$\varepsilon$ similar to the case of Fig. \ref{Fig3}. 
Without $p$-wave pair potential, $s_{1}$ vanishes as shown in 
curve (c). 
As compared to $p$-wave dominant cases [curves (a)-(c) in Fig. \ref{Fig3}], 
the magnitudes of $s_{1}$ are suppressed. 
The imaginary part of $s_{1}$ is always zero at $\varepsilon=0$. 
Since the magnitude of $s_{2}$ exceeds that of $s_{1}$ 
for all cases, proximity effect in this region is governed by even-frequency 
pairing even in the presence of $p$-wave pair potential.\par
Next, we discuss the charge conductance. 
From eqs. (\ref{relations1no1}) and (\ref{coefficientsCR}), 
we can show 
\begin{equation}
\hat{R}_{1\downarrow}\left( x \right)
=-\hat{\tau}_{2}\hat{R}_{1\uparrow}\left( x \right)
\hat{\tau}_{2}, \ \ 
\hat{C}_{R\downarrow}
=-\hat{\tau}_{2}\hat{C}_{R\uparrow}
\hat{\tau}_{2}.
\label{relationupanddown}
\end{equation}
In the following, using these relations, we calculate the 
Keldysh component of the Green's function. 
We denote $\hat{I}_{K}$ for each spin sector as 
$\hat{I}_{K\uparrow(\downarrow)}$. 
${\rm Trace}(\hat{I}_{K\uparrow(\downarrow)}\hat{\tau}_{3})$ is expressed as 
\begin{equation}
{\rm Trace}\left[ \hat{I}_{K\uparrow(\downarrow)} \hat{\tau}_{3} \right]
=\frac{2 \sigma_{1N}}
{\mid d_{R \uparrow\left(\downarrow \right)}\mid^{2}}
\left(S_{1\uparrow\left(\downarrow\right)} + 
S_{2\uparrow\left(\downarrow\right)} + S_{3\uparrow\left(\downarrow\right)} + 
S_{4\uparrow\left(\downarrow\right)} + S_{5\uparrow\left(\downarrow\right)} + 
S_{6\uparrow\left(\downarrow \right)}
\right).
\label{TracehatIK}
\end{equation}
The details of the calcuation of 
$S_{1\uparrow\left(\downarrow\right)}$,  
$S_{2\uparrow\left(\downarrow\right)}$, 
$S_{3\uparrow\left(\downarrow\right)}$, 
$S_{4\uparrow\left(\downarrow\right)}$, 
$S_{5\uparrow\left(\downarrow\right)}$, and  
$S_{6\uparrow\left(\downarrow \right)}$ 
in eq.(\ref{TracehatIK}) is shown in Appendix E. 
Summing up the contribution 
from both up and down spin sectors, 
we obtain the following relation
\begin{eqnarray}
&&
{\rm Trace}\left[ \hat{I}_{k\uparrow}\hat{\tau}_{3} \right]
+ 
{\rm Trace}\left[ \hat{I}_{k\downarrow}\hat{\tau}_{3} \right]
\nonumber
\\
&=&
\frac{4 \sigma_{1N} f_{3N}\left(x=0_{-}\right)}
{\mid d_{R} \mid^{2}}
\nonumber
\\
&\times&
{\rm Trace}
\left[
\left(\hat{R}_{1\uparrow} + \hat{R}^{\dagger}_{1\uparrow}\right)
\left(\hat{C}_{R\uparrow} + \hat{C}^{\dagger}_{R\uparrow}\right)
\left(1 + \sigma^{2}_{1N}\right)
+ 2\sigma_{1N}
\left( \hat{\mathbb{I}} 
+ \hat{C}_{R\uparrow}\hat{C}^{\dagger}_{R\uparrow} 
\right)
\right]
\label{spinsumKeldyshcomponent}.
\end{eqnarray}

\begin{figure}[t]
\begin{center}
\includegraphics[width=12cm,clip]{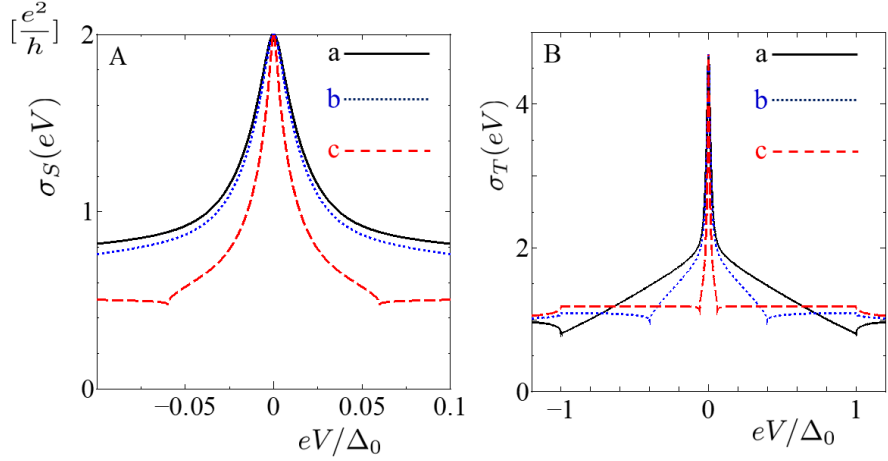}
\end{center}
\caption{Charge conductance per a spin $\sigma_{S}(eV)$ and 
normalized one by its value in the normal state $\sigma_{T}(eV)$ is plotted as a function of $\varepsilon$ in Figs.7A and 7B, respectively.  
$E_{Th}=0.02\Delta_{0}$, $R_{d}/R_{b}=0.5$, and $Z=0.75$.  
(a)$\Delta_{p}=\Delta_{0}$ and $\Delta_{s}=0$, 
(b)$\Delta_{p}=0.7\Delta_{0}$ and $\Delta_{s}=0.3\Delta_{0}$,
and (c)$\Delta_{p}=0.53\Delta_{0}$ and $\Delta_{s}=0.47\Delta_{0}$.
}
\label{Fig7}
\end{figure}

\begin{figure}[t]
\begin{center}
\includegraphics[width=12cm,clip]{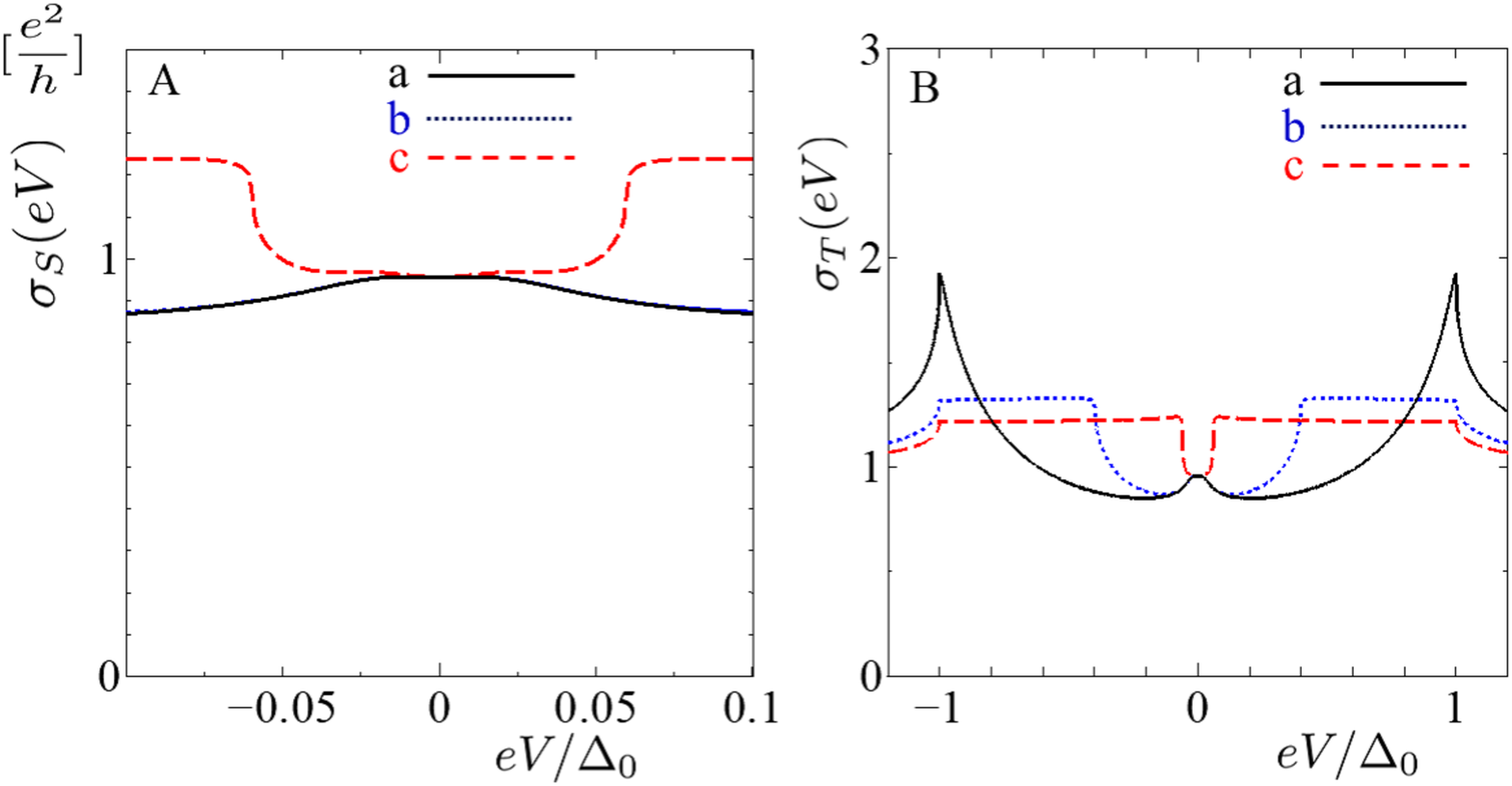}
\end{center}
\caption{$\sigma_{S}(eV)$ and 
$\sigma_{T}(eV)$ are plotted as a function of $\varepsilon$ in Figs.8A and 8B, 
respectively.  
$E_{Th}=0.02\Delta_{0}$, $R_{d}/R_{b}=0.5$, and $Z=0.75$. 
(a)$\Delta_{s}=\Delta_{0}$ and $\Delta_{p}=0$, 
(b)$\Delta_{s}=0.7\Delta_{0}$ and $\Delta_{p}=0.3\Delta_{0}$,
and 
(c)$\Delta_{s}=0.53\Delta_{0}$ and $\Delta_{p}=0.47\Delta_{0}$.
}
\label{Fig8}
\end{figure}

The boundary condition of the up spin sector of the Keldysh component at $x=0$ becomes 
\begin{equation}
\left.
\frac{L}{R_{d}}
{\rm Trace}
\left[ \hat{\tau}_{3}
\left(\hat{R}_{\uparrow}\left( x \right)
\frac{\partial}{\partial x}\hat{K}_{\uparrow}\left( x \right)
+ \hat{K}_{\uparrow}\left( x \right)\frac{\partial}{\partial x}
\hat{A}_{\uparrow}\left( x \right)
\right)
\right]
\right|_{x=0_{-}}
=-\frac{1}{R_{b}}{\rm Trace}
\left< \left[ \hat{\tau}_{3} \hat{I}_{K\uparrow} \right]
\right>
\label{boundarycondition1}.
\end{equation}
The left side of this boundary condition becomes 
\begin{equation}
\left.
-2 {\rm Imag}\left( s^{*}_{1\uparrow}s_{2\uparrow} \right)
\left( \frac{\partial f_{0N}}{\partial x} \right)
\right|_{x=0_{-}}
\left.
+2 \left( 1 + \mid s_{1\uparrow} \mid^{2} + 
\mid s_{2\uparrow} \mid^{2} + \mid s_{3\uparrow} \mid^{2}
\right) 
\left( \frac{\partial f_{3N}}{\partial x} \right)
\right|_{x=0_{-}}
\label{boundarycondition1left}.
\end{equation}
The corresponding boundary condition for down spin sector is 
\begin{equation}
\left.
\frac{L}{R_{d}}
{\rm Trace}
\left[ \hat{\tau}_{3}
\left(\hat{R}_{\downarrow}\left( x \right)
\frac{\partial}{\partial x}\hat{K}_{\downarrow}\left( x \right)
+ \hat{K}_{\downarrow}\left( x \right)\frac{\partial}{\partial x}\hat{A}_{\downarrow}
\right)
\right]
\right|_{x=0_{-}}
=-\frac{1}{R_{b}}
\left<
{\rm Trace}
\left[ \hat{\tau}_{3} \hat{I}_{K\downarrow} \right]
\right>.
\label{boundarycondition2}
\end{equation}
The left side of this boundary condition becomes 
\begin{equation}
\left.
-2 {\rm Imag}\left( s^{*}_{1\downarrow}s_{2\downarrow} \right)
\left( \frac{\partial f_{0N}}{\partial x} \right)\right|_{x=0_{-}}
\left.
+2 \left( 1 + \mid s_{1\downarrow} \mid^{2} + 
\mid s_{2\downarrow} \mid^{2} + \mid s_{3\downarrow} \mid^{2}
\right) 
\left( \frac{\partial f_{3N}}{\partial x} \right)\right|_{x=0_{-}}
\label{boundarycondition2left}.
\end{equation}
Since $s_{1\uparrow}=s_{1\downarrow}=s_{1}$ and 
$s_{2\uparrow}=-s_{2\downarrow}=s_{2}$ are satisfied, 
we obtain using eqs. (\ref{boundarycondition1left})
and (\ref{boundarycondition2left})
\begin{eqnarray}
&&
\left.
\frac{4L}{R_{d}}
\left( 1 + \mid s_{1} \mid^{2} + 
\mid s_{2} \mid^{2} + \mid s_{3} \mid^{2}
\right) 
\left( \frac{\partial f_{3N}}{\partial x} \right)
\right|_{x=0_{-}}
\nonumber
\\
&=& 
-\frac{1}{R_{b}}
\left<
{\rm Trace}
\left[ \hat{\tau}_{3} 
\left( \hat{I}_{K\uparrow} + \hat{I}_{K\downarrow} \right)
\right]
\right>
=-\frac{8}{R_{b}}
\left< I_{K} \right>,
%
\end{eqnarray}
with 
\begin{equation}
I_{K}
=
\frac{\sigma_{1N}f_{3N}\left(x=0_{-}\right)}
{2\mid d_{R} \mid^{2}}
{\rm Trace}
\left[
\left(\hat{R}_{1\uparrow} + \hat{R}^{\dagger}_{1\uparrow}\right)
\left(\hat{C}_{R\uparrow} + \hat{C}^{\dagger}_{R\uparrow}\right)
\left(1 + \sigma^{2}_{1N}\right)
+ 2\sigma_{1N}
\left( \hat{\mathbb{I}} 
+ \hat{C}_{R\uparrow}\hat{C}^{\dagger}_{R\uparrow} 
\right)
\right].
\label{spinKeldysh}
\end{equation}
The resulting boundary condition is given by 
\begin{equation}
\left.
\left[
{\rm cosh}^{2}\zeta_{im} + \mid \sin \zeta \mid^{2}
{\rm sinh}^{2}\psi_{im}\right]
\left( \frac{\partial f_{3N}}{\partial x} \right)
\right|_{x=0_{-}}
=-\frac{R_{d}}{R_{b}L}f_{3N}\left(x=0_{-}\right)
\left< I_{K} \right>
\end{equation}
with $\zeta_{im}={\rm Imag}\zeta$ 
and $\psi_{im}={\rm Imag}\psi$. 
Since we are considering 1d case, $\left<I_{K} \right>$ is simply given by 
\[
\left<I_{K}\right>=\frac{I_{K}}{\sigma_{N}}. 
\]
From the Keldysh part of the Usadel equation in the present case, 
we obtain
\begin{equation}
D\frac{ \partial}{\partial x}
{\rm Trace}
\left[
\hat{\tau}_{3}
\left(
\hat{R}_{\uparrow\left(\downarrow \right)}
\frac{\partial}{\partial x}
\hat{K}_{\uparrow\left(\downarrow \right)}
+
\hat{K}_{\uparrow\left(\downarrow \right)}
\frac{\partial}{\partial x}
\hat{A}_{\uparrow\left(\downarrow \right)}
\right)
\right]
=0.
\end{equation}
From this equation, we obtain 
\begin{equation} 
\left.
\left[
{\rm cosh}^{2}\zeta_{im} + 
\mid \sin \zeta \mid^{2} {\rm sinh}^{2} \psi_{im} 
\right]
\left( \frac{\partial f_{3N}\left( x=0_{-} \right)}
{\partial x}\right)
\right|_{x=0_{-}}
=-\frac{R_{d}}{R_{b}L}
f_{3N}\left( x=0_{-} \right)
\left< I_{K} \right>=C_{0}.
\label{conservationKeldysh}
\end{equation}
The electric current from both spin up and down components are 
given by 
\begin{eqnarray}
I_{e}&=&-\frac{L}{4eR_{d}}
\int^{\infty}_{0} d\varepsilon
{\rm Trace}
\left[\hat{\tau}_{3}
\left(
\hat{R}_{1\uparrow}\frac{\partial}{\partial x}\hat{K}_{1\uparrow}
+ \hat{R}_{1\downarrow}\frac{\partial}{\partial x}\hat{K}_{1\downarrow}
+ \hat{K}_{1\uparrow}\frac{\partial}{\partial x}\hat{A}_{1\uparrow}
+ \hat{K}_{1\downarrow}\frac{\partial}{\partial x}\hat{A}_{1\downarrow}
\right)
\right]
\nonumber
\\
&=&
-\frac{2L}{eR_{d}} 
\int^{\infty}_{0}
 d\varepsilon
\left( \frac{\partial f_{3N}\left( x=0_{-} \right)}
{\partial x}\right)F(\zeta,\psi)
\label{chargecurrentfull}
\end{eqnarray}
with
\begin{equation}
F(\zeta,\psi)=
{\rm cosh}^{2}\zeta_{im} + 
\mid \sin \zeta \mid^{2} {\rm sinh}^{2} \psi_{im}.
\end{equation}
From eq. (\ref{conservationKeldysh}), 
the following relation is satisfied 
\begin{equation}
\left( \frac{\partial f_{3N}\left( x=0_{-} \right)}
{\partial x}\right) 
= \frac{C_{0}}{F\left(\zeta,\psi \right)},
\label{constantC0}
\end{equation}
we obtain
\begin{equation}
f_{3N}\left(0 \right)
\left[ 1 + \frac{R_{d}\left<I_{K} \right>}
{R_{b}L} \int^{0}_{-L} \frac{dx}{F\left( \zeta,\phi \right) }
\right] = f_{3N}(x=-L)
\label{f3Nx=-L}
\end{equation}
From eqs. (\ref{conservationKeldysh}), (\ref{chargecurrentfull}),
(\ref{constantC0}),  and (\ref{f3Nx=-L}), 
$I_{e}$ is given by 
\begin{equation}
I_{e}=\frac{2}{e}
\int^{\infty}_{0}
 d\varepsilon
\frac{f_{3N}(x=-L)}{\frac{R_{b}}{\left< I_{K} \right>} 
+ \frac{R_{d}}{L}\int^{0}_{-L} \frac{dx}
{F\left(\zeta,\psi \right)}},
\end{equation}
with 
\[
f_{3N}\left(x=-L\right)
=\frac{1}{2}
\left[
{\rm tanh}\left[\left(\varepsilon+eV\right)/\left(2T \right)\right]
- {\rm tanh}\left[\left(\varepsilon-eV \right)/\left(2T \right)
\right]
\right].
\]
Here, we have used units with $k_{B}=1$. 
At sufficiently low temperatures, total resistance of the junction is 
given by 
\begin{equation}
R=\frac{1}{2} \left[
\frac{R_{b}}{\left< I_{K} \right>} + 
\frac{R_{d}}{L}\int^{0}_{-L} \frac{dx}{F\left( \zeta,\psi \right)}
\right].
\end{equation}
The corresponding resistance becomes 
\[
R_{N}=\frac{1}{2}\left(R_{d} + R_{b} \right)
\]
The normalized conductance of the junction by its value 
in the normal state is given by 
\begin{equation}
\sigma_{T}
\left(eV \right)
= \frac{R_{N}}{R}
=\frac{R_{b} + R_{d}}{
\frac{R_{b}}{ \left< I_{K} \right>} + 
\frac{R_{d}}{L}\int^{0}_{-L} \frac{dx}{F\left( \zeta,\psi \right)}
}
\label{sigmaTsp}.
\end{equation}                 
By denoting $\zeta(x=0)=\zeta_{N}$ and 
the imaginary part of $\zeta_{N}$ as $\zeta_{Ni}$, 
$I_{K}$ is given by 
\begin{equation}
I_{K}=\frac{\sigma_{N} 
\left[2\left(2 - \sigma_{N} \right) \Lambda_{c1} + \sigma_{N}\Lambda_{c2} \right]}
{2 \mid \Gamma_{+} \Gamma_{-} \mid
\mid \left(2 -\sigma_{N} \right)
\left(1 - \Gamma_{+}\Gamma_{-} \right) + 
\sigma_{N} \left[ \left(1 + \Gamma_{+}\Gamma_{-} \right) 
\cos \zeta_{N} + 2 i \sqrt{\Gamma_{+}\Gamma_{-}}\sin \zeta_{N} \right) \mid^{2} },
\end{equation}

\begin{eqnarray}
\Lambda_{c1} &=&
{\rm Real} \left[\left(\Gamma_{-} - \Gamma_{+} \right)
\left(1 - \Gamma^{*}_{+}\Gamma^{*}_{-}\right) 
\right]
{\rm Imag}
\left[\sqrt{\Gamma_{+}\Gamma_{-}}\left(\Gamma^{*}_{+} - \Gamma^{*}_{-}\right)
\sin^{*} \zeta_{N} \right]
\nonumber
\\
&+&{\rm Imag} \left[\left(\Gamma^{*}_{+} + \Gamma^{*}_{-} \right)
\left(1 - \Gamma_{+}\Gamma_{-}\right) \right]
{\rm Real}
\left[\sqrt{\Gamma_{+}\Gamma_{-}}\left(\Gamma^{*}_{+} + \Gamma^{*}_{-}\right)
\sin^{*} \zeta_{N} \right]
\nonumber
\\
&+&
2{\rm Real}\left( \cos\zeta_{N} \right)
\left(1 - \mid \Gamma_{+} \mid^{2} \mid \Gamma_{-} \mid^{2} \right)
\mid \Gamma_{+} \mid \mid \Gamma_{-}\mid 
\label{Lambdac1},
\end{eqnarray}

\begin{eqnarray}
\Lambda_{c2} &=&
\left(1 + \mid \Gamma_{+} \mid^{2} \right)
\left(1 + \mid \Gamma_{-} \mid^{2} \right)
\nonumber
\\
&\times&
\left[
4\left(1 + {\rm sinh}^{2} \zeta_{Ni} \right)
\mid \Gamma_{+} \mid \mid \Gamma_{-} \mid
+ 
\mid \sin \zeta_{N} \mid^{2}
\left( \mid \Gamma_{+} \mid^{2} -\mid \Gamma_{-} \mid^{2}
\right)
\right]
\nonumber
\\
&+&2
{\rm Imag}
\left[ \sqrt{\Gamma_{+}\Gamma_{-}} 
\left(\Gamma^{*}_{+} + \Gamma^{*}_{-} \right) \sin^{*}\zeta_{N} 
\cos\zeta_{N}
\right]
{\rm Real}
\left[ \left(\Gamma_{+}+\Gamma_{-} \right)
\left( 1 + \Gamma^{*}_{+}\Gamma^{*}_{-} \right)
\right]
\nonumber
\\
&+&2
{\rm Real}
\left[ \sqrt{\Gamma_{+}\Gamma_{-}} 
\left(\Gamma^{*}_{+} - \Gamma^{*}_{-} \right) \sin^{*}\zeta_{N} \cos\zeta_{N}
\right]
{\rm Imag}
\left[ \left(\Gamma_{+} - \Gamma_{-} \right)
\left( 1 + \Gamma^{*}_{+}\Gamma^{*}_{-} \right)
\right]
\nonumber
\\
&-&
2 \mid \sin \zeta_{N} \mid^{2}
\left( \mid \Gamma_{+} \mid^{2} - \mid \Gamma_{-} \mid^{2} \right)
{\rm Real}
\left[ \left(\Gamma_{+}-\Gamma_{-} \right)
\left(\Gamma^{*}_{+} + \Gamma^{*}_{-} \right) \right]
\label{Lambdac2},
\end{eqnarray}
using $\Gamma_{\pm}$ as defined in eq. (\ref{definitionGamma}). 
In Figs. \ref{Fig7} and \ref{Fig8}, we plot 
charge conductance per a spin $\sigma_{S}(eV)$ and 
normalized one $\sigma_{T}(eV)$  as a function of $eV$. 
$\sigma_{S}(eV)$ is given by 
\[
\sigma_{S}(eV)=\frac{\sigma_{T}\left(eV \right)}{R_{d} + R_{b}}. 
\]


In Fig. \ref{Fig7}, we plot $\sigma_{S}(eV)$ and $\sigma_{T}(eV)$ for 
$\Delta_{p}>\Delta_{s}$. 
In all cases, $\sigma_{S}(eV)$ has a sharp peak at $eV=0$ and 
$\sigma(eV=0)$ is always $2e^{2}/h$ independent of 
the magnitude of $\Delta_{s}$. We can explain this reason in the following. 
At $eV=\varepsilon=0$, 
$\Lambda_{c1}$ in eq. (\ref{Lambdac1}) becomes zero 
since $1=\Gamma_{+}\Gamma_{-}$ is satisfied. 
On the other hand, $\Lambda_{c2}$ becomes 
\begin{equation}
\Lambda_{c2}=
16\left[ {\rm cosh}^{2}\zeta_{Ni} + 
{\rm Imag}\left(\sin\zeta^{*}_{N}
 \cos\zeta_{N} \right)
\right]
=8\left[ 1 + \exp\left(-2\zeta_{Ni} \right)\right].
\end{equation}
Then, we get 
\begin{equation}
I_{K}= 1+ \exp\left(2\zeta_{Ni}\right).
\end{equation}
By using $\zeta$ calculated in 
eq.(\ref{analyticalzeta}) and $\psi=0$, we can show
\[
\int^{0}_{-L}dx \frac{1}{{\rm \cosh}^{2} 
\left[ \zeta_{Ni} \left(x + L \right) \right] } 
= \frac{\sigma_{N}R_{b}}{2}
{\rm tanh}\zeta_{Ni}
\]
in eq. (\ref{sigmaTsp}) with 
\[
\zeta_{Ni}=\frac{2R_{d}}{\sigma_{N}R_{b}L}.
\]
Then, we obtain 
\[
\sigma_{S}(eV=0)=\frac{1}{2}R_{b}\sigma_{N}=\frac{2e^{2}}{h}.
\]
This perfect resonance at zero voltage 
has been shown for the spin-triplet $p$-wave 
superconductor junction \cite{Proximityp,Proximityp2}
($\Delta_{s}=0$ in the present case) and 
its physical origin has been also interpreted by the index theorem
\cite{Ikegaya2016}. 
It is noted that the present perfect resonance remains 
even in the presence of $\Delta_{s}$. 
We also show the normalized value of charge conductance in its value in normal 
state $\sigma_{T}(eV)$ in Fig. \ref{Fig7}B for 
$-1.2\Delta_{0} < eV < 1.2\Delta_{0}$. 
$\sigma_{T}(eV)$ has a dip like structure at $\Delta_{p} \pm \Delta_{s}$.  

In Fig. \ref{Fig8}, we plot the $\sigma_{S}(eV)$ and $\sigma_{T}(eV)$ for 
$\Delta_{s}>\Delta_{p}$ where even-frequency pair amplitude is dominant.  
Around $eV=0$, both $\sigma_{S}(eV)$ and $\sigma_{T}(eV)$ 
are slightly enhanced for (a)$\Delta_{p}=0$ and (b)$\Delta_{p}=0.3\Delta_{0}$. 
This is due to the coherent Andreev reflection in 
DN by conventional proximity effect by even-frequency pairing 
\cite{Yip1995,Volkov1993,TGK}. 
For $\Delta_{p}=0.43\Delta_{0}$, this peak structure disappears.  
At the same time, odd-frequency pair amplitude $s_{1}$ is enhanced shown in 
curve (c) in Fig. \ref{Fig6}. 
We also show the normalized value of charge conductance in its value in normal 
state $\sigma_{T}(eV)$ in Fig. \ref{Fig8}B for 
$-1.2\Delta_{0} < eV < 1.2\Delta_{0}$. The derivative of each curve has a 
sharp change at $eV=\pm \Delta_{0}$ for $\Delta_{p}=0$ [curve (c) in Fig. \ref{Fig8}B],  $eV=\pm \Delta_{0}$ and $eV=\pm 0.4 \Delta_{0}$ for $\Delta_{p}=0.3\Delta_{0}$ [curve (b) in Fig. \ref{Fig8}B], and 
$eV=\pm \Delta_{0}$ and $eV=\pm 0.06 \Delta_{0}$ for 
$\Delta_{p}=0.47\Delta_{0}$ [curve (c) in Fig. \ref{Fig8}B]. 

\section{Conclusion}
\label{sec:Conclusion}
In this paper, we have revisited the boundary condition of 
the Nambu Keldysh Green's function in 
DN/unconventional superconductor junctions. 
We have derived a more compact expression of the 
boundary condition of the Nambu-Keldysh Green's function 
and shown that it is consistent with 
the formal boundary condition of Green's function derived 
by Zaitsev \cite{Zaitsev1984}. 
We have shown both retarded part and Keldysh part of the 
boundary condition available for general situation including
mixed parity cases. 
We have demonstrated a clearer and shorter way to 
derive the expression for the charge conductance
of the junction both for 
spin-singlet or spin-triplet superconductor cases 
studied before \cite{Proximityd2,Proximityp2}. 
By applying this formula to a one-dimensional $s+p$-wave 
superconductor model, 
we have calculated LDOS, pair amplitude and 
charge conductance. 
When the $s$-wave component of the pair potential 
is dominant, the dominant pairing in DN is even-frequency spin-singlet 
$s$-wave one and the local density of states (LDOS) of quasiparticle 
have a minimum at zero energy. On the other hand, when spin-triplet $p$-wave component is dominant, the dominant pairing in DN is odd-frequency spin-triplet $s$-wave and LDOS has a zero energy peak. 
We have shown  the robustness of the quantization of the 
conductance when the magnitude of the $p$-wave 
component of the pair potential is larger than 
that of $s$-wave one. These results show that 
the anomalous proximity effect owing to the odd-frequency pairing 
is robust with respect to the inclusion of $s$-wave 
component of the pair potential when the $p$-wave one is dominant. 
In this paper, in order to understand the essence of the 
crossover of the proximity effect from anomalous one due to the odd-frequency pairing to conventional one by the even-frequency pairing, 
we have used a one-dimensional model. 
Extension to two-dimensional model of $s+p_{x}$-wave superconductor is 
an important forthcoming work. As a $p$-wave pair potential, 
it is a challenging issue to 
choose chiral $p$-wave and helical $p$-wave pairing. 
Up to now, charge transport has been calculated in $s$ + chiral $p$-wave 
and $s$ + helical $p$-wave superconductor junction 
in the ballistic limit \cite{Bursetchiral,TYBN09}. 
It is timely to study the proximity effect in these junctions. 
Extension of the present work to diffusive ferromagnet (DF) / 
$s+p$-wave junction is also an interesting topic 
\cite{Yokoyama2007} 
from the view point of superconducting spintronics
\cite{Eschrig2015,Linder2015}. 

\label{sec:AppendixD}
\par

\begin{acknowledgments}
This work was supported by 
Scientific Research (A) (KAKENHIGrant No. JP20H00131), 
and 
Scientific Research (B) (KAKENHIGrants No. JP18H01176 and No. JP20H01857).  
\end{acknowledgments}

\section{Appendix A}
\label{sec:AppendixA}
Here, we explain the derivation of $\check{I}_{n}$ in 
\cite{Proximityd} step by step \cite{TextTanaka2021}. 
We choose the basis in order to diagonalize 
$\bar{Q}$. 
Then, $\bar{G}_{2}$ can be written as 
\[
\bar{G}_{2}=
\begin{pmatrix}
\check{H}_{+} & \check{0} \\
\check{0} & \check{H}_{+}
\end{pmatrix}
+ 
\begin{pmatrix}
\check{H}_{-} & \check{0} \\
\check{0} & \check{H}_{-}
\end{pmatrix}
\bar{\Sigma}^{z}.
\]
Using the basis which diagonalizes $\bar{Q}$, 
$\bar{g}_{2}$ is obtained as shown in 
(\ref{barg2unconventional})\cite{Proximityd,Proximityd2}
with 
\[
\bar{\Sigma}_{z}=
\begin{pmatrix}
\check{0} & \check{\mathbb{I}} \\
\check{\mathbb{I}} & \check{0}
\end{pmatrix}.
\]
The matrix current $\check{I}_{n}$
\[
\check{I}_{n} \equiv {\rm Tr}\left[ \bar{\Sigma}^{z}\bar{g}_{2} \right]
\]
is obtained as  
\begin{align}
\check{I}_{n}
=&
\left[\check{H}_{-}-\left(q_{n}^{-1}\check{H}_{+}+\check{G}_{1}\right)
\check{H}_{-}^{-1}\left(q_{n}\check{H}_{+} + \check{G}_{1}\right)\right]^{-1}
\left(2\check{\mathbb{I}} - \check{H}_{-}\right)
\nonumber\displaybreak[0]
\\
+&
\left[\check{H}_{-}-\left(q_{n}\check{H}_{+}+\check{G}_{1} \right)
\check{H}_{-}^{-1}
\left(q_{n}^{-1}\check{H}_{+} + \check{G}_{1}\right)\right]^{-1}
\left(2\check{\mathbb{I}} - \check{H}_{-}\right)
\nonumber\displaybreak[0]
\\
+&
\left[\left(q_{n}\check{H}_{+}+\check{G}_{1}\right)
-\check{H}_{-}\left(q_{n}^{-1}\check{H}_{+}+\check{G}_{1}\right)^{-1}
\check{H}_{-} \right]^{-1}
\left( \check{G}_{1} -q_{n}\check{H}_{+} \right)
\nonumber\displaybreak[0]
\\
+&
\left[\left(q_{n}^{-1}\check{H}_{+}+\check{G}_{1}\right)
-\check{H}_{-}\left(q_{n}\check{H}_{+}+\check{G}_{1}\right)^{-1}
\check{H}_{-} \right]^{-1}
\left( \check{G}_{1} -q_{n}^{-1}\check{H}_{+} \right)
\nonumber\displaybreak[0]
\\
=&
\left[\check{H}_{-}-\left(q_{n}\check{H}_{+}+\check{G}_{1}\right)
\check{H}_{-}^{-1}\left(q_{n}^{-1}\check{H}_{+} + \check{G}_{1}\right)\right]^{-1}
\nonumber\displaybreak[0]\\
\times&\left[
\left(2\check{\mathbb{I}} - \check{H}_{-} \right)
-\left(q_{n}\check{H}_{+}+\check{G}_{1}\right)
\check{H}_{-}^{-1}
\left( \check{G}_{1} -q_{n}^{-1}\check{H}_{+} \right)
\right]
\nonumber
\\
+&
\left[\check{H}_{-}-\left(q_{n}^{-1}\check{H}_{+}+\check{G}_{1}\right)
\check{H}_{-}^{-1}\left(q_{n}\check{H}_{+} + \check{G}_{1}\right)\right]^{-1}
\nonumber\\
\times&\left[
\left(2\check{\mathbb{I}} - \check{H}_{-} \right)
-\left(q_{n}^{-1}\check{H}_{+}+\check{G}_{1}\right)
\check{H}_{-}^{-1}\left( \check{G}_{1} -q_{n}\check{H}_{+} \right)
\right].
\end{align}
By using the following relations between $\check{H}_{-}$ and $\check{H}_{+}$,  
\[
\check{H}_{-}-\check{H}_{+}\check{H}^{-1}_{-}\check{H}_{+}=\check{H}^{-1}_{-}, 
\ \ 
\check{H}_{-}\check{H}_{+} =-\check{H}_{+}\check{H}_{-},
\]
$\check{I}_{n}$ is transformed as
\begin{align}
\check{I}_{n}
&=\left[
\check{H}_{-}^{-1} -\check{G}_{1}\check{H}_{-}^{-1}\check{G}_{1}
-\frac{1}{q_{n}}\check{G}_{1}\check{H}_{-}^{-1}\check{H}_{+}
- q_{n}\check{H}_{+}\check{H}_{-}^{-1}\check{G}_{1}
\right]^{-1}
\nonumber\displaybreak[0]\\
&\times
\left[ 2\check{\mathbb{I}}-\check{H}_{-}^{-1}
-\check{G}_{1}\check{H}_{-}^{-1}\check{G}_{1} + 
\frac{1}{q_{n}}\check{G}_{1}\check{H}_{-}^{-1}\check{H}_{+}
+ q_{n}\check{H}_{-}^{-1}\check{H}_{+}\check{G}_{1}
\right]
\nonumber
\displaybreak[0]\\
&+ \left[
\check{H}_{-}^{-1} -\check{G}_{1}\check{H}_{-}^{-1}\check{G}_{1}
-q_{n}\check{G}_{1}\check{H}_{-}^{-1}\check{H}_{+}
-\frac{1}{q_{n}}\check{H}_{+}\check{H}_{-}^{-1}\check{G}_{1}
\right]^{-1}
\nonumber\displaybreak[0]\\
&\times
\left[
2\check{\mathbb{I}}-\check{H}_{-}^{-1}
-\check{G}_{1}\check{H}_{-}^{-1}\check{G}_{1} + 
q_{n}\check{G}_{1}\check{H}_{-}^{-1}\check{H}_{+}
+ \frac{1}{q_{n}}\check{H}_{-}^{-1}\check{H}_{+}\check{G}_{1}
\right]
\nonumber
\displaybreak[0]\\
&=\frac{1}{2}
\left[
\check{H}_{-}^{-1} -\check{G}_{1}\check{H}_{-}^{-1}\check{G}_{1}
-\frac{1}{q_{n}}\check{G}_{1}\check{H}_{-}^{-1}\check{H}_{+}
- q_{n}\check{H}_{+}\check{H}_{-}^{-1}\check{G}_{1}
\right]^{-1}
\nonumber\displaybreak[0]\\
&\times
\biggl[ 
-\left(\check{H}_{-}^{-1} -\check{G}_{1}\check{H}_{-}^{-1}\check{G}_{1}
-\frac{1}{q_{n}}\check{G}_{1}\check{H}_{-}^{-1}\check{H}_{+}
- q_{n}\check{H}_{+}\check{H}_{-}^{-1}\check{G}_{1} \right)
\nonumber\displaybreak[0]\\
&\phantom{\times}+ 2\left( \check{\mathbb{I}}
 -\check{G}_{1}\check{H}_{-}^{-1}\check{G}_{1} + 
q_{n}\check{H}_{-}^{-1}\check{H}_{+}\check{G}_{1} \right)
\biggr]
\nonumber
\displaybreak[0]\\
&+\frac{1}{2}
\left[
\check{H}_{-}^{-1} -\check{G}_{1}\check{H}_{-}^{-1}\check{G}_{1}
-\frac{1}{q_{n}}\check{G}_{1}\check{H}_{-}^{-1}\check{H}_{+}
- q_{n}\check{H}_{+}\check{H}_{-}^{-1}\check{G}_{1}
\right]^{-1}
\nonumber\displaybreak[0]\\
&\times
\biggl[ 
\left(\check{H}_{-}^{-1} -\check{G}_{1}\check{H}_{-}^{-1}\check{G}_{1}
-\frac{1}{q_{n}}\check{G}_{1}\check{H}_{-}^{-1}\check{H}_{+}
- q_{n}\check{H}_{+}\check{H}_{-}^{-1}\check{G}_{1} \right)
\nonumber\displaybreak[0]\\
&\phantom{\times}+ 2\left( \check{\mathbb{I}}
-\check{H}_{-}^{-1} + 
\frac{1}{q_{n}}\check{G}_{1}\check{H}_{-}^{-1}\check{H}_{+} \right)
\biggr]
\nonumber
\displaybreak[0]\\
&+ \frac{1}{2}
\left[
\check{H}_{-}^{-1} -\check{G}_{1}\check{H}_{-}^{-1}\check{G}_{1}
-q_{n}\check{G}_{1}\check{H}_{-}^{-1}\check{H}_{+}
-\frac{1}{q_{n}}\check{H}_{+}\check{H}_{-}^{-1}\check{G}_{1}
\right]^{-1}
\nonumber\displaybreak[0]\\
&\times
\biggl[
-\left(
\check{H}_{-}^{-1} -\check{G}_{1}\check{H}_{-}^{-1}\check{G}_{1}
-q_{n}\check{G}_{1}\check{H}_{-}^{-1}\check{H}_{+}
-\frac{1}{q_{n}}\check{H}_{+}\check{H}_{-}^{-1}\check{G}_{1}
\right)
\nonumber\displaybreak[0]\\
&\phantom{\times}+2\left(\check{\mathbb{I}}-
\check{G}_{1}\check{H}_{-}^{-1}\check{G}_{1} 
+ \frac{1}{q_{n}}\check{H}_{-}^{-1}\check{H}_{+}\check{G}_{1}
\right)
\biggr]
\nonumber
\displaybreak[0]\\
&+\frac{1}{2} \left[
\check{H}_{-}^{-1} -\check{G}_{1}\check{H}_{-}^{-1}\check{G}_{1}
-q_{n}\check{G}_{1}\check{H}_{-}^{-1}\check{H}_{+}
-\frac{1}{q_{n}}\check{H}_{+}\check{H}_{-}^{-1}\check{G}_{1}
\right]^{-1}
\nonumber\displaybreak[0]\\
&\times
\biggl[
\left(
\check{H}_{-}^{-1} -\check{G}_{1}\check{H}_{-}^{-1}\check{G}_{1}
-q_{n}\check{G}_{1}\check{H}_{-}^{-1}\check{H}_{+}
-\frac{1}{q_{n}}\check{H}_{+}\check{H}_{-}^{-1}\check{G}_{1} \right)
\nonumber\displaybreak[0]\\
&\phantom{\times}+2\left( 
\check{\mathbb{I}}-\check{H}_{-}^{-1}
+ q_{n}\check{G}_{1}\check{H}_{-}^{-1}\check{H}_{+} \right)
\biggr].
\end{align}
Applying the normalization condition of $\check{G}_{1}$ 
given by $\check{G}_{1}^{2}=\check{\mathbb{I}}$, 
we obtain 
\begin{align}
&\check{I}_{n}
=
\nonumber\\
&\left[
\check{H}_{-}^{-1} -\check{G}_{1}\check{H}_{-}^{-1}
\check{G}_{1} - 
\frac{1}{q_{n}}\check{G}_{1}\check{H}_{-}^{-1}\check{H}_{+}
+ q_{n}\check{H}_{-}^{-1}\check{H}_{+}\check{G}_{1}
\right]^{-1}
\!
\left[
\check{\mathbb{I}}-
\check{G}_{1}\check{H}_{-}^{-1}\check{G}_{1}  + 
q_{n}\check{H}_{-}^{-1}\check{H}_{+}\check{G}_{1} 
\right]
\nonumber\\
&+
\!\left[
\check{H}_{-}^{-1} -\check{G}_{1}\check{H}_{-}^{-1}\check{G}_{1}
-\frac{1}{q_{n}}\check{G}_{1}\check{H}_{-}^{-1}\check{H}_{+}
+ q_{n}\check{H}_{-}^{-1}\check{H}_{+}\check{G}_{1}
\right]^{-1}
\!\!
\left[
\check{\mathbb{I}}-
\check{H}_{-}^{-1} + 
\frac{1}{q_{n}}\check{G}_{1}\check{H}_{-}^{-1}\check{H}_{+} 
\right]
\nonumber\\
&+ 
\!\left[
\check{H}_{-}^{-1} -\check{G}_{1}\check{H}_{-}^{-1}\check{G}_{1}
-q_{n}\check{G}_{1}\check{H}_{-}^{-1}\check{H}_{+}
+ \frac{1}{q_{n}}\check{H}_{-}^{-1}\check{H}_{+}\check{G}_{1}
\right]^{-1}
\!\!
\left[
\check{\mathbb{I}}-
\check{G}_{1}\check{H}_{-}^{-1}\check{G}_{1} + 
\frac{1}{q_{n}}\check{H}_{-}^{-1}\check{H}_{+}\check{G}_{1}
\!
\right]
\nonumber
\\
&+
\!\left[
\check{H}_{-}^{-1} -\check{G}_{1}\check{H}_{-}^{-1}\check{G}_{1}
-q_{n}\check{G}_{1}\check{H}_{-}^{-1}\check{H}_{+}
+ \frac{1}{q_{n}}\check{H}_{-}^{-1}\check{H}_{+}\check{G}_{1}
\right]^{-1}
\!
\left[
\check{\mathbb{I}}-
\check{H}_{-}^{-1} + 
q_{n}\check{G}_{1}\check{H}_{-}^{-1}\check{H}_{+} 
\right]
\nonumber
\\
&=
\left\{
q_{n}\!\left[\check{G}_{1},\check{H}_{-}^{-1} \right]
- \check{H}_{-}^{-1}\check{H}_{+}
+ q_{n}^{2}
\check{G}_{1}\check{H}_{-}^{-1}
\check{H}_{+}\check{G}_{1}
\right\}^{-1}
\!\left[
q_{n}\!
\left(\check{G}_{1} - \check{H}_{-}^{-1}
\check{G}_{1} \right)
+ q_{n}^{2}\check{G}_{1}\check{H}_{-}^{-1}
\check{H}_{+}\check{G}_{1}
\right]
\nonumber
\\
&+
\left\{
q_{n}\!\left[\check{G}_{1},\check{H}_{-}^{-1} \right]
- \check{H}_{-}^{-1}\check{H}_{+}
+ q_{n}^{2}
\check{G}_{1}\check{H}_{-}^{-1}
\check{H}_{+}\check{G}_{1}
\right\}^{-1}
\!\left[
q_{n}\!
\left(\check{G}_{1} - \check{G}_{1}\check{H}_{-}^{-1}
 \right)
+ \check{H}_{-}^{-1}\check{H}_{+}
\right]
\nonumber
\\
&+
\!\left\{
\frac{1}{q_{n}}
\left[\check{G}_{1},\check{H}_{-}^{-1} \right]
- \check{H}_{-}^{-1}\check{H}_{+}
+ \frac{1}{q_{n}^{2}}
\check{G}_{1}\check{H}_{-}^{-1}
\check{H}_{+}\check{G}_{1}
\right\}^{-1}
\nonumber\\
&\phantom{+}\times
\!\left[
\frac{1}{q_{n}}
\left(\check{G}_{1} - \check{H}_{-}^{-1}\check{G}_{1}
 \right)
+ \frac{1}{q_{n}^{2}}
\check{G}_{1}\check{H}_{-}^{-1}
\check{H}_{+}\check{G}_{1}
\right]
\nonumber
\\
&+
\!\left\{
\frac{1}{q_{n}}\left[\check{G}_{1},\check{H}_{-}^{-1} \right]
- \check{H}_{-}^{-1}\check{H}_{+}
+ \frac{1}{q_{n}^{2}}
\check{G}_{1}\check{H}_{-}^{-1}
\check{H}_{+}\check{G}_{1}
\right\}^{-1}
\!\left[
\frac{1}{q_{n}}
\left(\check{G}_{1} - \check{G}_{1}\check{H}_{-}^{-1}
 \right)
+ \check{H}^{-1}_{-}\check{H}_{+}
\right].
\end{align}
Since $q_{n}$ is expressed by $\sigma_{N}$,
\[
q_{n}=\sigma_{N}/(1 - \sqrt{1 - \sigma_{N}})^{2},
\]
the relations 
\[
\frac{1}{\left(1 + q_{n} \right)^{2}}=\frac{1}{4}
\left( 1 - \sqrt{1 - \sigma_{N}} \right)^{2}, \quad 
\frac{q_{n}}{\left(1 + q_{n} \right)^{2}}=\frac{\sigma_{N}}{4}, \quad 
\frac{q_{n}^{2}}{\left(1 + q_{n} \right)^{2}}=\frac{1}{4}
\left( 1 + \sqrt{1 - \sigma_{N}} \right)^{2}
\]
are satisfied. 
By using these relations, $\check{I}_{n}$ is given by 
\begin{align*}
&\check{I}_{n}=
\left\{\sigma_{N}
\left[\check{G}_{1},\check{H}_{-}^{-1}\right]
- (1 - \sqrt{1 - \sigma_{N}})^{2}
\check{H}_{-}^{-1}\check{H}_{+} 
+ (1 + \sqrt{1 - \sigma_{N}})^{2}
\check{G}_{1}\check{H}_{-}^{-1}\check{H}_{+}
\check{G}_{1}
\right\}^{-1}
\nonumber
\\
&\times\!
\left[\sigma_{N}\!
\left(2\check{G}_{1} - 
\left[ \check{H}^{-1}_{-}, \check{G}_{1}\right]_{+}
\right)\!
+ \left( 1 + \sqrt{1 -\sigma_{N}}\right)^{2}
\check{G}_{1}\check{H}_{-}^{-1}
\check{H}_{+}\check{G}_{1}
+ \left( 1 - \sqrt{1  - \sigma_{N}}\right)^{2}\!
\check{H}_{-}^{-1}\check{H}_{+}
\right]
\nonumber
\\
&+
\left\{\sigma_{N}
\left[\check{G}_{1},\check{H}_{-}^{-1}\right]
- (1 + \sqrt{1 - \sigma_{N}})^{2}
\check{H}_{-}^{-1}\check{H}_{+} 
+ (1 - \sqrt{1 - \sigma_{N}})^{2}
\check{G}_{1}\check{H}_{-}^{-1}\check{H}_{+}
\check{G}_{1}
\right\}^{-1}
\nonumber
\\
&\times\!
\left[
\sigma_{N}\!
\left(2\check{G}_{1} - 
\left[ \check{H}_{-}^{-1}, \check{G}_{1}\right]_{+}
\right)\!
+ \left( 1 - \sqrt{1 -\sigma_{N}}\right)^{2}
\check{G}_{1}\check{H}_{-}^{-1}
\check{H}_{+}\check{G}_{1}
+ \left( 1 + \sqrt{1  - \sigma_{N}}\right)^{2}\!
\check{H}_{-}^{-1}\check{H}_{+}
\right]
\nonumber
\end{align*}
Here, by using  
$\sigma_{1N}=
\sigma_{N}/(1+\sqrt{1-\sigma_{N}})^{2}$, 
\[\sigma_{1N}^{2}
=\left( 
\frac{1-\sqrt{1-\sigma_{N}}}{1+ \sqrt{1-\sigma_{N}}}
\right)^{2}
\]
is satisfied and 
$\check{I}_{n}$ is expressed using $\sigma_{1N}$ as follows. 

\begin{align}
\check{I}_{n}
&=
\left\{\sigma_{1N}
\left[\check{G}_{1},\check{H}_{-}^{-1}\right]
- \sigma_{1N}^{2}\check{H}_{-}^{-1}\check{H}_{+} 
+ \check{G}_{1}\check{H}_{-}^{-1}\check{H}_{+}
\check{G}_{1}
\right\}^{-1}
\nonumber
\displaybreak[0]\\
&\phantom{=}\times
\left\{
\sigma_{1N}
\left( 2\check{G}_{1}- 
\left[\check{H}_{-}^{-1}, \check{G}_{1}\right]_{+}
\right)
+ \sigma_{1N}^{2}\check{H}_{-}^{-1}\check{H}_{+}
+ \check{G}_{1}\check{H}_{-}^{-1}
\check{H}_{+}\check{G}_{1}
\right\}
\nonumber
\displaybreak[0]\\
&+
\left\{\sigma_{1N}
\left[\check{G}_{1},\check{H}_{-}^{-1} \right]
- \check{H}_{-}^{-1}\check{H}_{+} 
+ \sigma_{1N}^{2}
\check{G}_{1}\check{H}_{-}^{-1}\check{H}_{+}
\check{G}_{1}
\right\}^{-1}
\nonumber
\\
&\phantom{=}\times
\left\{
\sigma_{1N}
\left( 2\check{G}_{1}- 
\left[\check{H}_{-}^{-1}, \check{G}_{1}\right]_{+}
\right)
+ \check{H}_{-}^{-1}\check{H}_{+}
+ \sigma_{1N}^{2}
\check{G}_{1}\check{H}_{-}^{-1}
\check{H}_{+}\check{G}_{1}
\right\}
\nonumber
\displaybreak[0]\\
&=
\left\{\sigma_{1N}
\left[\check{G}_{1},\check{H}_{-}^{-1}\right]
- \sigma_{1N}^{2}\check{H}_{-}^{-1}\check{H}_{+} 
+ \check{G}_{1}\check{H}_{-}^{-1}\check{H}_{+}
\check{G}_{1}
\right\}^{-1}
\nonumber
\displaybreak[0]\\
&\phantom{=}\times
\{
\left[ \sigma_{1N}
\left[\check{G}_{1},\check{H}_{-}^{-1}\right]
- \sigma_{1N}^{2}\check{H}_{-}^{-1}\check{H}_{+} 
+ \check{G}_{1}\check{H}_{-}^{-1}\check{H}_{+}
\check{G}_{1} \right]
\nonumber\displaybreak[0]\\
&\phantom{=}+2\sigma_{1N}
\left(\check{G}_{1}- \check{G}_{1}\check{H}_{-}^{-1}
\right)
+ 2\sigma_{1N}^{2}
\check{H}^{-1}_{-}\check{H}_{+}
\}
\nonumber
\displaybreak[0]\\
&+
\left\{\sigma_{1N}
\left[\check{G}_{1},\check{H}_{-}^{-1}\right]
- \check{H}_{-}^{-1}\check{H}_{+} 
+ \sigma_{1N}^{2}
\check{G}_{1}\check{H}_{-}^{-1}\check{H}_{+}
\check{G}_{1}
\right\}^{-1}
\nonumber
\displaybreak[0]\\
&\phantom{=}\times
\{
-\left[ \sigma_{1N}
\left[\check{G}_{1},\check{H}_{-}^{-1}\right]
- \check{H}_{-}^{-1}\check{H}_{+} 
+ \sigma_{1N}^{2}
\check{G}_{1}\check{H}_{-}^{-1}\check{H}_{+}
\check{G}_{1}
\right]
\displaybreak[0]\nonumber\\
&\phantom{=}+ 2\sigma_{1N}
\left(
\check{G}_{1}- \check{H}_{-}^{-1}\check{G}_{1}
\right)
+ 2 \sigma_{1N}^{2}
\check{G}_{1}\check{H}_{-}^{-1}
\check{H}_{+}\check{G}_{1}
\}
\nonumber
\displaybreak[0]\\
&=
2\left\{
\sigma_{1N}
\left[\check{G}_{1},\check{H}_{-}^{-1}\right]
- \sigma_{1N}^{2}\check{H}_{-}^{-1}\check{H}_{+} 
+ \check{G}_{1}\check{H}_{-}^{-1}\check{H}_{+}
\check{G}_{1}
\right\}^{-1}
\nonumber
\displaybreak[0]\\
&\phantom{=}\times
\left\{
\sigma_{1N}
\left(\check{G}_{1}- \check{G}_{1}\check{H}_{-}^{-1}
\right)
+ \sigma_{1N}^{2}
\check{H}^{-1}_{-}\check{H}_{+}
\right\}
\nonumber
\displaybreak[0]\\
&+
2\left\{\sigma_{1N}
\left[\check{G}_{1},\check{H}_{-}^{-1}\right]
- \check{H}_{-}^{-1}\check{H}_{+} 
+ \sigma_{1N}^{2}
\check{G}_{1}\check{H}_{-}^{-1}\check{H}_{+}
\check{G}_{1}
\right\}^{-1}
\nonumber
\displaybreak[0]\\
&\phantom{=}\times
\left\{
\sigma_{1N}
\left(\check{G}_{1}- \check{H}_{-}^{-1}\check{G}_{1}
\right)
+  \sigma_{1N}^{2}
\check{G}_{1}\check{H}_{-}^{-1}
\check{H}_{+}\check{G}_{1}
\right\}
\label{TanakaNazarovmatrix1}
\end{align}
If we define 
$\check{A}_{c}$ and $\check{A}_{a}$ 
as
\begin{align}
\check{A}_{c} \equiv &
-\left\{
\sigma_{1N}
\left[\check{G}_{1},\check{H}_{-}^{-1}\right]
- \sigma_{1N}^{2}\check{H}_{-}^{-1}\check{H}_{+} 
+ \check{G}_{1}\check{H}_{-}^{-1}\check{H}_{+}
\check{G}_{1}
\right\}^{-1}
\nonumber\\
&\times
\left\{
\sigma_{1N}
\left(\check{G}_{1}- \check{G}_{1}\check{H}_{-}^{-1}
\right)
+ \sigma_{1N}^{2}
\check{H}^{-1}_{-}\check{H}_{+}
\right
\}
\nonumber
\displaybreak[0]\\
\check{A}_{a} \equiv & 
-\left\{\sigma_{1N}
\left[\check{G}_{1},\check{H}_{-}^{-1}\right]
- \check{H}_{-}^{-1}\check{H}_{+} 
+ \sigma_{1N}^{2}
\check{G}_{1}\check{H}_{-}^{-1}\check{H}_{+}
\check{G}_{1}
\right\}^{-1}
\nonumber\\
&\times
\left\{
\sigma_{1N}
\left(\check{G}_{1}- \check{H}_{-}^{-1}\check{G}_{1}
\right)
+  \sigma_{1N}^{2}
\check{G}_{1}\check{H}_{-}^{-1}
\check{H}_{+}\check{G}_{1}
\right\}
\nonumber
\end{align}
$
\check{A}_{c}=-\check{G}_{1}\check{A}_{a}\check{G}_{1}$
is satisfied. 
Here, defining 
\[
\check{B} \equiv \check{A}_{a}\check{G}_{1}
\]
$\check{I}_{n}$ is expressed by 
\[
\check{I}_{n}=2 \left[\check{G}_{1},\check{B}\right]
\]
with 
\begin{equation}
\check{B}=
\left(\check{H}_{-}^{-1}\check{H}_{+}
-\sigma_{1N}\left[\check{G}_{1},\check{H}_{-}^{-1} \right]
-\sigma_{1N}^{2}\check{G}_{1}\check{H}_{-}^{-1}\check{H}_{+}
\check{G}_{1}\right)^{-1}
\left\{
\sigma_{1N}\left(\check{\mathbb{I}}-\check{H}_{-}^{-1} \right)
+\sigma_{1N}^{2}\check{G}_{1}\check{H}_{-}^{-1}\check{H}_{+}\right\} .
\label{TanakaNazarov2003Appendix}
\end{equation}%
\section{Appendix B} 
\label{sec:Zaitsev}
\begin{figure}[b]
\begin{center}
\includegraphics[width=5.5cm,clip]{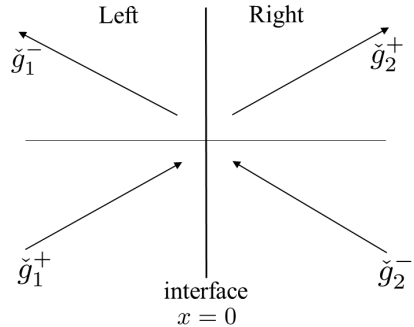}
\end{center}
\caption{Schematic picture showing trajectory of 
quasiclassical Green's function, 
$\check{g}_{1}^{+}$, $\check{g}_{1}^{-}$, $\check{g}_{2}^{+}$, 
and $\check{g}_{2}^{-}$. 
Left and right corresponds to normal metal or superconductor. 
}
\label{fig:81}
\end{figure}

In this section, we show that our obtained boundary condition is 
consistent with Zaitsev's one  \cite{Zaitsev1984}
although we are considering unconventional superconductor junctions.
Zaitsev disussed the boundary condition of quasiclassical Green's function 
as shown in Fig. \ref{fig:81} 
assuming a spin-singlet $s$-wave pair potential. 
He defined 
$\check{g}_{1}^{s}$, $\check{g}_{1}^{a}$, 
$\check{g}_{2}^{s}$, $\check{g}_{2}^{a}$ as
\begin{equation}
\check{g}_{1}^{s}=\frac{1}{2}\left(\check{g}_{1}^{+} + \check{g}_{1}^{-}\right),\ \
\check{g}_{1}^{a}=\frac{1}{2}\left(\check{g}_{1}^{+} - \check{g}_{1}^{-}\right),\end{equation}
and 
\begin{equation}
\check{g}_{2}^{s}=\frac{1}{2}\left(\check{g}_{2}^{+} + \check{g}_{2}^{-}\right),\ 
\check{g}_{2}^{a}=\frac{1}{2}\left(\check{g}_{2}^{+} - \check{g}_{2}^{-}\right),\end{equation}
both left and right sides. 
By using these functions, 
$\check{g}_{s}^{+}$ and $\check{g}_{s}^{-}$ are defined at the interface as 
\begin{equation}
\check{g}_{s}^{+} \equiv \frac{1}{2}\left(\check{g}_{1}^{s}+\check{g}_{2}^{s}\right), \\ 
\check{g}_{s}^{-} \equiv \frac{1}{2}\left(\check{g}_{1}^{s}-\check{g}_{2}^{s}\right). 
\end{equation}
The Zaitsev's boundary condition is given by 
\begin{equation}
\check{g}_{a}=\check{g}_{1}^{a}=\check{g}_{2}^{a}, \ \ 
\label{Zaitsev1}
\end{equation}
\begin{equation}
\check{g}_{a}
\left[ \left(1-\sigma_{N}\right) \left( \check{g}_{s}^{+} \right)^{2} + \left( \check{g}_{s}^{-} \right)^{2}\right]
= \sigma_{N} \check{g}_{s}^{-} \check{g}_{s}^{+}
\label{Zaitsev2}
\end{equation}
by using the transmissivity at the interface $\sigma_{N}$. 
In the following,it is shown that 
our obtained boundary condition of Nambu-Keldysh Green's 
function is consistent with Zaitsev's one. \par
For this purpose, we must start from so called interface matrix 
$\bar{g}_{2}$ defined in eq.(\ref{barg2unconventional}).  
It is written as 
\begin{equation}
\bar{g}_{2} = \left(
\begin{array}{cc}
q_{n}\check{H}_{+} + \check{G}_{1} & q_{n}\check{H}_{-} \\
q_{n}^{-1}\check{H}_{-} & q_{n}^{-1}\check{H}_{+} + \check{G}_{1}%
\end{array}
\right)^{-1} 
\left(
\begin{array}{cc}
q_{n}\left(2\check{\mathbb{I}} 
- \check{H}_{-} \right) & \ \check{G}_{1}-q_{n}\check{H}_{+} \\
\check{G}_{1} - q_{n}^{-1}\check{H}_{+} & \ q_{n}^{-1}
\left(2\check{\mathbb{I}} - \check{H}_{-} \right)%
\end{array}
\right)
\label{barg2unconventionalappendix}
\end{equation}
The matrix current 
is defined by 
\[
\check{I}_{n}={\rm Trace}
\left[ \bar{\Sigma}^{z} \bar{g}_{2} \right].
\]
As shown in eq. (\ref{matrixcurrentTanakanew}), 
$\check{I}_{n}$ is expressed by 
\begin{eqnarray}
\check{I}_{n}=
2\sigma_{N}
\left[\left( 4-2\sigma_{N} \right)\check{\mathbb{I}} 
+ \sigma_{N}[\check{C},\check{G}_{1}]_{+}\right]^{-1}
\left[\check{C},\check{G}_{1}\right]
\label{matrixcurrentTanakaNazarovAppendix}
\end{eqnarray}
Next, we calculate ${\rm Trace}[\bar{g}_{2}]$. 
After a straightforward transformation, 
\begin{eqnarray}
&&{\rm Trace}\left[\bar{g}_{2} \right] 
\noindent
\\
&=&
2\left\{ \sigma_{1N}\left[\check{G}_{1},\check{H}_{-}^{-1} \right]
-\check{H}_{-}^{-1}\check{H}_{+} 
-\sigma_{1N}^{2} \check{G}_{1}\check{H}_{+}\check{H}_{-}^{-1} \check{G}_{1} 
\right\}
^{-1} 
\noindent
\\
&\times&
\left[ 
\check{\mathbb{I}}
- \sigma_{1N}\check{G}_{1}\check{H}_{+}\check{H}_{-}^{-1}
-\check{H}_{-}^{-1} \right] 
\noindent
\\
&+&
2\sigma_{1N}
\left\{ \sigma_{1N}\left[\check{G}_{1},\check{H}_{-}^{-1} \right]
-\sigma_{1N}^{2} \check{H}_{-}^{-1}\check{H}_{+} 
-\check{G}_{1}\check{H}_{+}\check{H}_{-}^{-1} \check{G}_{1} 
\right\}^{-1} 
\noindent
\\
&\times&
\left[ 
\sigma_{1N}\check{\mathbb{I}} - \check{G}_{1}\check{H}_{+}\check{H}_{-}^{-1}
-\sigma_{1N} \check{H}_{-}^{-1} \right]
\end{eqnarray}

Following the same procedure used in the calculation of 
${\rm Trace}[\bar{g}_{2}\Sigma^{z}]$, 
we obtain ${\rm Trace}[\bar{g}_{2}]$ 
\begin{eqnarray}
&&{\rm Trace}
\left[ \bar{g}_{2} \right] 
\noindent
\\
&=&
2\left\{
\left(1+\sigma^{2}_{1N}\right)\check{\mathbb{I}}
 + \sigma_{1N} 
\left[\check{C},\check{G}_{1} \right]_{+} 
\right\}^{-1} 
\left[ \check{C} \left( 1 + \sigma_{1N}^{2} \right)
+ 2 \sigma_{1N}\check{G}_{1}
\right] 
\noindent
\\
&=& 
4
\left\{ 2\left( 2-\sigma_{N} \right)\check{\mathbb{I}} + \sigma_{N} 
\left[\check{C},\check{G}_{1} \right]_{+} 
\right\}^{-1} 
\left[ \left(2 - \sigma_{N} \right) \check{C}
+ \sigma_{N}\check{G}_{1}
\right]
\label{traceg2bar}
\end{eqnarray}
Then, let us look at $\bar{g}_{1}$. 
It can be calculated similar to $\bar{g}_{1}$ and is 
given by 
\begin{equation}
\bar{g}_{1}=
\begin{pmatrix}
\frac{1}{q_{n}}\left( 2 \check{\mathbb{I}} + \check{H}_{-} \right) 
& \check{H}_{+} + q_{n} \check{G}_{1} \\
\check{H}_{+} + \frac{1}{q_{n}}\check{G}_{1} & 
q_{n}\left(2 \check{\mathbb{I}}+ \check{H}_{-} \right)
\end{pmatrix}
\begin{pmatrix}
\check{H}_{+} + \frac{1}{q_{n}} \check{G}_{1} & 
q_{n}\}\check{G}_{1} \\
\frac{1}{q_{n}}\check{G}_{1} & \check{H}_{+} + q_{n} \check{G}_{1}
\end{pmatrix}
\end{equation}
${\rm Trace}[\bar{g}_{1}]$ and ${\rm Trace}[\bar{g}_{1}\Sigma^{z}]$ can be calculated in the similar way. 
\begin{eqnarray}
&&{\rm Trace}\left[\bar{g}_{1} \Sigma^{z} \right] 
\nonumber
\\ 
&=&2\sigma_{N}
\left[\left( 4-2\sigma_{N} \right)\check{\mathbb{I}} 
+ \sigma_{N}[\check{C},\check{G}_{1}]_{+}\right]^{-1}
\left[\check{C},\check{G}_{1}\right]
\label{matrixcurrentTanakaNazarov5}
\end{eqnarray}
\begin{eqnarray}
&&{\rm Trace}
\left[ \bar{g}_{1} \right] 
\nonumber
\\
&=& 
4
\left\{ 2\left( 2-\sigma_{N} \right)\check{\mathbb{I}} + \sigma_{N} 
\left[\check{C},\check{G}_{1} \right]_{+} 
\right\}^{-1} 
\left[ \left(2 - \sigma_{N} \right) \check{G}_{1}
+ \sigma_{N}\check{C}
\right]
\label{traceg1bar}
\end{eqnarray}
Let us discuss about the Zaitsev's condition. 
The following relations are useful.
\begin{equation}
\check{g}_{1}^{s}
=\frac{1}{2}\left(\check{g}_{1}^{+} + \check{g}_{1}^{-}\right)
=\frac{1}{2}{\rm Trace}\left[ \bar{g}_{1} \right], 
\label{g1s}
\end{equation}
\begin{equation}
\check{g}_{2}^{s}=\frac{1}{2}\left(\check{g}_{2}^{+} + \check{g}_{2}^{-}\right)
=\frac{1}{2}{\rm Trace}
\left[ \bar{g}_{2} \right],
\label{g2s}
\end{equation}
\begin{equation}
\check{g}_{1}^{a}=\frac{1}{2}\left(\check{g}_{1}^{+} - \check{g}_{1}^{-}\right)
=\frac{1}{2}{\rm Trace}\left[\bar{g}_{1}\Sigma^{z} \right]
\label{g1a}
\end{equation}
\begin{equation}
\check{g}_{2}^{a}=\frac{1}{2}\left(\check{g}_{2}^{+} - \check{g}_{2}^{-}\right),=\frac{1}{2}{\rm Trace}\left[\bar{g}_{2}\Sigma^{z} \right]. 
\label{g2a}
\end{equation}
Then, we can express $\check{g}_{1}^{s}$, $\check{g}_{2}^{s}$, 
$\check{g}_{1}^{a}$, and $\check{g}_{2}^{a}$ as 
\begin{equation}
\check{g}_{1}^{s}
= 
2
\left\{ 2\left( 2-\sigma_{N} \right)\check{\mathbb{I}} + \sigma_{N} 
\left[\check{C},\check{G}_{1} \right]_{+} 
\right\}^{-1} 
\left[ \left(2 - \sigma_{N} \right) \check{G}_{1}
+ \sigma_{N}\check{C}
\right]
\end{equation}
\begin{equation}
\check{g}_{2}^{s}
= 
2
\left\{ 2\left( 2-\sigma_{N} \right)\check{\mathbb{I}} + \sigma_{N} 
\left[\check{C},\check{G}_{1} \right]_{+} 
\right\}^{-1} 
\left[ \left(2 - \sigma_{N} \right) \check{C}
+ \sigma_{N}\check{G}_{1}
\right]
\end{equation}
\begin{equation}
\check{g}_{1}^{a}
= 
\sigma_{N}
\left\{ 2\left( 2-\sigma_{N} \right)\check{\mathbb{I}} + \sigma_{N} 
\left[\check{C},\check{G}_{1} \right]_{+} 
\right\}^{-1} 
\left[\check{C},\check{G}_{1}\right]
\end{equation}

\begin{equation}
\check{g}_{2}^{a}
= 
\sigma_{N}
\left\{ 2\left( 2-\sigma_{N} \right)\check{\mathbb{I}} + \sigma_{N} 
\left[\check{C},\check{G}_{1} \right]_{+} 
\right\}^{-1} 
\left[\check{C},\check{G}_{1}\right]
\end{equation}
Then, $g_{s}^{+}$ and $g_{s}^{-}$ are obtained as 
\begin{equation}
\check{g}_{s}^{+}=\frac{1}{2}\left(\check{g}_{1}^{s}+\check{g}_{2}^{s}\right)
=2
\left\{ 2\left( 2-\sigma_{N} \right)\check{\mathbb{I}} + \sigma_{N} 
\left[\check{C},\check{G}_{1} \right]_{+} 
\right\}^{-1} 
\left[ \check{C} + \check{G}_{1} \right]
\end{equation}
\begin{equation}
\check{g}_{s}^{-}=\frac{1}{2}\left(\check{g}_{1}^{s}-\check{g}_{2}^{s}\right)
=2\left(1 - \sigma_{N} \right)
\left\{ 2\left( 2-\sigma_{N} \right)\check{\mathbb{I}} + \sigma_{N} 
\left[\check{C},\check{G}_{1} \right]_{+} 
\right\}^{-1} 
\left[ -\check{C} + \check{G}_{1} \right]
\end{equation}
Since $\check{C}^{2}=\check{I}$ and $\check{G}_{1}^{2}=\check{I}$ 
are satisfied, we obtain following relations. 
\begin{eqnarray}
\left(1 - \sigma_{N} \right)
\left( \check{g}_{s+} \right)^{2}&=&4\left(1 - \sigma_{N} \right)
\left[ 2\check{\mathbb{I}} + \left[\check{G}_{1},\check{C}\right]_{+} \right]
\left\{ 2\left( 2-\sigma_{N} \right)\check{\mathbb{I}} + \sigma_{N} 
\left[\check{C},\check{G}_{1} \right]_{+} 
\right\}^{-2} \\
\left( \check{g}_{s-}\right)^{2}&=&4\left(1 - \sigma_{N} \right)^{2}
\left[ 2\check{\mathbb{I}} - \left[\check{G}_{1},\check{C}\right]_{+} \right]
\left\{ 2\left( 2-\sigma_{N} \right)\check{\mathbb{I}} + \sigma_{N} 
\left[\check{C},\check{G}_{1} \right]_{+} 
\right\}^{-2}
\end{eqnarray}
\begin{eqnarray}
\left(1 -\sigma_{N} \right) \left(\check{g}_{s+}\right)^{2} + 
\left( \check{g}_{s-} \right)^{2}
&=&4\left(1 -\sigma_{N} \right)
\left\{ 2\left( 2-\sigma_{N} \right)\check{\mathbb{I}} + \sigma_{N} 
\left[\check{C},\check{G}_{1} \right]_{+} 
\right\}^{-1} \\
\check{g}_{s+}\check{g}_{s-}
&=&4\left( 1 - \sigma_{N} \right)
\left[ \check{C},\check{G}_{1} \right]
\left\{ 2\left( 2-\sigma_{N} \right)\check{\mathbb{I}} + \sigma_{N} 
\left[\check{C},\check{G}_{1} \right]_{+} 
\right\}^{-2}
\end{eqnarray}
Based on these results, we can verify Zaitsev's condition as 
\begin{eqnarray}
&&\check{g}^{a} \left[\left(1 -\sigma_{N} \right)
\left( \check{g}_{s+}\right)^{2} + \left(\check{g}_{s-}\right)^{2} \right] 
\nonumber
\\
&=& 4 \sigma_{N} \left(1 -\sigma_{N} \right) \left[\check{C}, \check{G}_{1} \right]
\left\{ 2\left( 2-\sigma_{N} \right)\check{\mathbb{I}} + \sigma_{N} 
\left[\check{C},\check{G}_{1} \right]_{+} 
\right\}^{-2} 
\nonumber
\\
&=&\sigma_{N} \check{g}_{s-}\check{g}_{s+}
\label{Zaitsevboundary}
\end{eqnarray}

\section{Appendix C}
\label{sec:AppendixC}
In this appendix, we show the details of the calculation of 
$\hat{I}_{K}$and ${\rm Trace}[\hat{I}_{K}\tau_{3}]$. 
$\hat{I}_{K}$ defined in eq. (\ref{Keldysh1}) 
is given by 
\begin{eqnarray}
\hat{I}_{K}&=&
\frac{ 2\sigma_{1N}}
{\mid d_{R} \mid^{2}}
\left[
\left(1 + \sigma_{1N}^{2} \right)
\hat{\Lambda}_{1} 
+ 2\sigma_{1N} \hat{\Lambda}_{2}
\right],
\label{Keldysh1a}
\\
\hat{\Lambda}_{1}&=&
\left[
\hat{C}_{R}\hat{K}_{1} + \hat{C}_{K}\hat{A}_{1}
- \hat{R}_{1}\hat{C}_{K} - \hat{K}_{1}\hat{C}_{A}
\right],
\label{Keldysh2a}
\\
\hat{\Lambda}_{2}&=&
\left[
\left(\hat{C}_{R}\hat{K}_{1} + \hat{C}_{K}\hat{A}_{1} \right)
\hat{A}_{1}\hat{C}_{A} 
- \left(\hat{R}_{1}\hat{C}_{K} + \hat{K}_{1} \hat{C}_{A}\right)
\hat{C}_{A}\hat{A}_{1}
\right].
\label{Keldysh3a}
\end{eqnarray}
After some transformation, 
$\hat{\Lambda}_{1}$ and $\hat{\Lambda}_{2}$ are given by 
\begin{eqnarray}
\hat{\Lambda}_{1}
&=&
\left[\hat{C}_{R}\left(\hat{R}_{1}-\hat{A}_{1} \right)
- \left(\hat{R}_{1}-\hat{A}_{1}\right) \hat{C}_{A} \right]f_{0N}\left(x \right) \nonumber
\\
&+& \left[ \left(\hat{C}_{R}-\hat{C}_{A} \right)\hat{A}_{1} - 
\hat{R}_{1} \left( \hat{C}_{R} - \hat{C}_{A} \right) \right]
f_{S}\left( x \right) 
\nonumber
\\
&+& 
\left[ \hat{C}_{R} \left(\hat{R}_{1}\hat{\tau}_{3} - \hat{\tau}_{3}
\hat{A}_{1} \right) - 
\left( \hat{R}_{1}\hat{\tau}_{3} - \hat{\tau}_{3}\hat{A}_{1} \right)
\hat{C}_{A} \right] f_{3N}\left(x \right)
\nonumber
\\
&=&
\left[\hat{C}_{R}\left(\hat{R}_{1}+\hat{\tau}_{3}\hat{R}^{\dagger}_{1}\hat{\tau}_{3} \right)
+ \left(\hat{R}_{1}+ \hat{\tau}_{3} \hat{R}^{\dagger}_{1}\hat{\tau}_{3}\right) 
\hat{\tau}_{3}\hat{C}^{\dagger}_{R}\hat{\tau}_{3} \right]
f_{0N}\left(x \right) 
\nonumber
\\
&+& 
\left[-\left(\hat{C}_{R} + \hat{\tau}_{3} \hat{C}^{\dagger}_{R}\hat{\tau}_{3} 
\right)\hat{\tau}_{3}\hat{R}_{1}^{\dagger}\hat{\tau}_{3} - 
\hat{R}_{1} \left( \hat{C}_{R} + \hat{\tau}_{3} \hat{C}^{\dagger}_{R}\hat{\tau}_{3} \right) \right]
f_{S}\left( x \right) 
\nonumber
\\
&+& 
\left[ \hat{C}_{R} 
\left(\hat{R}_{1}\hat{\tau}_{3} + 
\hat{R}^{\dagger}_{1}\hat{\tau}_{3} \right) +  
\left( \hat{R}_{1}\hat{\tau}_{3} + \hat{R}^{\dagger}_{1} \hat{\tau}_{3} \right)
\hat{\tau}_{3}\hat{C}^{\dagger}_{R}\hat{\tau}_{3} 
\right] 
f_{3N}
\left(x \right),
\label{lambda1a}
\end{eqnarray}
\begin{eqnarray}
\hat{\Lambda}_{2}
&=&
\left[ \hat{C}_{R}\left(\hat{R}_{1} - \hat{A}_{1}\right)
\hat{A}_{1}\hat{C}_{A}
-\left(\hat{R}_{1}-\hat{A}_{1}\right)\hat{A}_{1} \right] f_{0N}\left(x\right)
\nonumber
\\
&+&
\left[
\left(\hat{C}_{R}-\hat{C}_{A}\right) \hat{C}_{A} - \hat{R}_{1}
\left(\hat{C}_{R}-\hat{C}_{A}\right) \hat{C}_{A}\hat{A}_{1}\right] f_{S}
\left(x \right)
\nonumber
\\
&+&
\left[\hat{C}_{R}\left(\hat{R}_{1}\tau_{3} - \hat{\tau}_{3}\hat{A}_{1}
\right) \hat{A}_{1}\hat{C}_{A}
-\left( \hat{R}_{1}\hat{\tau}_{3} - \hat{\tau}_{3}\hat{A}_{1} \right)
\hat{A}_{1} \right]f_{3N}\left(x \right)
\nonumber
\\
&=&
\left( 
\hat{\mathbb{I}} 
+ \hat{R}_{1}\hat{\tau}_{3}\hat{R}^{\dagger}_{1}\hat{\tau}_{3} 
+\hat{C}_{R}\hat{\tau}_{3}\hat{C}^{\dagger}_{R}\hat{\tau}_{3}
+\hat{C}_{R}\hat{R}_{1}\hat{\tau}_{3}\hat{R}^{\dagger}_{1}
\hat{C}^{\dagger}_{R}\hat{\tau}_{3}
\right)f_{0N}\left(x\right)
\nonumber
\\
&-&
\left( \hat{\mathbb{I}} + \hat{C}_{R}\hat{\tau}_{3}\hat{C}^{\dagger}_{R}\hat{\tau}_{3}
+ \hat{R}_{1}\hat{\tau}_{3}\hat{R}^{\dagger}_{1}\hat{\tau}_{3}
+ \hat{R}_{1}\hat{C}_{R}\hat{\tau}_{3}\hat{C}^{\dagger}_{R}\hat{R}^{\dagger}_{1}\hat{\tau}_{3}
\right)f_{S}\left(x \right)
\nonumber
\\
&+&
\left(
\hat{\tau}_{3} + \hat{R}_{1}\hat{R}^{\dagger}_{1}\hat{\tau}_{3}
+ \hat{C}_{R}\hat{C}^{\dagger}_{R}\hat{\tau}_{3}
+ \hat{C}_{R}\hat{R}_{1}\hat{R}^{\dagger}_{1}
\hat{C}^{\dagger}_{R}\hat{\tau}_{3}
\right)
f_{3N} \left( x \right)
\label{lambda2a}.
\end{eqnarray}
To obtain the charge current, 
we focus on the boundary condition given by eq. (\ref{KeldyshKLNazarov}). 
The Keldysh part of the left side of this equation is proportional to 
\begin{eqnarray}
&&
\hat{R}\left(x \right) \frac{\partial }{\partial x}
\hat{K}\left( x \right) 
+ \hat{K}\frac{\partial }{\partial x}\hat{A}\left( x \right)
\nonumber
\\
&=&
\hat{R}\left(x \right)
\frac{\partial \hat{R}\left( x \right)}{\partial x}\hat{f}\left( x \right)
+\frac{\partial \hat{f}\left( x \right)}{\partial x} 
-\hat{R}\left( x \right)\frac{\partial \hat{f}\left( x \right)}
{\partial x}\hat{A}\left( x \right) 
-\hat{f}\left( x \right)
\hat{A}\left( x \right)
\frac{\partial \hat{A}\left( x \right)}{\partial x}
\nonumber
\\
&=&
\hat{R}\left(x \right)\frac{\partial \hat{R}\left(x\right)}{\partial x}
\left[ f_{0N}\left( x \right) + \hat{\tau}_{3}f_{3N}\left( x \right)
\right]
- 
\left[ f_{0N}\left( x \right) + \hat{\tau}_{3}f_{3N}\left( x \right)
\right]
\hat{\tau}_{3}\hat{R}^{\dagger}\left( x \right)
\frac{\partial \hat{R}^{\dagger}\left( x \right)}{\partial x}
\hat{\tau}_{3}
\nonumber
\\
&+&
\frac{\partial f_{0N}\left( x \right)}{\partial x}
\left[ \hat{\mathbb{I}} + \hat{R}\left( x \right)
\hat{\tau}_{3} \hat{R}^{\dagger}\left( x \right) \hat{\tau}_{3}
\right]
+ \frac{\partial f_{3N}\left( x \right)}{\partial x}
\left[ \hat{\mathbb{I}} + \hat{R}\left( x \right)
\hat{R}^{\dagger}\left( x \right) 
\right] \hat{\tau}_{3}.
\label{derivationKeldysha}
\end{eqnarray}
In order to obtain the charge conductance, the 
following calculation is needed. 
From eq. (\ref{KeldyshKLNazarov}), the 
$\hat{\tau}_{3}$ component of the boundary condition of the Keldysh
component is given by 
\begin{equation}
\frac{L}{R_{d}}
\left.
{\rm Trace}
\left[
\left(\hat{R}\frac{\partial}{\partial x}\hat{K} + 
\hat{K}\frac{\partial }{\partial x} \hat{A} \right)
\hat{\tau}_{3}
\right] \right|_{x=0_{-}}
=
-\frac{1}{R_{b}} \left< {\rm Trace}
\left[ \hat{I}_{K} \hat{\tau}_{3} \right]
\right>.
\end{equation}
By using eq. (\ref{derivationKeldysha}), 
the left side of the boundary condition 
is proportional to 
\begin{equation}
\left.
{\rm Trace}
\left[
\left(\hat{R}\frac{\partial}{\partial x}\hat{K} + 
\hat{K}\frac{\partial }{\partial x} \hat{A} \right)
\hat{\tau}_{3}
\right] \right|_{x=0_{-}}
=\left.
4 \left(
\frac{\partial f_{3N}\left( x \right) }
{\partial x}
\right)
{\rm cosh}^{2} \zeta_{im} \right|_{x=0_{-}}.
\end{equation}
with the imaginary part of $\zeta$ denoted by $\zeta_{im}$. 
Then, the boundary condition is expressed by 
\begin{equation}
\left.
4 \left(
\frac{L}{R_{d}}
\right)
\left(
\frac{ \partial f_{3} \left(x \right)}
{\partial x}
\right)
{\rm cosh}^{2} \zeta_{im} \right|_{x=0_{-}}
=
-\frac{1}{R_{b}}
\left< 
{\rm Trace}
\left[ \hat{I}_{K}\hat{\tau}_{3}
\right] 
\right>.
\end{equation}
Below, we calculate 
\begin{equation}
\left.
\left(
\frac{L}{R_{d}}
\right)
\left(
\frac{ \partial f_{3} \left(x \right)}
{\partial x}
\right)
{\rm cosh}^{2} \zeta_{im} \right|_{x=0_{-}}
=
-\frac{1}{4R_{b}}
\left< 
{\rm Trace}
\left[ \hat{I}_{K}\hat{\tau}_{3}
\right] 
\right>
=
-\frac{1}{R_{b}}
\left< I_{K} \right>.
\end{equation}
From eqs.(\ref{Keldysh1a}), (\ref{Keldysh2a}) and 
(\ref{Keldysh3a}), 
\begin{eqnarray}
{\rm Trace}
\left( \hat{\Lambda}_{1} \hat{\tau}_{3} \right)
&=&
{\rm Trace}
\left[
\left(\hat{C}_{R}\hat{R}_{1} + \hat{R}_{1}^{\dagger} \hat{C}^{\dagger}_{R}
\right) \hat{\tau}_{3}
+ \left( \hat{C}_{R}\hat{\tau}_{3}\hat{R}^{\dagger}_{1}
+ \hat{R}_{1}\hat{\tau}_{3}\hat{C}^{\dagger}_{R}
\right)
\right]
 f_{0N}\left(x\right)
\nonumber
\\
&-&
{\rm Trace}\left[
\left(\hat{C}^{\dagger}_{R}\hat{R}^{\dagger}_{1} 
+ \hat{R}_{1} \hat{C}_{R}
\right) \hat{\tau}_{3}
+ \left( \hat{C}_{R}\hat{\tau}_{3}\hat{R}^{\dagger}_{1}
+ \hat{R}_{1}\hat{\tau}_{3}\hat{C}^{\dagger}_{R}
\right)
\right]
f_{S}\left(x \right)
\nonumber
\\
&+&
{\rm Trace}
\left[
\left(\hat{R}_{1} + \hat{R}^{\dagger}_{1} \right) 
\left( \hat{C}_{R} + \hat{C}^{\dagger}_{R} \right)
\right]
f_{3N}\left(x \right),
\label{Tanakamatrixcurrentcharge1a}
\end{eqnarray}

\begin{eqnarray}
{\rm Trace}
\left( \hat{\Lambda}_{2} \hat{\tau}_{3} \right)
&=&
{\rm Trace}
\left[
\left(
\hat{\mathbb{I}} + \hat{C}^{\dagger}_{R}\hat{C}_{R}
+ \hat{R}^{\dagger}_{1} \hat{R}_{1} 
+ \hat{R}^{\dagger}_{1}\hat{C}^{\dagger}_{R}
\hat{C}_{R}\hat{R}_{1} \right) 
\hat{\tau}_{3}
\right]
f_{0N}\left(x\right)
\nonumber
\\
&-&
{\rm Trace}
\left[
\left(\hat{\mathbb{I}} + \hat{C}^{\dagger}_{R}\hat{C}_{R}
+ \hat{R}^{\dagger}_{1} \hat{R}_{1} 
+ \hat{C}^{\dagger}_{R}\hat{R}^{\dagger}_{1}
\hat{R}_{1} \hat{C}_{R} 
\right)\hat{\tau}_{3}
\right]
f_{S}(x)
\nonumber
\\
&+&
{\rm Trace}
\left[
\left(\hat{\mathbb{I}} + \hat{C}^{\dagger}_{R}\hat{C}_{R}
+ \hat{R}^{\dagger}_{1} \hat{R}_{1} 
+ \hat{R}^{\dagger}_{1}\hat{C}^{\dagger}_{R}
\hat{C}_{R}\hat{R}_{1} \right) 
\right]
f_{3N}\left(x \right).
\label{Tanakamatrixcurrentcharge2a}
\end{eqnarray}
Then, ${\rm Trace}[\hat{I}_{K}\hat{\tau}_{3}]$ becomes 
eq. (\ref{Tanakamatrixcurrentcharge3}).

\section{Appendix D}

\label{sec:AppendixD}
In this Appendix, we show how eq.(\ref{Tanakamatrixcurrentcharge3parity}) 
is obtained. 
We decompose matrices in 
eq. (\ref{Tanakamatrixcurrentcharge3}) as 
\begin{eqnarray}
\hat{C}_{R}\hat{R}_{1}
& =& c_{11}\left(\theta\right)\hat{\tau}_{1} + c_{12}\left(\theta\right)\hat{\tau}_{2}
+ c_{13}\left(\theta\right)\hat{\tau}_{3}
+ c_{0}\left(\theta\right),
\label{coefficienta1d}
\\
\hat{R}^{\dagger}_{1}\hat{C}^{\dagger}_{R}
&=& c^{*}_{11}\left(\theta\right)\hat{\tau}_{1} + c^{*}_{12}\left(\theta\right)\hat{\tau}_{2}
+ c^{*}_{13}\left(\theta\right)\hat{\tau}_{3}
+ c^{*}_{0}\left(\theta\right),
\label{coefficienta2d}
\\
\hat{R}^{\dagger}_{1}\hat{C}_{R}
&=& c_{21}\left(\theta\right)\hat{\tau}_{1} + c_{22}\left(\theta\right)\hat{\tau}_{2}
+ c_{23}\left(\theta\right)\hat{\tau}_{3}
+ \bar{c}_{0}\left(\theta\right),
\label{coefficienta3d}
\\
\hat{C}^{\dagger}_{R}\hat{R}_{1}
&=& c^{*}_{21}\left(\theta\right)\hat{\tau}_{1} 
+ c^{*}_{22}\left(\theta\right)\hat{\tau}_{2}
+ c^{*}_{23}\left(\theta\right)\hat{\tau}_{3}
+ \bar{c}^{*}_{0}\left(\theta\right),
\label{coefficienta4d}
\\
\hat{R}_{1}\hat{C}_{R}
&=& -\left[ c_{11}\left(\theta\right)\hat{\tau}_{1} + c_{12}\left(\theta\right)\hat{\tau}_{2}
+ c_{13}\left(\theta\right)\hat{\tau}_{3} \right]
+ c_{0}\left(\theta\right),
\label{coefficienta5d}
\\
\hat{C}^{\dagger}_{R}\hat{R}^{\dagger}_{1}
&=& -\left[ c^{*}_{11}\left(\theta\right)\hat{\tau}_{1} 
+ c^{*}_{12}\left(\theta\right)\hat{\tau}_{2}
+ c^{*}_{13}\left(\theta\right)\hat{\tau}_{3} \right]
+ c^{*}_{0}\left(\theta\right).
\label{coefficienta6d}
\end{eqnarray}
These coefficients satisfy following relations
\[
c_{0}\left(\theta \right)=c_{0}\left(-\theta \right), \ \ 
\bar{c}_{0}\left(\theta \right)=\bar{c}_{0}\left(-\theta \right)
\]
for both spin-singlet and spin-triplet superconductors. 
\begin{eqnarray}
c_{11}\left(\theta \right)&=&c_{11}\left(-\theta \right), \  
c_{12}\left(\theta \right)=-c_{12}\left(-\theta \right), \  
c_{13}\left(\theta \right)=-c_{13}\left(-\theta \right), \  
c_{21}\left(\theta \right)=c_{21}\left(-\theta \right),  \
\nonumber
\\
c_{22}\left(\theta \right)&=&-c_{22}\left(-\theta \right), \  
c_{23}\left(\theta \right)=-c_{23}\left(-\theta \right) 
\label{coefficients4d}
\end{eqnarray}
for a spin-singlet superconductor 
and 
\begin{eqnarray}
c_{11}\left(\theta \right)&=&-c_{11}\left(-\theta \right), \ 
c_{12}\left(\theta \right)=c_{12}\left(-\theta \right), \  
c_{13}\left(\theta \right)=-c_{13}\left(-\theta \right), \  
c_{21}\left(\theta \right)=-c_{21}\left(-\theta \right),  \
\nonumber
\\
c_{22}\left(\theta \right)&=&c_{22}\left(-\theta \right), \  
c_{23}\left(\theta \right)=-c_{23}\left(-\theta \right) 
\label{coefficients5d}
\end{eqnarray}
for a spin-triplet one. 
In addition, $d_{R}$ becomes 
\[
d_{R}=1 + \sigma_{1N}^{2} 
+ 2 \sigma_{1N}
\left[s_{2}c_{2}\left(\theta\right) + s_{3}c_{3}\left(\theta\right)
\right]
\]
for a spin-singlet superconductor and 
\[
d_{R}=1 + \sigma_{1N}^{2} 
+ 2 \sigma_{1N}
\left[s_{1}c_{1}\left(\theta\right) + s_{3}c_{3}\left(\theta\right)
\right]
\]
for a spin-triplet one. 
From eqs. (\ref{coefficients2}) and (\ref{coefficients3}), 
$d_{R}=d_{R}(\theta)$ 
satisfies $d_{R}(\theta)=d_{R}(-\theta)$ for both two cases. 
Then, we can show that
\begin{equation}
\left<
\frac{2 \sigma_{1N}}{\mid d_{R} \mid^{2}}
{\rm Trace}
\left[ \left\{
\left( \hat{C}_{R} + \hat{C}^{\dagger}_{R} \right)\hat{R}_{1}
+ 
\hat{R}^{\dagger}_{1}
\left( \hat{C}_{R} + \hat{C}^{\dagger}_{R} \right)
\right\}
\hat{\tau}_{3} \right]
\left(1 + \sigma^{2}_{1N} \right)
\right>
=0
\end{equation}
and 
\begin{equation}
\left<
\frac{2 \sigma_{1N}}{\mid d_{R} \mid^{2}}
{\rm Trace}
\left[ 
\left\{
\left( \hat{R}_{1} + \hat{R}^{\dagger}_{1} \right)\hat{C}_{R}
+ \hat{C}^{\dagger}_{R}
\left( \hat{R}_{1} + \hat{R}^{\dagger}_{1} \right) 
\right\}
\hat{\tau}_{3}
\right]
\left( 1 + \sigma_{1N}^{2}\right)
\right>=0.
\end{equation}

From eqs.(\ref{R1structure1}) and (\ref{R1structure2}),  
$\hat{R}_{1}\hat{R}^{\dagger}_{1}$ 
and $\hat{R}^{\dagger}_{1}\hat{R}_{1}$ satisfy 
\begin{equation}
{\rm Trace}
\left[
\hat{R}_{1}\hat{R}^{\dagger}_{1}\hat{\tau}_{3}
\right]
=0, \ 
{\rm Trace}
\left[
\hat{R}^{\dagger}_{1}\hat{R}_{1}\hat{\tau}_{3}
\right]
=0
\label{derivation1d}.
\end{equation}
Using eqs. (\ref{coefficients2}) and (\ref{coefficients3}), 
\[
c_{1}\left(\theta \right)c^{*}_{2}\left(\theta \right)-
c_{2}\left(\theta \right)c^{*}_{1}\left(\theta \right)
=
-\left(
c_{1}\left(-\theta \right)c^{*}_{2}\left(-\theta \right)-
c_{2}\left(-\theta \right)c^{*}_{1}\left(-\theta \right)
\right)
\]
is satisfied both for spin-singlet and spin-triplet cases. 
Here the relation 
\begin{equation}
{\rm Trace}
\left[
\hat{C}^{\dagger}_{R}\hat{C}_{R}\hat{\tau}_{3}
\right]
=2i\left(c_{1}c^{*}_{2}-c_{2}c^{*}_{1} \right)
\label{derivation20d}
\end{equation}
is satisfied. Since it is an odd function of $\theta$  we obtain 
\begin{equation}
\left<
\frac{4 \sigma^{2}_{1N}}{\mid d_{R} \mid^{2}}
{\rm Trace}
\left[
\hat{C}^{\dagger}_{R}\hat{C}_{R}\hat{\tau}_{3}
\right]
\right>
=0.
\label{derivation2d}
\end{equation}
Using eqs. (\ref{coefficienta5d}) and (\ref{coefficienta6d}), 
we obtain
\begin{equation}
{\rm Trace}
\left[\left(\hat{C}^{\dagger}_{R}\hat{R}^{\dagger}_{1}
\hat{R}_{1}\hat{C}_{R} \right)\tau_{3} \right]
=2i\left[
c^{*}_{11}\left(\theta \right)c_{12}\left(\theta \right)
-c^{*}_{12}\left(\theta \right)c_{11}\left(\theta \right)
\right]
-2\left[
c^{*}_{13}\left(\theta \right)c_{0}\left(\theta \right)
+ c^{*}_{0}\left(\theta \right)c_{13}\left(\theta \right)
\right].
\end{equation}
It becomes an odd-function of $\theta$ 
both for spin-singlet and spin-triplet cases 
from eqs. (\ref{coefficients4d}) and (\ref{coefficients5d}). 
Then, we obtain 
\begin{equation}
\left< 
\frac{4 \sigma^{2}_{1N}}{\mid d_{R} \mid^{2}}
{\rm Trace}
\left[\left(\hat{C}^{\dagger}_{R}\hat{R}^{\dagger}_{1}
\hat{R}_{1}\hat{C}_{R} \right)\tau_{3} \right]
\sigma_{1N} 
\right>=0
\label{derivation3d}.
\end{equation}
Similarly, since 
\begin{equation}
{\rm Trace}
\left[\left(\hat{R}^{\dagger}_{1}\hat{C}^{\dagger}_{R}
\hat{C}_{R}\hat{R}_{1} \right)\tau_{3} \right]
=2i\left[
c^{*}_{11}\left(\theta \right)c_{12}\left(\theta \right)
-c^{*}_{12}\left(\theta \right)c_{11}\left(\theta \right)
\right]
+2\left[
c^{*}_{13}\left(\theta \right)c_{0}\left(\theta \right)
+c^{*}_{0}\left(\theta \right)c_{13}\left(\theta \right)
\right]
\label{derivation40d}
\end{equation}
we obtain 
\begin{equation}
\left< 
\frac{4 \sigma^{2}_{1N}}{\mid d_{R} \mid^{2}}
{\rm Trace}
\left[\left(\hat{R}^{\dagger}_{1}\hat{C}^{\dagger}_{R}
\hat{C}_{R}\hat{R}_{1} \right)\tau_{3} \right]
\right>=0
\label{derivation4d}.
\end{equation}
From eqs. (\ref{derivation1d}), (\ref{derivation2d}), 
(\ref{derivation3d}), and (\ref{derivation4d}), 
we obtain 
\begin{equation}
\left<
\frac{4 \sigma^{2}_{1N}}{\mid d_{R} \mid^{2}}
{\rm Trace}
\left[ \left\{
\hat{\mathbb{I}} + \hat{C}^{\dagger}_{R}\hat{C}_{R} 
+ \hat{R}^{\dagger}_{1}
\left( \hat{\mathbb{I}} + \hat{C}^{\dagger}_{R} \hat{C}_{R} \right) 
\hat{R}_{1}
\right\}
\hat{\tau}_{3} \right]
\right>f_{0N}\left(x\right) =0
\label{derivation5}
\end{equation}
and 
\begin{equation}
\left<
\frac{4 \sigma^{2}_{1N}}{\mid d_{R} \mid^{2}}
{\rm Trace}
\left[ 
\left\{
\hat{\mathbb{I}} + \hat{R}^{\dagger}_{1}\hat{R}_{1} 
+ \hat{C}^{\dagger}_{R}
\left( \hat{\mathbb{I}} + \hat{R}^{\dagger}_{1} \hat{R}_{1} \right) 
\hat{C}_{R}
\right\}
\hat{\tau}_{3} \right]
\right>
f_{S}\left(x \right)=0.
\label{derivation6}
\end{equation}
Using these relations in eq. (\ref{Tanakamatrixcurrentcharge3})
immediatly results in eq.  (\ref{Tanakamatrixcurrentcharge3parity}). 

\section{Appendix E}
\label{sec:AppendixE}
In this Appendix, we calculate 
$S_{1\uparrow\left(\downarrow\right)}$, 
$S_{2\uparrow\left(\downarrow\right)}$, $S_{3\uparrow\left(\downarrow\right)}$, $S_{4\uparrow\left(\downarrow\right)}$, 
$S_{5\uparrow\left(\downarrow\right)}$ and 
$S_{6\uparrow\left(\downarrow \right)}$
which appear in eq. (\ref{TracehatIK}). 

\begin{eqnarray}
S_{1\uparrow\left(\downarrow\right)}
&=& f_{0N\uparrow\left(\downarrow\right)}\left(x=0_{-} \right)
\left( 1 + \sigma_{1N}^{2}\right)
\nonumber
\\
& \times &
{\rm Trace}
\left[ \left\{
\left( \hat{C}_{R\uparrow\left(\downarrow\right)} 
+ \hat{C}^{\dagger}_{R\uparrow\left(\downarrow\right)} \right)
\hat{R}_{1\uparrow\left(\downarrow\right)}
+ 
\hat{R}^{\dagger}_{1\uparrow\left(\downarrow\right)}
\left( \hat{C}_{R\uparrow\left(\downarrow\right)} 
+ \hat{C}^{\dagger}_{R\uparrow\left(\downarrow\right)} \right) \right\}
\hat{\tau}_{3} \right]
\label{S1updowne}
\\
S_{2\uparrow\left(\downarrow\right)}
&=&2\sigma_{1N}
f_{0N\uparrow\left(\downarrow\right)}\left(x=0_{-} \right)
\nonumber
\\
& \times &
{\rm Trace}
\left[ \left\{
\hat{\mathbb{I}} + \hat{C}^{\dagger}_{R\uparrow\left(\downarrow\right)}
\hat{C}_{R\uparrow\left(\downarrow\right)} 
+ \hat{R}^{\dagger}_{1\uparrow\left(\downarrow\right)}
\left( \hat{\mathbb{I}} + \hat{C}^{\dagger}_{R\uparrow\left(\downarrow\right)} 
\hat{C}_{R\uparrow\left(\downarrow\right)} \right) 
\hat{R}_{1\uparrow\left(\downarrow\right)}
\right\}
\hat{\tau}_{3} \right]
\label{S2updowne}
\\
S_{3\uparrow\left(\downarrow\right)}
&=&-
\left( 1 + \sigma_{1N}^{2}\right) f_{S\uparrow\left(\downarrow\right)}\left(x=0_{+} \right)
\nonumber
\\
&\times&
{\rm Trace}
\left[ \left\{
\left( \hat{R}_{1\uparrow\left(\downarrow\right)} 
+ \hat{R}^{\dagger}_{1\uparrow\left(\downarrow\right)} \right)
\hat{C}_{R\uparrow\left(\downarrow\right)}
+ \hat{C}^{\dagger}_{R\uparrow\left(\downarrow\right)}
\left( \hat{R}_{1\uparrow\left(\downarrow\right)} 
+ \hat{R}^{\dagger}_{1\uparrow\left(\downarrow\right)} \right) \right\}
\hat{\tau}_{3} \right]
\label{S3updowne}
\\
S_{4\uparrow\left(\downarrow\right)}
&=&
-2\sigma_{1N}
f_{S\uparrow\left(\downarrow\right)}\left(x=0_{+} \right)
\nonumber
\\
&\times&
{\rm Trace}
\left[
\left\{
\hat{\mathbb{I}} + \hat{R}^{\dagger}_{1\uparrow\left(\downarrow\right)}\hat{R}_{1\uparrow\left(\downarrow\right)} 
+ \hat{C}^{\dagger}_{R\uparrow\left(\downarrow\right)}
\left( \hat{\mathbb{I}} + \hat{R}^{\dagger}_{1\uparrow\left(\downarrow\right)} 
\hat{R}_{1\uparrow\left(\downarrow\right)} \right) 
\hat{C}_{R\uparrow\left(\downarrow\right)}
\right\}
\hat{\tau}_{3} \right]
\label{S4updowne}
\\
S_{5\uparrow\left(\downarrow\right)}
&=&\left( 1 + \sigma_{1N}^{2}\right) 
\nonumber
\\
&\times&
f_{3N\uparrow\left(\downarrow\right)}\left(x=0_{-} \right)
{\rm Trace}
\left[ 
\left( \hat{R}_{1\uparrow\left(\downarrow\right)} 
+ \hat{R}^{\dagger}_{1\uparrow\left(\downarrow\right)} \right)
\left( \hat{C}_{R\uparrow\left(\downarrow\right)} 
+ \hat{C}^{\dagger}_{R\uparrow\left(\downarrow\right)} \right) 
\right]
\label{S5updowne}
\\
S_{6\uparrow\left(\downarrow\right)}
&=&2\sigma_{1N}f_{3N\uparrow\left(\downarrow\right)}\left(x=0_{-} \right)
\nonumber
\\
&\times&
{\rm Trace}
\left[
\left( 
\hat{\mathbb{I}} + \hat{R}^{\dagger}_{1\uparrow\left(\downarrow\right)}
\hat{R}_{1\uparrow\left(\downarrow\right)} 
\right)
\left( \hat{\mathbb{I}}
+ \hat{C}^{\dagger}_{R\uparrow\left(\downarrow\right)}\hat{C}_{R\uparrow\left(\downarrow\right)} \right)
\right].
\label{S6updowne}
\end{eqnarray}

Since $\zeta_{\uparrow}(x)=\zeta_{\downarrow}(x)$ is satisfied, it is 
plausible to assume 
\begin{eqnarray}
f_{0N\uparrow}\left(x \right)&=&f_{0N\downarrow}\left(x \right)=f_{0N}\left(x \right)
\nonumber
\\
f_{3N\uparrow}\left(x \right)&=&f_{3N\downarrow}\left(x \right)=f_{3N}\left(x \right)
\nonumber
\\
f_{S\uparrow}\left(x \right)&=&f_{S\downarrow}\left(x \right)=f_{S}\left(x \right)
\label{assumptiondistributione}
\end{eqnarray}
in the following calculations. 
By using eqs. (\ref{S1updowne}) to (\ref{S6updowne}) and 
(\ref{assumptiondistributione}), 
we can derive the following relations: 
\begin{eqnarray}
S_{1\downarrow}
&=& 
f_{0N}\left(x=0_{-} \right)
\left( 1 + \sigma_{1N}^{2}\right)
{\rm Trace}
\left[ \left\{
\left( \hat{C}_{R\downarrow} 
+ \hat{C}^{\dagger}_{R\downarrow} \right)
\hat{R}_{1\downarrow}
+ 
\hat{R}^{\dagger}_{1\downarrow}
\left( \hat{C}_{R\downarrow} 
+ \hat{C}^{\dagger}_{R\downarrow} \right) \right\}
\hat{\tau}_{3} \right]
\nonumber
\\
&=& 
f_{0N}\left(x=0_{-} \right)
\left( 1 + \sigma_{1N}^{2}\right)
{\rm Trace}
\left[ \left\{ \hat{\tau}_{2}
\left( \hat{C}_{R\uparrow} 
+ \hat{C}^{\dagger}_{R\uparrow} \right)
\hat{R}_{1\uparrow}\hat{\tau}_{2}
+ \hat{\tau}_{2}
\hat{R}^{\dagger}_{1\uparrow}
\left( \hat{C}_{R\uparrow} 
+ \hat{C}^{\dagger}_{R\uparrow} \right)\hat{\tau}_{2} \right\}
\hat{\tau}_{3} \right]
\nonumber
\\
&=&-S_{1\uparrow}
\\
S_{2\downarrow}
&=& 2\sigma_{1N}
f_{0N}\left(x=0_{-} \right)
{\rm Trace}
\left[ \left\{
\hat{\mathbb{I}} + \hat{C}^{\dagger}_{R\downarrow}
\hat{C}_{R\downarrow} 
+ \hat{R}^{\dagger}_{1\downarrow}
\left( \hat{\mathbb{I}} + \hat{C}^{\dagger}_{R\downarrow} 
\hat{C}_{R\downarrow} \right) 
\hat{R}_{1\downarrow}
\right\}
\hat{\tau}_{3} \right]
\nonumber
\\
&=& 2\sigma_{1N}
f_{0N}\left(x=0_{-} \right)
{\rm Trace}
\left[ \left\{
\hat{\mathbb{I}} + \hat{\tau}_{2}\hat{C}^{\dagger}_{R\uparrow}
\hat{C}_{R\uparrow}\hat{\tau}_{2} 
+ \hat{\tau}_{2}\hat{R}^{\dagger}_{1\uparrow}\hat{\tau}_{2}
\left( \hat{\mathbb{I}} + \hat{\tau}_{2}
\hat{C}^{\dagger}_{R\uparrow} 
\hat{C}_{R\uparrow}\hat{\tau}_{2} \right) 
\hat{\tau}_{2}
\hat{R}_{1\uparrow}\hat{\tau}_{2}
\right\}
\hat{\tau}_{3} \right]
\nonumber
\\
&=&
-S_{2\uparrow}
\\
S_{3\downarrow}
&=&
-\left( 1 + \sigma_{1N}^{2}\right) f_{S}\left(x=0_{+} \right)
{\rm Trace}
\left[ \left\{
\left( \hat{R}_{1\downarrow} 
+ \hat{R}^{\dagger}_{1\downarrow}\right)
\hat{C}_{R\downarrow}
+ \hat{C}^{\dagger}_{R\downarrow}
\left( \hat{R}_{1\downarrow} 
+ \hat{R}^{\dagger}_{\downarrow} \right) 
\right\}
\hat{\tau}_{3} \right]
\nonumber
\\
&=&
-\left( 1 + \sigma_{1N}^{2}\right) f_{S}\left(x=0_{+} \right)
{\rm Trace}
\left[ \left\{
\hat{\tau}_{2}
\left( \hat{R}_{1\uparrow} 
+ \hat{R}^{\dagger}_{1\uparrow} \right)\hat{C}_{R\uparrow}\hat{\tau}_{2}
+ \hat{\tau}_{2}\hat{C}^{\dagger}_{R\uparrow}
\left( \hat{R}_{1\uparrow} 
+ \hat{R}^{\dagger}_{1\uparrow} \right) \hat{\tau}_{2} \right\}
\hat{\tau}_{3} \right]
\nonumber
\\
&=&
-S_{3\uparrow}
\\
S_{4\downarrow}
&=&
-2\sigma_{1N}
f_{S}\left(x=0_{+} \right)
{\rm Trace}
\left[ \left\{
\hat{\mathbb{I}} + \hat{R}^{\dagger}_{1\downarrow}\hat{R}_{1\downarrow} 
+ \hat{C}^{\dagger}_{R\downarrow}
\left( \hat{\mathbb{I}} + \hat{R}^{\dagger}_{1\downarrow} 
\hat{R}_{1\downarrow} \right) 
\hat{C}_{R\downarrow}
\right\}
\hat{\tau}_{3} \right]
\nonumber
\\
&=&
-2\sigma_{1N}
f_{S}\left(x=0_{+} \right)
{\rm Trace}
\left[ \left\{
\hat{\mathbb{I}} + 
\hat{\tau}_{2}
\hat{R}^{\dagger}_{1\uparrow}\hat{R}_{1\uparrow}\hat{\tau}_{2} 
+ \hat{\tau}_{2}\hat{C}^{\dagger}_{R\uparrow}
\left( \hat{\mathbb{I}} + \hat{R}^{\dagger}_{1\uparrow} 
\hat{R}_{1\uparrow} \right) 
\hat{C}_{R\uparrow}\hat{\tau}_{2}
\right\}
\hat{\tau}_{3} \right]
\nonumber
\\
&=&-S_{4\uparrow}
\\
S_{5\downarrow}
&=& \left( 1 + \sigma_{1N}^{2}\right) 
f_{3N}\left(x=0_{-} \right)
{\rm Trace}
\left[ 
\left( \hat{R}_{1\downarrow} 
+ \hat{R}^{\dagger}_{1\downarrow} \right)
\left( \hat{C}_{R\downarrow} 
+ \hat{C}^{\dagger}_{R\downarrow} \right) 
\right]
\nonumber
\\
&=&
\left( 1 + \sigma_{1N}^{2}\right) 
f_{3N}\left(x=0_{-} \right)
{\rm Trace}
\left[ \hat{\tau}_{2}
\left( \hat{R}_{1\uparrow} 
+ \hat{R}^{\dagger}_{1\uparrow} \right)
\left( \hat{C}_{R\uparrow} 
+ \hat{C}^{\dagger}_{R\uparrow} \right) 
\hat{\tau}_{2}
\right]
\nonumber
\\
&=&
S_{5\uparrow}
\\
S_{6\downarrow}
&=&
2\sigma_{1N}f_{3N}\left(x=0_{-} \right)
{\rm Trace}
\left[
\left( 
\hat{\mathbb{I}} + \hat{R}^{\dagger}_{1\downarrow}
\hat{R}_{1\downarrow} 
\right)
\left( \hat{\mathbb{I}}
+ \hat{C}^{\dagger}_{R\downarrow}\hat{C}_{R\downarrow} \right)
\right]
\nonumber
\\
&=&
2\sigma_{1N}f_{3N}\left(x=0_{-} \right)
{\rm Trace}
\left[
\left( 
\hat{\mathbb{I}} + \hat{\tau}_{2}
\hat{R}^{\dagger}_{1\uparrow}
\hat{R}_{1\uparrow} \hat{\tau}_{2}
\right)
\left( \hat{\mathbb{I}}
+ \hat{\tau}_{2} \hat{C}^{\dagger}_{R\uparrow}\hat{C}_{R\uparrow} 
\hat{\tau}_{2}
\right)
\right]
\nonumber
\\
&=&
S_{6\uparrow}
\\
d_{R\downarrow} \hat{\mathbb{I}}
&=&  \left( 1 + \sigma^{2}_{1N} \right)\hat{\mathbb{I}}
+ \sigma_{1N} \left( \hat{C}_{R\downarrow}\hat{R}_{1\downarrow} 
+ \hat{R}_{1\downarrow}\hat{C}_{R\downarrow} \right)
\nonumber
\\
&=& \left( 1 + \sigma^{2}_{1N}\right) \hat{\mathbb{I}}
+ \sigma_{1N} \hat{\tau}_{2}
\left( \hat{C}_{R\downarrow}\hat{R}_{1\downarrow} 
+ \hat{R}_{1\downarrow}\hat{C}_{R\downarrow} \right)
\hat{\tau}_{2}
\nonumber
\\
&=& \left( 1 + \sigma^{2}_{1N} \right)\hat{\mathbb{I}}
+ \sigma_{1N} 
\left( \hat{C}_{R\uparrow}\hat{R}_{1\uparrow} 
+ \hat{R}_{1\uparrow}\hat{C}_{R\uparrow} \right)
=d_{R\uparrow}\hat{\mathbb{I}}
\nonumber
\\
&\equiv& d_{R}\hat{\mathbb{I}}.
\end{eqnarray}
Substituting the above equations into eq. (\ref{TracehatIK}), 
one obtains eq. (\ref{spinsumKeldyshcomponent}).
%
\bibliography{TopologicalSC}

\begin{thebibliography}{66}
\expandafter\ifx\csname natexlab\endcsname\relax\def\natexlab#1{#1}\fi
\expandafter\ifx\csname bibnamefont\endcsname\relax
  \def\bibnamefont#1{#1}\fi
\expandafter\ifx\csname bibfnamefont\endcsname\relax
  \def\bibfnamefont#1{#1}\fi
\expandafter\ifx\csname citenamefont\endcsname\relax
  \def\citenamefont#1{#1}\fi
\expandafter\ifx\csname url\endcsname\relax
  \def\url#1{\texttt{#1}}\fi
\expandafter\ifx\csname urlprefix\endcsname\relax\def\urlprefix{URL }\fi
\providecommand{\bibinfo}[2]{#2}
\providecommand{\eprint}[2][]{\url{#2}}

\bibitem[{\citenamefont{de~Gennes}(1969)}]{deGennes1}
\bibinfo{author}{\bibfnamefont{P.~G.} \bibnamefont{de~Gennes}},
  \emph{\bibinfo{title}{Superconductivity of Metals and Alloys}}
  (\bibinfo{publisher}{Benjamin, New York}, \bibinfo{year}{1969}).

\bibitem[{\citenamefont{de~Gennes}(1964)}]{deGennes2}
\bibinfo{author}{\bibfnamefont{P.~G.} \bibnamefont{de~Gennes}},
  \bibinfo{journal}{Rev. Mod. Phys.} \textbf{\bibinfo{volume}{36}},
  \bibinfo{pages}{225} (\bibinfo{year}{1964}).

\bibitem[{\citenamefont{van Wees et~al.}(1992)\citenamefont{van Wees, de~Vries,
  Magn\'ee, and Klapwijk}}]{vanWees}
\bibinfo{author}{\bibfnamefont{B.~J.} \bibnamefont{van Wees}},
  \bibinfo{author}{\bibfnamefont{P.}~\bibnamefont{de~Vries}},
  \bibinfo{author}{\bibfnamefont{P.}~\bibnamefont{Magn\'ee}}, \bibnamefont{and}
  \bibinfo{author}{\bibfnamefont{T.~M.} \bibnamefont{Klapwijk}},
  \bibinfo{journal}{Phys. Rev. Lett.} \textbf{\bibinfo{volume}{69}},
  \bibinfo{pages}{510} (\bibinfo{year}{1992}).

\bibitem[{\citenamefont{Kastalsky et~al.}(1991)\citenamefont{Kastalsky,
  Kleinsasser, Greene, Bhat, Milliken, and Harbison}}]{Kastalsky}
\bibinfo{author}{\bibfnamefont{A.}~\bibnamefont{Kastalsky}},
  \bibinfo{author}{\bibfnamefont{A.~W.} \bibnamefont{Kleinsasser}},
  \bibinfo{author}{\bibfnamefont{L.~H.} \bibnamefont{Greene}},
  \bibinfo{author}{\bibfnamefont{R.}~\bibnamefont{Bhat}},
  \bibinfo{author}{\bibfnamefont{F.~P.} \bibnamefont{Milliken}},
  \bibnamefont{and} \bibinfo{author}{\bibfnamefont{J.~P.}
  \bibnamefont{Harbison}}, \bibinfo{journal}{Phys. Rev. Lett.}
  \textbf{\bibinfo{volume}{67}}, \bibinfo{pages}{3026} (\bibinfo{year}{1991}).

\bibitem[{\citenamefont{Larkin and Ovchinikov}(1977)}]{Larkin1977}
\bibinfo{author}{\bibfnamefont{A.~I.} \bibnamefont{Larkin}} \bibnamefont{and}
  \bibinfo{author}{\bibfnamefont{Y.~N.} \bibnamefont{Ovchinikov}},
  \bibinfo{journal}{Zh. Eksp. Teor. Fiz.} \textbf{\bibinfo{volume}{73}},
  \bibinfo{pages}{299} (\bibinfo{year}{1977}), \bibinfo{note}{[Sov. Phys. JETP
  46, 155 (1977)]}.

\bibitem[{\citenamefont{Volkov et~al.}(1993)\citenamefont{Volkov, Zaitsev, and
  Klapwijk}}]{Volkov1993}
\bibinfo{author}{\bibfnamefont{A.~F.} \bibnamefont{Volkov}},
  \bibinfo{author}{\bibfnamefont{A.~V.} \bibnamefont{Zaitsev}},
  \bibnamefont{and} \bibinfo{author}{\bibfnamefont{T.~M.}
  \bibnamefont{Klapwijk}}, \bibinfo{journal}{Physica C}
  \textbf{\bibinfo{volume}{210}}, \bibinfo{pages}{21} (\bibinfo{year}{1993}).

\bibitem[{\citenamefont{Nazarov}(1994)}]{Nazarov1994}
\bibinfo{author}{\bibfnamefont{Y.~V.} \bibnamefont{Nazarov}},
  \bibinfo{journal}{Phys. Rev. Lett.} \textbf{\bibinfo{volume}{73}},
  \bibinfo{pages}{1420} (\bibinfo{year}{1994}).

\bibitem[{\citenamefont{Yip}(1995)}]{Yip1995}
\bibinfo{author}{\bibfnamefont{S.}~\bibnamefont{Yip}}, \bibinfo{journal}{Phys.
  Rev. B} \textbf{\bibinfo{volume}{52}}, \bibinfo{pages}{15504}
  (\bibinfo{year}{1995}).

\bibitem[{\citenamefont{Usadel}(1970)}]{Usadel}
\bibinfo{author}{\bibfnamefont{K.~D.} \bibnamefont{Usadel}},
  \bibinfo{journal}{Phys. Rev. Lett.} \textbf{\bibinfo{volume}{25}},
  \bibinfo{pages}{507} (\bibinfo{year}{1970}).

\bibitem[{\citenamefont{Kopnin}(2001)}]{kopnin}
\bibinfo{author}{\bibfnamefont{N.}~\bibnamefont{Kopnin}},
  \emph{\bibinfo{title}{Theory of Nonequilibrium Superconductivity}}
  (\bibinfo{publisher}{Oxford University Press}, \bibinfo{year}{2001}).

\bibitem[{\citenamefont{Kuprianov and Lukichev}(1988)}]{KLboundary}
\bibinfo{author}{\bibfnamefont{M.~Y.} \bibnamefont{Kuprianov}}
  \bibnamefont{and} \bibinfo{author}{\bibfnamefont{V.~F.}
  \bibnamefont{Lukichev}}, \bibinfo{journal}{Zh. Eksp. Teor. Fiz.}
  \textbf{\bibinfo{volume}{94}}, \bibinfo{pages}{129} (\bibinfo{year}{1988}),
  \bibinfo{note}{[Sov. Phys. JETP 67, 1163 (1988)]}.

\bibitem[{\citenamefont{Blonder et~al.}(1982)\citenamefont{Blonder, Tinkham,
  and Klapwijk}}]{BTK82}
\bibinfo{author}{\bibfnamefont{G.~E.} \bibnamefont{Blonder}},
  \bibinfo{author}{\bibfnamefont{M.}~\bibnamefont{Tinkham}}, \bibnamefont{and}
  \bibinfo{author}{\bibfnamefont{T.~M.} \bibnamefont{Klapwijk}},
  \bibinfo{journal}{Phys. Rev. B} \textbf{\bibinfo{volume}{25}},
  \bibinfo{pages}{4515} (\bibinfo{year}{1982}).

\bibitem[{\citenamefont{Nazarov}(1999)}]{Nazarov1999}
\bibinfo{author}{\bibfnamefont{Y.~V.} \bibnamefont{Nazarov}},
  \bibinfo{journal}{Superlattices and Microstructures}
  \textbf{\bibinfo{volume}{25}}, \bibinfo{pages}{1221} (\bibinfo{year}{1999}).

\bibitem[{\citenamefont{Lambert et~al.}(1997)\citenamefont{Lambert, Raimondi,
  Sweeney, and Volkov}}]{Lambert}
\bibinfo{author}{\bibfnamefont{C.~J.} \bibnamefont{Lambert}},
  \bibinfo{author}{\bibfnamefont{R.}~\bibnamefont{Raimondi}},
  \bibinfo{author}{\bibfnamefont{V.}~\bibnamefont{Sweeney}}, \bibnamefont{and}
  \bibinfo{author}{\bibfnamefont{A.~F.} \bibnamefont{Volkov}},
  \bibinfo{journal}{Phys. Rev. B} \textbf{\bibinfo{volume}{55}},
  \bibinfo{pages}{6015} (\bibinfo{year}{1997}).

\bibitem[{\citenamefont{Laikhtman and Luryi}(1994)}]{Laikhtman}
\bibinfo{author}{\bibfnamefont{B.}~\bibnamefont{Laikhtman}} \bibnamefont{and}
  \bibinfo{author}{\bibfnamefont{S.}~\bibnamefont{Luryi}},
  \bibinfo{journal}{Phys. Rev. B} \textbf{\bibinfo{volume}{49}},
  \bibinfo{pages}{17177} (\bibinfo{year}{1994}).

\bibitem[{\citenamefont{Tanaka et~al.}(2003{\natexlab{a}})\citenamefont{Tanaka,
  Nazarov, and Kashiwaya}}]{Proximityd}
\bibinfo{author}{\bibfnamefont{Y.}~\bibnamefont{Tanaka}},
  \bibinfo{author}{\bibfnamefont{Y.~V.} \bibnamefont{Nazarov}},
  \bibnamefont{and}
  \bibinfo{author}{\bibfnamefont{S.}~\bibnamefont{Kashiwaya}},
  \bibinfo{journal}{Phys. Rev. Lett.} \textbf{\bibinfo{volume}{90}},
  \bibinfo{pages}{167003} (\bibinfo{year}{2003}{\natexlab{a}}).

\bibitem[{\citenamefont{Tanaka et~al.}(2004)\citenamefont{Tanaka, Nazarov,
  Golubov, and Kashiwaya}}]{Proximityd2}
\bibinfo{author}{\bibfnamefont{Y.}~\bibnamefont{Tanaka}},
  \bibinfo{author}{\bibfnamefont{Y.~V.} \bibnamefont{Nazarov}},
  \bibinfo{author}{\bibfnamefont{A.~A.} \bibnamefont{Golubov}},
  \bibnamefont{and}
  \bibinfo{author}{\bibfnamefont{S.}~\bibnamefont{Kashiwaya}},
  \bibinfo{journal}{Phys. Rev. B} \textbf{\bibinfo{volume}{69}},
  \bibinfo{pages}{144519} (\bibinfo{year}{2004}).

\bibitem[{\citenamefont{Buchholtz and Zwicknagl}(1981)}]{ABS}
\bibinfo{author}{\bibfnamefont{L.~J.} \bibnamefont{Buchholtz}}
  \bibnamefont{and}
  \bibinfo{author}{\bibfnamefont{G.}~\bibnamefont{Zwicknagl}},
  \bibinfo{journal}{Phys. Rev. B} \textbf{\bibinfo{volume}{23}},
  \bibinfo{pages}{5788} (\bibinfo{year}{1981}).

\bibitem[{\citenamefont{Hara and Nagai}(1986)}]{Hara}
\bibinfo{author}{\bibfnamefont{J.}~\bibnamefont{Hara}} \bibnamefont{and}
  \bibinfo{author}{\bibfnamefont{K.}~\bibnamefont{Nagai}},
  \bibinfo{journal}{Prog. Theor. Phys.} \textbf{\bibinfo{volume}{76}},
  \bibinfo{pages}{1237} (\bibinfo{year}{1986}).

\bibitem[{\citenamefont{L{\"o}fwander et~al.}(2001)\citenamefont{L{\"o}fwander,
  Shumeiko, and Wendin}}]{ABSR2}
\bibinfo{author}{\bibfnamefont{T.}~\bibnamefont{L{\"o}fwander}},
  \bibinfo{author}{\bibfnamefont{V.~S.} \bibnamefont{Shumeiko}},
  \bibnamefont{and} \bibinfo{author}{\bibfnamefont{G.}~\bibnamefont{Wendin}},
  \bibinfo{journal}{Supercond. Sci. Technol.} \textbf{\bibinfo{volume}{14}},
  \bibinfo{pages}{R53} (\bibinfo{year}{2001}).

\bibitem[{\citenamefont{Bruder}(1990)}]{Bruder90}
\bibinfo{author}{\bibfnamefont{C.}~\bibnamefont{Bruder}},
  \bibinfo{journal}{Phys. Rev. B} \textbf{\bibinfo{volume}{41}},
  \bibinfo{pages}{4017} (\bibinfo{year}{1990}).

\bibitem[{\citenamefont{Hu}(1994)}]{Hu94}
\bibinfo{author}{\bibfnamefont{C.-R.} \bibnamefont{Hu}},
  \bibinfo{journal}{Phys. Rev. Lett.} \textbf{\bibinfo{volume}{72}},
  \bibinfo{pages}{1526} (\bibinfo{year}{1994}).

\bibitem[{\citenamefont{Tanaka and Kashiwaya}(1995)}]{TK95}
\bibinfo{author}{\bibfnamefont{Y.}~\bibnamefont{Tanaka}} \bibnamefont{and}
  \bibinfo{author}{\bibfnamefont{S.}~\bibnamefont{Kashiwaya}},
  \bibinfo{journal}{Phys. Rev. Lett.} \textbf{\bibinfo{volume}{74}},
  \bibinfo{pages}{3451} (\bibinfo{year}{1995}).

\bibitem[{\citenamefont{Kashiwaya and Tanaka}(2000)}]{Kashiwaya00}
\bibinfo{author}{\bibfnamefont{S.}~\bibnamefont{Kashiwaya}} \bibnamefont{and}
  \bibinfo{author}{\bibfnamefont{Y.}~\bibnamefont{Tanaka}},
  \bibinfo{journal}{Rep. Prog. Phys.} \textbf{\bibinfo{volume}{63}},
  \bibinfo{pages}{1641} (\bibinfo{year}{2000}).

\bibitem[{\citenamefont{Tanaka and Kashiwaya}(2004)}]{Proximityp}
\bibinfo{author}{\bibfnamefont{Y.}~\bibnamefont{Tanaka}} \bibnamefont{and}
  \bibinfo{author}{\bibfnamefont{S.}~\bibnamefont{Kashiwaya}},
  \bibinfo{journal}{Phys. Rev. B} \textbf{\bibinfo{volume}{70}},
  \bibinfo{pages}{012507} (\bibinfo{year}{2004}).

\bibitem[{\citenamefont{Tanaka et~al.}(2005)\citenamefont{Tanaka, Kashiwaya,
  and Yokoyama}}]{Proximityp2}
\bibinfo{author}{\bibfnamefont{Y.}~\bibnamefont{Tanaka}},
  \bibinfo{author}{\bibfnamefont{S.}~\bibnamefont{Kashiwaya}},
  \bibnamefont{and} \bibinfo{author}{\bibfnamefont{T.}~\bibnamefont{Yokoyama}},
  \bibinfo{journal}{Phys. Rev. B} \textbf{\bibinfo{volume}{71}},
  \bibinfo{pages}{094513} (\bibinfo{year}{2005}).

\bibitem[{\citenamefont{Asano et~al.}(2006)\citenamefont{Asano, Tanaka, and
  Kashiwaya}}]{Proximityp3}
\bibinfo{author}{\bibfnamefont{Y.}~\bibnamefont{Asano}},
  \bibinfo{author}{\bibfnamefont{Y.}~\bibnamefont{Tanaka}}, \bibnamefont{and}
  \bibinfo{author}{\bibfnamefont{S.}~\bibnamefont{Kashiwaya}},
  \bibinfo{journal}{Phys. Rev. Lett.} \textbf{\bibinfo{volume}{96}},
  \bibinfo{pages}{097007} (\bibinfo{year}{2006}).

\bibitem[{\citenamefont{Golubov and Kuprianov}(1988)}]{Golubov88}
\bibinfo{author}{\bibfnamefont{A.~A.} \bibnamefont{Golubov}} \bibnamefont{and}
  \bibinfo{author}{\bibfnamefont{M.~Y.} \bibnamefont{Kuprianov}},
  \bibinfo{journal}{J. Low Temp. Phys.} \textbf{\bibinfo{volume}{70}},
  \bibinfo{pages}{83} (\bibinfo{year}{1988}).

\bibitem[{\citenamefont{Belzig et~al.}(1996)\citenamefont{Belzig, Bruder, and
  Sch{\"o}n}}]{Belzig96}
\bibinfo{author}{\bibfnamefont{W.}~\bibnamefont{Belzig}},
  \bibinfo{author}{\bibfnamefont{C.}~\bibnamefont{Bruder}}, \bibnamefont{and}
  \bibinfo{author}{\bibfnamefont{G.}~\bibnamefont{Sch{\"o}n}},
  \bibinfo{journal}{Phys. Rev. B} \textbf{\bibinfo{volume}{54}},
  \bibinfo{pages}{9443} (\bibinfo{year}{1996}).

\bibitem[{\citenamefont{Asano and Tanaka}(2013)}]{Asano2013}
\bibinfo{author}{\bibfnamefont{Y.}~\bibnamefont{Asano}} \bibnamefont{and}
  \bibinfo{author}{\bibfnamefont{Y.}~\bibnamefont{Tanaka}},
  \bibinfo{journal}{Phys. Rev. B} \textbf{\bibinfo{volume}{87}},
  \bibinfo{pages}{104513} (\bibinfo{year}{2013}).

\bibitem[{\citenamefont{Ikegaya et~al.}(2016)\citenamefont{Ikegaya, Suzuki,
  Tanaka, and Asano}}]{Ikegaya2016}
\bibinfo{author}{\bibfnamefont{S.}~\bibnamefont{Ikegaya}},
  \bibinfo{author}{\bibfnamefont{S.~I.} \bibnamefont{Suzuki}},
  \bibinfo{author}{\bibfnamefont{Y.}~\bibnamefont{Tanaka}}, \bibnamefont{and}
  \bibinfo{author}{\bibfnamefont{Y.}~\bibnamefont{Asano}},
  \bibinfo{journal}{Phys. Rev. B} \textbf{\bibinfo{volume}{94}},
  \bibinfo{pages}{054512} (\bibinfo{year}{2016}).

\bibitem[{\citenamefont{Tanaka et~al.}(2012)\citenamefont{Tanaka, Sato, and
  Nagaosa}}]{tanaka12}
\bibinfo{author}{\bibfnamefont{Y.}~\bibnamefont{Tanaka}},
  \bibinfo{author}{\bibfnamefont{M.}~\bibnamefont{Sato}}, \bibnamefont{and}
  \bibinfo{author}{\bibfnamefont{N.}~\bibnamefont{Nagaosa}},
  \bibinfo{journal}{J. Phys. Soc. Jpn.} \textbf{\bibinfo{volume}{81}},
  \bibinfo{pages}{011013} (\bibinfo{year}{2012}).

\bibitem[{\citenamefont{Suzuki et~al.}(2019)\citenamefont{Suzuki, Golubov,
  Asano, and Tanaka}}]{Suzuki2019}
\bibinfo{author}{\bibfnamefont{S.-I.} \bibnamefont{Suzuki}},
  \bibinfo{author}{\bibfnamefont{A.~A.} \bibnamefont{Golubov}},
  \bibinfo{author}{\bibfnamefont{Y.}~\bibnamefont{Asano}}, \bibnamefont{and}
  \bibinfo{author}{\bibfnamefont{Y.}~\bibnamefont{Tanaka}},
  \bibinfo{journal}{Phys. Rev. B} \textbf{\bibinfo{volume}{100}},
  \bibinfo{pages}{024511} (\bibinfo{year}{2019}).

\bibitem[{\citenamefont{Tanaka and Golubov}(2007)}]{odd1}
\bibinfo{author}{\bibfnamefont{Y.}~\bibnamefont{Tanaka}} \bibnamefont{and}
  \bibinfo{author}{\bibfnamefont{A.~A.} \bibnamefont{Golubov}},
  \bibinfo{journal}{Phys. Rev. Lett.} \textbf{\bibinfo{volume}{98}},
  \bibinfo{pages}{037003} (\bibinfo{year}{2007}).

\bibitem[{\citenamefont{Berezinskii}(1974)}]{Berezinskii}
\bibinfo{author}{\bibfnamefont{V.~L.} \bibnamefont{Berezinskii}},
  \bibinfo{journal}{Pis'ma Zh. Eksp. Teor. Fiz.} \textbf{\bibinfo{volume}{20}},
  \bibinfo{pages}{628} (\bibinfo{year}{1974}), \bibinfo{note}{[JETP Lett.20 287
  (1974)]}.

\bibitem[{\citenamefont{Bergeret et~al.}(2005)\citenamefont{Bergeret, Volkov,
  and Efetov}}]{Efetov2}
\bibinfo{author}{\bibfnamefont{F.~S.} \bibnamefont{Bergeret}},
  \bibinfo{author}{\bibfnamefont{A.~F.} \bibnamefont{Volkov}},
  \bibnamefont{and} \bibinfo{author}{\bibfnamefont{K.~B.}
  \bibnamefont{Efetov}}, \bibinfo{journal}{Rev. Mod. Phys.}
  \textbf{\bibinfo{volume}{77}}, \bibinfo{pages}{1321} (\bibinfo{year}{2005}).

\bibitem[{\citenamefont{Linder and Balatsky}(2019)}]{LinderBalatsky}
\bibinfo{author}{\bibfnamefont{J.}~\bibnamefont{Linder}} \bibnamefont{and}
  \bibinfo{author}{\bibfnamefont{A.~V.} \bibnamefont{Balatsky}},
  \bibinfo{journal}{Rev. Mod. Phys.} \textbf{\bibinfo{volume}{91}},
  \bibinfo{pages}{045005} (\bibinfo{year}{2019}).

\bibitem[{\citenamefont{Cayao et~al.}(2020)\citenamefont{Cayao, Triola, and
  Black-Schaffer}}]{Cayao2020}
\bibinfo{author}{\bibfnamefont{J.}~\bibnamefont{Cayao}},
  \bibinfo{author}{\bibfnamefont{C.}~\bibnamefont{Triola}}, \bibnamefont{and}
  \bibinfo{author}{\bibfnamefont{A.~M.} \bibnamefont{Black-Schaffer}},
  \bibinfo{journal}{Eur. Phys. J. Special Topics}
  \textbf{\bibinfo{volume}{229}}, \bibinfo{pages}{545} (\bibinfo{year}{2020}).

\bibitem[{\citenamefont{Triola et~al.}(2020)\citenamefont{Triola, Cayao, and
  Black-Schaffer}}]{Triola2020}
\bibinfo{author}{\bibfnamefont{C.}~\bibnamefont{Triola}},
  \bibinfo{author}{\bibfnamefont{J.}~\bibnamefont{Cayao}}, \bibnamefont{and}
  \bibinfo{author}{\bibfnamefont{A.~M.} \bibnamefont{Black-Schaffer}},
  \bibinfo{journal}{Ann. Phys.} \textbf{\bibinfo{volume}{532}},
  \bibinfo{pages}{1900298} (\bibinfo{year}{2020}).

\bibitem[{\citenamefont{Tanaka et~al.}(2007{\natexlab{a}})\citenamefont{Tanaka,
  Golubov, Kashiwaya, and Ueda}}]{odd3}
\bibinfo{author}{\bibfnamefont{Y.}~\bibnamefont{Tanaka}},
  \bibinfo{author}{\bibfnamefont{A.~A.} \bibnamefont{Golubov}},
  \bibinfo{author}{\bibfnamefont{S.}~\bibnamefont{Kashiwaya}},
  \bibnamefont{and} \bibinfo{author}{\bibfnamefont{M.}~\bibnamefont{Ueda}},
  \bibinfo{journal}{Phys. Rev. Lett.} \textbf{\bibinfo{volume}{99}},
  \bibinfo{pages}{037005} (\bibinfo{year}{2007}{\natexlab{a}}).

\bibitem[{\citenamefont{Tanaka et~al.}(2007{\natexlab{b}})\citenamefont{Tanaka,
  Tanuma, and Golubov}}]{odd3b}
\bibinfo{author}{\bibfnamefont{Y.}~\bibnamefont{Tanaka}},
  \bibinfo{author}{\bibfnamefont{Y.}~\bibnamefont{Tanuma}}, \bibnamefont{and}
  \bibinfo{author}{\bibfnamefont{A.~A.} \bibnamefont{Golubov}},
  \bibinfo{journal}{Phys. Rev. B} \textbf{\bibinfo{volume}{76}},
  \bibinfo{pages}{054522} (\bibinfo{year}{2007}{\natexlab{b}}).

\bibitem[{\citenamefont{Eschrig et~al.}(2007)\citenamefont{Eschrig,
  L{\"o}fwander, Champel, Cuevas, and Sch{\"o}n}}]{Eschrig2007}
\bibinfo{author}{\bibfnamefont{M.}~\bibnamefont{Eschrig}},
  \bibinfo{author}{\bibfnamefont{T.}~\bibnamefont{L{\"o}fwander}},
  \bibinfo{author}{\bibfnamefont{T.}~\bibnamefont{Champel}},
  \bibinfo{author}{\bibfnamefont{J.}~\bibnamefont{Cuevas}}, \bibnamefont{and}
  \bibinfo{author}{\bibfnamefont{G.}~\bibnamefont{Sch{\"o}n}},
  \bibinfo{journal}{J. Low Temp. Phys.} \textbf{\bibinfo{volume}{147}},
  \bibinfo{pages}{457} (\bibinfo{year}{2007}).

\bibitem[{\citenamefont{Balatsky and Abrahams}(1992)}]{Balatsky}
\bibinfo{author}{\bibfnamefont{A.}~\bibnamefont{Balatsky}} \bibnamefont{and}
  \bibinfo{author}{\bibfnamefont{E.}~\bibnamefont{Abrahams}},
  \bibinfo{journal}{Phys. Rev. B} \textbf{\bibinfo{volume}{45}},
  \bibinfo{pages}{13125} (\bibinfo{year}{1992}).

\bibitem[{\citenamefont{Abrahams et~al.}(1995)\citenamefont{Abrahams, Balatsky,
  Scalapino, and Schrieffer}}]{Balatsky2}
\bibinfo{author}{\bibfnamefont{E.}~\bibnamefont{Abrahams}},
  \bibinfo{author}{\bibfnamefont{A.}~\bibnamefont{Balatsky}},
  \bibinfo{author}{\bibfnamefont{D.~J.} \bibnamefont{Scalapino}},
  \bibnamefont{and} \bibinfo{author}{\bibfnamefont{J.~R.}
  \bibnamefont{Schrieffer}}, \bibinfo{journal}{Phys. Rev. B}
  \textbf{\bibinfo{volume}{52}}, \bibinfo{pages}{1271} (\bibinfo{year}{1995}).

\bibitem[{\citenamefont{Coleman et~al.}(1997)\citenamefont{Coleman, Georges,
  and Tsvelik}}]{Coleman}
\bibinfo{author}{\bibfnamefont{P.}~\bibnamefont{Coleman}},
  \bibinfo{author}{\bibfnamefont{A.}~\bibnamefont{Georges}}, \bibnamefont{and}
  \bibinfo{author}{\bibfnamefont{A.~M.} \bibnamefont{Tsvelik}},
  \bibinfo{journal}{J. Phys. Condens. Matter} \textbf{\bibinfo{volume}{9}},
  \bibinfo{pages}{345} (\bibinfo{year}{1997}).

\bibitem[{\citenamefont{Vojta and Dagotto}(1999)}]{Vojta}
\bibinfo{author}{\bibfnamefont{M.}~\bibnamefont{Vojta}} \bibnamefont{and}
  \bibinfo{author}{\bibfnamefont{E.}~\bibnamefont{Dagotto}},
  \bibinfo{journal}{Phys. Rev. B} \textbf{\bibinfo{volume}{59}},
  \bibinfo{pages}{R713} (\bibinfo{year}{1999}).

\bibitem[{\citenamefont{Fuseya et~al.}(2003)\citenamefont{Fuseya, Kohno, and
  Miyake}}]{Fuseya}
\bibinfo{author}{\bibfnamefont{Y.}~\bibnamefont{Fuseya}},
  \bibinfo{author}{\bibfnamefont{H.}~\bibnamefont{Kohno}}, \bibnamefont{and}
  \bibinfo{author}{\bibfnamefont{K.}~\bibnamefont{Miyake}},
  \bibinfo{journal}{J. Phys. Soc. Jpn.} \textbf{\bibinfo{volume}{72}},
  \bibinfo{pages}{2914} (\bibinfo{year}{2003}).

\bibitem[{\citenamefont{Bergeret et~al.}(2001)\citenamefont{Bergeret, Volkov,
  and Efetov}}]{Efetov1}
\bibinfo{author}{\bibfnamefont{F.~S.} \bibnamefont{Bergeret}},
  \bibinfo{author}{\bibfnamefont{A.~F.} \bibnamefont{Volkov}},
  \bibnamefont{and} \bibinfo{author}{\bibfnamefont{K.~B.}
  \bibnamefont{Efetov}}, \bibinfo{journal}{Phys. Rev. Lett.}
  \textbf{\bibinfo{volume}{86}}, \bibinfo{pages}{4096} (\bibinfo{year}{2001}).

\bibitem[{\citenamefont{Asano et~al.}(2007)\citenamefont{Asano, Tanaka, and
  Golubov}}]{Asano2007PRL}
\bibinfo{author}{\bibfnamefont{Y.}~\bibnamefont{Asano}},
  \bibinfo{author}{\bibfnamefont{Y.}~\bibnamefont{Tanaka}}, \bibnamefont{and}
  \bibinfo{author}{\bibfnamefont{A.~A.} \bibnamefont{Golubov}},
  \bibinfo{journal}{Phys. Rev. Lett.} \textbf{\bibinfo{volume}{98}},
  \bibinfo{pages}{107002} (\bibinfo{year}{2007}).

\bibitem[{\citenamefont{Chiu et~al.}(2021)\citenamefont{Chiu, Tsuei, Yeh,
  Zhang, Kirchner, and Lin}}]{Lin2021}
\bibinfo{author}{\bibfnamefont{S.-P.} \bibnamefont{Chiu}},
  \bibinfo{author}{\bibfnamefont{C.~C.} \bibnamefont{Tsuei}},
  \bibinfo{author}{\bibfnamefont{S.-S.} \bibnamefont{Yeh}},
  \bibinfo{author}{\bibfnamefont{F.-C.} \bibnamefont{Zhang}},
  \bibinfo{author}{\bibfnamefont{S.}~\bibnamefont{Kirchner}}, \bibnamefont{and}
  \bibinfo{author}{\bibfnamefont{J.-J.} \bibnamefont{Lin}},
  \bibinfo{journal}{Science Advances} \textbf{\bibinfo{volume}{7}},
  \bibinfo{pages}{eabg6569} (\bibinfo{year}{2021}).

\bibitem[{\citenamefont{Tanaka}(2021)}]{TextTanaka2021}
\bibinfo{author}{\bibfnamefont{Y.}~\bibnamefont{Tanaka}},
  \emph{\bibinfo{title}{Physics of Superconducting junctions}}
  (\bibinfo{publisher}{Nagoya University Press}, \bibinfo{year}{2021}).

\bibitem[{\citenamefont{Bauer et~al.}(2004)\citenamefont{Bauer, Hilscher,
  Michor, Paul, Scheidt, Gribanov, Seropegin, No{\"e}l, Sigrist, and
  Rogl}}]{Bauer}
\bibinfo{author}{\bibfnamefont{E.}~\bibnamefont{Bauer}},
  \bibinfo{author}{\bibfnamefont{G.}~\bibnamefont{Hilscher}},
  \bibinfo{author}{\bibfnamefont{H.}~\bibnamefont{Michor}},
  \bibinfo{author}{\bibfnamefont{C.}~\bibnamefont{Paul}},
  \bibinfo{author}{\bibfnamefont{E.~W.} \bibnamefont{Scheidt}},
  \bibinfo{author}{\bibfnamefont{A.}~\bibnamefont{Gribanov}},
  \bibinfo{author}{\bibfnamefont{Y.}~\bibnamefont{Seropegin}},
  \bibinfo{author}{\bibfnamefont{H.}~\bibnamefont{No{\"e}l}},
  \bibinfo{author}{\bibfnamefont{M.}~\bibnamefont{Sigrist}}, \bibnamefont{and}
  \bibinfo{author}{\bibfnamefont{P.}~\bibnamefont{Rogl}},
  \bibinfo{journal}{Phys. Rev. Lett.} \textbf{\bibinfo{volume}{92}},
  \bibinfo{pages}{027003} (\bibinfo{year}{2004}).

\bibitem[{\citenamefont{Gor'kov and Rashba}(2001)}]{Gorkov}
\bibinfo{author}{\bibfnamefont{L.~P.} \bibnamefont{Gor'kov}} \bibnamefont{and}
  \bibinfo{author}{\bibfnamefont{E.~I.} \bibnamefont{Rashba}},
  \bibinfo{journal}{Phys. Rev. Lett.} \textbf{\bibinfo{volume}{87}},
  \bibinfo{pages}{037004} (\bibinfo{year}{2001}).

\bibitem[{\citenamefont{Frigeri et~al.}(2004)\citenamefont{Frigeri, Agterberg,
  Koga, and Sigrist}}]{Frigeri}
\bibinfo{author}{\bibfnamefont{P.~A.} \bibnamefont{Frigeri}},
  \bibinfo{author}{\bibfnamefont{D.~F.} \bibnamefont{Agterberg}},
  \bibinfo{author}{\bibfnamefont{A.}~\bibnamefont{Koga}}, \bibnamefont{and}
  \bibinfo{author}{\bibfnamefont{M.}~\bibnamefont{Sigrist}},
  \bibinfo{journal}{Phys. Rev. Lett.} \textbf{\bibinfo{volume}{92}},
  \bibinfo{pages}{097001} (\bibinfo{year}{2004}).

\bibitem[{\citenamefont{Fujimoto}(2007)}]{Fujimoto1}
\bibinfo{author}{\bibfnamefont{S.}~\bibnamefont{Fujimoto}},
  \bibinfo{journal}{J. Phys. Soc. Jpn.} \textbf{\bibinfo{volume}{76}},
  \bibinfo{pages}{051008} (\bibinfo{year}{2007}).

\bibitem[{\citenamefont{Vorontsov et~al.}(2008)\citenamefont{Vorontsov,
  Vekhter, and Eschrig}}]{VVE08}
\bibinfo{author}{\bibfnamefont{A.~B.} \bibnamefont{Vorontsov}},
  \bibinfo{author}{\bibfnamefont{I.}~\bibnamefont{Vekhter}}, \bibnamefont{and}
  \bibinfo{author}{\bibfnamefont{M.}~\bibnamefont{Eschrig}},
  \bibinfo{journal}{Phys. Rev. Lett.} \textbf{\bibinfo{volume}{101}},
  \bibinfo{pages}{127003} (\bibinfo{year}{2008}).

\bibitem[{\citenamefont{Tanaka et~al.}(2009)\citenamefont{Tanaka, Yokoyama,
  Balatsky, and Nagaosa}}]{TYBN09}
\bibinfo{author}{\bibfnamefont{Y.}~\bibnamefont{Tanaka}},
  \bibinfo{author}{\bibfnamefont{T.}~\bibnamefont{Yokoyama}},
  \bibinfo{author}{\bibfnamefont{A.~V.} \bibnamefont{Balatsky}},
  \bibnamefont{and} \bibinfo{author}{\bibfnamefont{N.}~\bibnamefont{Nagaosa}},
  \bibinfo{journal}{Phys. Rev. B} \textbf{\bibinfo{volume}{79}},
  \bibinfo{pages}{060505} (\bibinfo{year}{2009}).

\bibitem[{\citenamefont{Annunziata et~al.}(2012)\citenamefont{Annunziata,
  Manske, and Linder}}]{Gaetano}
\bibinfo{author}{\bibfnamefont{G.}~\bibnamefont{Annunziata}},
  \bibinfo{author}{\bibfnamefont{D.}~\bibnamefont{Manske}}, \bibnamefont{and}
  \bibinfo{author}{\bibfnamefont{J.}~\bibnamefont{Linder}},
  \bibinfo{journal}{Phys. Rev. B} \textbf{\bibinfo{volume}{86}},
  \bibinfo{pages}{174514} (\bibinfo{year}{2012}).

\bibitem[{\citenamefont{Mishra et~al.}(2021)\citenamefont{Mishra, Li, Zhang,
  and Kirchner}}]{Mishra}
\bibinfo{author}{\bibfnamefont{V.}~\bibnamefont{Mishra}},
  \bibinfo{author}{\bibfnamefont{Y.}~\bibnamefont{Li}},
  \bibinfo{author}{\bibfnamefont{F.-C.} \bibnamefont{Zhang}}, \bibnamefont{and}
  \bibinfo{author}{\bibfnamefont{S.}~\bibnamefont{Kirchner}},
  \bibinfo{journal}{Phys. Rev. B} \textbf{\bibinfo{volume}{103}},
  \bibinfo{pages}{184505} (\bibinfo{year}{2021}).

\bibitem[{\citenamefont{Zaitsev}(1984)}]{Zaitsev1984}
\bibinfo{author}{\bibfnamefont{A.~V.} \bibnamefont{Zaitsev}},
  \bibinfo{journal}{Zh. Eksp. Teor. Fiz.} \textbf{\bibinfo{volume}{86}},
  \bibinfo{pages}{1742} (\bibinfo{year}{1984}), \bibinfo{note}{[Sov. Phys. JETP
  59, 1163 (1984)]}.

\bibitem[{\citenamefont{Kashiwaya et~al.}(1996)\citenamefont{Kashiwaya, Tanaka,
  Koyanagi, and Kajimura}}]{KT96}
\bibinfo{author}{\bibfnamefont{S.}~\bibnamefont{Kashiwaya}},
  \bibinfo{author}{\bibfnamefont{Y.}~\bibnamefont{Tanaka}},
  \bibinfo{author}{\bibfnamefont{M.}~\bibnamefont{Koyanagi}}, \bibnamefont{and}
  \bibinfo{author}{\bibfnamefont{K.}~\bibnamefont{Kajimura}},
  \bibinfo{journal}{Phys. Rev. B} \textbf{\bibinfo{volume}{53}},
  \bibinfo{pages}{2667} (\bibinfo{year}{1996}).

\bibitem[{\citenamefont{Tanaka et~al.}(2003{\natexlab{b}})\citenamefont{Tanaka,
  Golubov, and Kashiwaya}}]{TGK}
\bibinfo{author}{\bibfnamefont{Y.}~\bibnamefont{Tanaka}},
  \bibinfo{author}{\bibfnamefont{A.~A.} \bibnamefont{Golubov}},
  \bibnamefont{and}
  \bibinfo{author}{\bibfnamefont{S.}~\bibnamefont{Kashiwaya}},
  \bibinfo{journal}{Phys. Rev. B} \textbf{\bibinfo{volume}{68}},
  \bibinfo{pages}{054513} (\bibinfo{year}{2003}{\natexlab{b}}).

\bibitem[{\citenamefont{Burset et~al.}(2014)\citenamefont{Burset, Keidel,
  Tanaka, Nagaosa, and Trauzettel}}]{Bursetchiral}
\bibinfo{author}{\bibfnamefont{P.}~\bibnamefont{Burset}},
  \bibinfo{author}{\bibfnamefont{F.}~\bibnamefont{Keidel}},
  \bibinfo{author}{\bibfnamefont{Y.}~\bibnamefont{Tanaka}},
  \bibinfo{author}{\bibfnamefont{N.}~\bibnamefont{Nagaosa}}, \bibnamefont{and}
  \bibinfo{author}{\bibfnamefont{B.}~\bibnamefont{Trauzettel}},
  \bibinfo{journal}{Phys. Rev. B} \textbf{\bibinfo{volume}{90}},
  \bibinfo{pages}{085438} (\bibinfo{year}{2014}).

\bibitem[{\citenamefont{Yokoyama et~al.}(2007)\citenamefont{Yokoyama, Tanaka,
  and Golubov}}]{Yokoyama2007}
\bibinfo{author}{\bibfnamefont{T.}~\bibnamefont{Yokoyama}},
  \bibinfo{author}{\bibfnamefont{Y.}~\bibnamefont{Tanaka}}, \bibnamefont{and}
  \bibinfo{author}{\bibfnamefont{A.~A.} \bibnamefont{Golubov}},
  \bibinfo{journal}{Phys. Rev. B} \textbf{\bibinfo{volume}{75}},
  \bibinfo{pages}{134510} (\bibinfo{year}{2007}).

\bibitem[{\citenamefont{Eschrig}(2015)}]{Eschrig2015}
\bibinfo{author}{\bibfnamefont{M.}~\bibnamefont{Eschrig}},
  \bibinfo{journal}{Rep. Prog. Phys.} \textbf{\bibinfo{volume}{78}},
  \bibinfo{pages}{104501} (\bibinfo{year}{2015}).

\bibitem[{\citenamefont{Linder and Robinson}(2015)}]{Linder2015}
\bibinfo{author}{\bibfnamefont{J.}~\bibnamefont{Linder}} \bibnamefont{and}
  \bibinfo{author}{\bibfnamefont{J.~W.~A.} \bibnamefont{Robinson}},
  \bibinfo{journal}{Nat. Phys.} \textbf{\bibinfo{volume}{11}},
  \bibinfo{pages}{307} (\bibinfo{year}{2015}).

\end{thebibliography}
\end{document}